\documentclass[12pt]{article}
\usepackage{amsmath,amssymb}
\usepackage{authblk}
\usepackage{hyperref}
\usepackage[top=26truemm,bottom=26truemm,left=24truemm,right=24truemm]{geometry}
\usepackage{setspace}
\usepackage{amsmath,amssymb,mathrsfs,amsbsy,latexsym,amsfonts,amsthm, ascmac}

\usepackage[Symbol]{upgreek}

\pdfoutput=1 
\usepackage{graphicx}
\usepackage{color}
\usepackage{braket}
\usepackage{here}

\usepackage[utf8]{inputenc}

\newcommand{\nn}{\nonumber\\}
\newcommand{\vx}{\vec{x}}

\DeclareMathOperator{\tr}{tr}
\DeclareMathOperator{\sgn}{sgn}

\usepackage{epsf}

\numberwithin{equation}{section}

\begin{document}
\setlength{\abovedisplayskip}{15pt}
\setlength{\belowdisplayskip}{15pt}

\title{\textbf{
Target space entanglement\\
in quantum mechanics of fermions\\
at finite temperature
}
}

\author[1]{
	Temma~Hanyuda\thanks{\tt hanyuda.temma.i3(at)s.mail.nagoya-u.ac.jp}
}
\author[1]{
	Soichiro~Mori\thanks{\tt mori.soichiro.f1(at)s.mail.nagoya-u.ac.jp}
}
\author[1, 2]{
	Sotaro~Sugishita\thanks{
	\tt sugishita.sotaro.r6(at)f.mail.nagoya-u.ac.jp}
	\vspace{5mm}
}
\affil[1]{\it\normalsize Department of Physics, 
Nagoya University,
Nagoya, Aichi 464-8602, Japan}
\affil[2]{\it\normalsize Institute for Advanced Research, 
Nagoya University,
Nagoya, Aichi 464-8601, Japan}
\setcounter{Maxaffil}{0}

\date{ }

\maketitle
\begin{abstract}
We consider the target space entanglement in quantum mechanics of non-interacting fermions at finite temperature. Unlike pure states investigated in \cite{Sugishita:2021vih}, the (R\'enyi) entanglement entropy for thermal states does not follow a simple bound because all states in the infinite-dimensional Hilbert space are involved. 
We investigate a general formula of the target space R\'enyi entropy for $N$ fermions at finite temperature, and present numerical results of the entropy in a one-dimensional model.
We also argue the large $N$ behaviors with a comparison to the grand canonical ensemble.
\end{abstract}
\newpage
\setcounter{tocdepth}{2}
\tableofcontents
\newpage
\section{Introduction}
In quantum field theories, it is interesting to consider the entanglement entropy for subregions in the base space of the QFTs. 
In particular in the AdS/CFT correspondence,
if theories have the gravity duals, the (base space) entanglement entropy is well approximated by area of surfaces in the bulk by the Ryu-Takayanagi formula \cite{Ryu:2006bv, Ryu:2006ef}.
The formula relates entanglement in QFTs to the  geometry in the gravitational theories.

However, we do not have the notion of base space entanglement for an important class of holographic theories, \textit{i.e.}, matrix models.
For example, the BFSS model \cite{Banks:1996vh}, which is believed to have a gravitational dual, is $(1+0)$-dimensional matrix quantum mechanics and does not have the spatial base space.
We cannot use the Ryu-Takayanagi formula for theories without base space. 
On the other hand, it is expected that we generally have a relation between geometry and entanglement in quantum gravity \cite{Sorkin:2014kta, Srednicki:1993im, Jacobson:2012yt, Bianchi:2012ev, Myers:2013lva, Balasubramanian:2013lsa, Myers:2014jia, Balasubramanian:2014sra, Jacobson:2015hqa, Anous:2019rqb}.
To discuss such a relation for holographic matrix models, we need another concept of entanglement rather than base space entanglement.

One proposal is the target space entanglement 
\cite{Mazenc:2019ety, Das:2020jhy, Das:2020xoa, Hampapura:2020hfg, Sugishita:2021vih, Frenkel:2021yql, Tsuchiya:2022ffu, Das:2022mtb}.\footnote{See also \cite{Gautam:2022akq} for another proposal of entanglement in matrix models.} 
In holographic matrix models, the target spaces of matrix elements can be interpreted as bulk space, e.g. the space where D0-branes moves. 
Thus, it is natural to consider entanglement associated with partition of target spaces. 
A problem to define the target space entanglement is that the Hilbert space is generally not tensor-factorized with respect to the target space. 
Nevertheless, we can associate a subalgebra of operators to any subregion in the target space \cite{Mazenc:2019ety}. 
We can then define the reduced density matrix associated with the subalgebra, and the entanglement entropy.  

Hence, it is interesting to investigate the target space entanglement of matrix models. 
A simple toy model is one-matrix models.
It is known that a two-dimensional string theory is dual to a one-matrix model with a class of potential (see, e.g., \cite{Klebanov:1991qa, Polchinski:1994mb}). 
The dynamics of one-matrix models is reduced to that of non-interacting non-relativistic fermions by diagonalizing the matrix (see also, e.g., \cite{Klebanov:1991qa, Polchinski:1994mb}).
Thus, by considering the entanglement of non-relativistic fermions, we can understand the entanglement structure of a matrix model and a string theory \cite{Das:1995vj, Hartnoll:2015fca}.

In \cite{Sugishita:2021vih}, one of the authors developed the target space entanglement of fermions in pure states, especially in the ground states. 
It is shown that, for pure states with the Slater determinant wave functions, the target space entanglement entropy is bounded as $S\leq N \log 2$ independently of the  details of the subregion and the states, where $N$ is the number of fermions.
The derivation cannot be applied to mixed states. 
It is also shown in \cite{Sugishita:2021vih} that the target space R\'enyi entropy for the ground state of $N$ free fermions on one-dimensional circle can be computed analytically in the large $N$ limit. 
It is not sure whether we can perform such an analytical computations for other states. 

Thus, it is worth extending the analysis in \cite{Sugishita:2021vih} to mixed states. 
One important class of mixed states is thermal states. 
The aim of this paper is to investigate the target space entanglement for non-interacting $N$ fermions at finite temperature.
We will see that the target space entanglement entropy for thermal states does not follow a simple formula in \cite{Sugishita:2021vih} which that for pure states does. 
Because of this difficulty, it is hard to analytically investigate general properties of the target space  entanglement for thermal states. 
Hence, instead of analytical computations, we will develop  formulae of the target space R\'enyi entropy which are useful for numerical computations.
Using the formulae, we will numerically study the target space R\'enyi entropy for thermal states in a simple model.

The paper is organized as follows. 
In section~\ref{sec:basic}, we summarize the basic setup for quantum mechanics of fermions and  the definition of the target space R\'enyi entropy. 
In section~\ref{sec:Renyi_T}, we develop a formula of the target space R\'enyi entropy, in particular the 2nd R\'enyi entropy, for non-interacting $N$ fermions at finite temperature. 
In section~\ref{sec:circle}, we apply the obtained formula to a concrete model which is free $N$ fermions on a one-dimensional circle, and present the numerical results of the R\'enyi entropy.
\section{Basic setup}\label{sec:basic}

\subsection{Quantum mechanics of $N$ fermions}\label{sec:QM-fermion}

We consider  quantum mechanics of non-interacting $N$ fermions. 
Let $\mathcal{H}$ be the Hilbert space of a single particle. The Hilbert space of $N$ particles, $\mathcal{H}_N$, is obtained by anti-symmetrizing $N$ tensor products of $\mathcal{H}$, i.e., $\mathcal{H}_N$ is the $N$-th exterior power of $\mathcal{H}$ as
\begin{align}
    \mathcal{H}_N=\bigwedge\nolimits^{\!N} \mathcal{H}.
\end{align}
We introduce the projection $P^-$ from $\mathcal{H}^{\otimes N}$ onto $\bigwedge\nolimits^{\!N} \mathcal{H}$ as 
\begin{align}
\label{eq:P-}
    P^{-}\left(\ket{\psi_1}\otimes\cdots\otimes\ket{\psi_N}\right) &:=\frac{1}{N!}\sum_{\sigma\in S_N}(-)^{\sigma}\ket{\psi_{\sigma(1)}}\otimes\cdots\otimes\ket{\psi_{\sigma(N)}},
\end{align}
where $(-)^\sigma$ denotes the sign of the permutation $\sigma$, i.e., $(-)^\sigma=\sgn\sigma$.
We also represent the anti-symmetrized states using the wedge product as 
\begin{align}
    P^{-}\left(\ket{\psi_1}\otimes\cdots\otimes\ket{\psi_N}\right) &:=\ket{\psi_1}\wedge\cdots\wedge\ket{\psi_N}.
\end{align}

We label energy eigenstates of a single particle by integers $n$, as $\ket{n} \in \mathcal{H}$. 
Then, energy eigenstates of non-interacting $N$ fermions are labeled by sets of $N$ distinct integers $I=\{n_1, \cdots, n_N\}$, where we suppose that $n_1<\cdots<n_N$, as
\begin{align}
\label{def:I-basis}
    \ket{I}=\frac{1}{\sqrt{N!}}\sum_{\sigma \in  S_N}(-)^\sigma \ket{n_{\sigma(1)}} \otimes \cdots \otimes  \ket{n_{\sigma(N)}}
    =\sqrt{N!}\ket{n_1}\wedge\cdots\wedge\ket{n_N},
\end{align}
which are normalized as $\braket{I|J}=\delta_{I,J}$.

It is also convenient to introduce the anti-symmetrized operators. 
Let $A_1, \cdots, A_N$ be operators on $\mathcal{H}$. We then define a class of operators on $\bigwedge\nolimits^{\!N} \mathcal{H}$ as 
\begin{align}
\label{def:wedge-op}
    A_1 \wedge \cdots \wedge A_N:= P^- (A_1 \otimes \cdots \otimes A_N) P^-.
\end{align}
The matrix elements in the basis \eqref{def:I-basis} are given by
\begin{align}
   (A_1 \wedge \dots \wedge A_N)_{I,I'}
   &:= \bra{I}A_1 \wedge \cdots \wedge A_N\ket{I'}
   \nn
   & =\frac{1}{N!} \sum_{\sigma, \sigma' \in S_N}(-)^{\sigma\sigma'}
    (A_1)_{n_{\sigma(1)},n'_{\sigma'(1)}}
    \dots (A_N)_{n_{\sigma(N)},n'_{\sigma'(N)}}.
\end{align}
We summarize useful formulae of the wedge products, e.g. the trace and the product of \eqref{def:wedge-op}, in appendix~\ref{app:wedge}.

\subsubsection{Fermions at finite temperature}
Our aim is to develop the target space entanglement for $N$ fermions at finite temperature $1/\beta$. 
Let $\ket{n}$ be energy eigenstates of the Hamiltonian $H_1$ for a single particle. 
Then, the states \eqref{def:I-basis} are energy eigenstates for non-interacting $N$ fermions with energy $E_I=E_{n_1}+\cdots E_{n_N}$.
Using this basis, the thermal density matrix for $N$ fermions is given by
\begin{align}
\label{rhoN-sumI}
    \rho_{N}=\frac{1}{Z_N}\sum_{I}e^{-\beta E_{I}}\ket{I}\bra{I}
\end{align}
with $Z_N(\beta)=\sum_I e^{-\beta E_{I}}$.
The density matrix has a basis-independent expression using the wedge product as 
\begin{align}
    \rho_{N}=\frac{1}{Z_N(\beta)}\left(e^{-\beta H_1}\wedge \cdots \wedge e^{-\beta H_1}\right)
    \label{th-den-mat}
\end{align}
with
\begin{align}
\label{ZN_wedge}
    Z_N(\beta)=\tr\left(e^{-\beta H_1}\wedge \cdots \wedge e^{-\beta H_1}\right).
\end{align}
From the formula \eqref{tr-wedge_same}, $Z_N(\beta)$ can be computed from the single-particle partition function $Z_1(\beta)$ as
\begin{align}
    Z_N(\beta)=
\frac{1}{N!}\left|\begin{array}{ccccc}
Z_1(\beta) & N-1 & 0 & \cdots & \\
Z_1(2\beta) & Z_1(\beta) & N-2 & \cdots & \\
\vdots & \vdots & & \ddots & \vdots \\
Z_1((N-1)\beta) &Z_1((N-2)\beta) & & \cdots & 1 \\
Z_1(N\beta) & Z_1((N-1)\beta)  & & \cdots & Z_1(\beta)
\end{array}\right|.
\label{ZN-det}
\end{align}

The thermal $n$-th R\'enyi entropy of $\rho_{N}$ is given by 
\begin{align}
    S_{th,(N)}^{(n)}:=\frac{\log \tr\left( \rho_{N}^n\right)}{1-n}.\label{thermal renyi entropy}
\end{align}
This is reduced to the thermal entropy $-\tr(\rho_{N}\log\rho_{N})$ in the limit $n\to 1$. By using the product formula \eqref{prod-wedge}, we have
\begin{align}
    \rho_{N}^n=\frac{1}{Z_N(\beta)^n}\left(e^{-n\beta H_1}\wedge \cdots \wedge e^{-n\beta H_1}\right).
\end{align}
It means 
\begin{align}
    \tr\left( \rho_{N}^n\right)
    =\frac{Z_N(n\beta)}{Z_N(\beta)^n}.
\end{align}
Accordingly, the thermal $n$-th R\'enyi entropy \eqref{thermal renyi entropy} is written as 
\begin{align}
\label{n-renyi-thermal}
   S_{th,(N)}^{(n)}=\frac{\log Z_N(n\beta)-n\log Z_N(\beta)}{1-n}. 
\end{align}
In particular, taking the limit $n\to 1$, the thermal entropy is given by
\begin{align}
  S_{th,(N)}^{(1)}  =\log Z_N(\beta)-\beta\partial_\beta \log Z_N(\beta),
\end{align}
which represents the standard thermal relation $S=\beta (-F+E)$.

\subsection{Definition of the target space R\'enyi entanglement entropy}
Here we briefly review the definition of the target space entanglement for quantum mechanics of $N$ fermions (see \cite{Sugishita:2021vih} for details).

Let $M$ be the entire position space where fermions live, i.e., the target space of fermions. 
We take a subregion $A$ in $M$, and define the R\'enyi entanglement entropy of $A$ for a given total density matrix $\rho_N$.
A difficulty defining the entropy is that the total Hilbert space $\mathcal{H}_N$ (and also the single particle space $\mathcal{H}$) are not tensor-factorized with respect to the position space in the first quantized picture.
However, by adopting an algebraic definition of entanglement, we can define the target space  R\'enyi entanglement entropy.

The key point is decomposing $\mathcal{H}_N$ into the direct sum of subsectors as $\mathcal{H}_N=\bigoplus_{k=0}^N \mathcal{H}_{k,N}$ where $k$-th sector $\mathcal{H}_{k,N}$ consists of states with $k$ fermions in $A$ and the other $(N-k)$ fermions in the complement $\bar{A}$.
Let $\Pi_A,\Pi_{\bar{A}}$ be projections for a single particle onto $A,\bar{A}$ respectively as 
\begin{align}
    \Pi_{A}:=\int_{A}dx\ket{x}\bra{x},\qquad \Pi_{\bar{A}}:=\int_{\bar{A}}dx\ket{x}\bra{x}.
\end{align}
Then, the projection onto $k$-th sector,  $\Pi_k:\mathcal{H}\rightarrow\mathcal{H}_{k,N}$, is given by
\begin{align}
    \Pi_k:=\binom{N}{k}P^{-}\left(\Pi_A^{\otimes k}\otimes\Pi_{\bar{A}}^{\otimes(N-k)}\right)P^{-}
   = \binom{N}{k} \underbrace{\Pi_A\wedge\cdots \wedge \Pi_A}_{k}\wedge\underbrace{\Pi_{\bar{A}}\cdots\wedge\Pi_{\bar{A}}}_{N-k},
    \label{projection}
\end{align}
where $P^-$ is the anti-symmetrizing projection defined in \eqref{eq:P-}.

For a given density matrix $\rho_N$, the probability resulting in sector $\mathcal{H}_{k,N}$ is 
\begin{align}
    p_k=\tr(\Pi_k\rho_N\Pi_k)
    \label{probability}
\end{align}
which satisfies $0\leq p_k\leq 1$ and  $\sum_{k=0}^Np_k=\tr(\rho_N\sum_k\Pi_k)=\tr(\rho_N)=1$.
We also define the projected density matrix on $\mathcal{H}_{k,N}$ as 
\begin{align}
    \rho_k:=\frac{1}{p_k}\Pi_k\rho_N\Pi_k,
    \label{densitymatrixofeachk}
\end{align}
which is normalized as $\tr\rho_k=1$. 
We can compute the reduced density matrix of $\rho_k$ by taking the partial trace over $\bar{A}$,
\begin{align}
\label{def:rho_k}
    \rho_{k,A}=\tr_{\bar{A}}\rho_k,
\end{align}
because, in each sector $\mathcal{H}_{k,N}$, the degrees of freedom in $A$ are factorized from those in $\bar{A}$ up to permutations.\footnote{See subsec.~\ref{subsec:partial} for the concrete computations of the reduced density matrix.}

Then, the (von Neumann) entanglement entropy for subregion $A$ is defined as
\begin{align}
    S^{(1)}:=-\sum_{k=0}^N p_k\log p_k
    +\sum_{k=0}^N p_kS_{A_k}(\rho_k),
    \label{entropy}
\end{align}
where $S_{A_k}(\rho_k)$ is the entanglement entropy of the projected density matrix on $k$-th sector $\mathcal{H}_{k,N}$ given by
\begin{align}
\label{def:S_Ak}
    S_{A_k}(\rho_k)=-\tr_{A}\rho_{k,A}\log\rho_{k,A}.
\end{align}
Eq.~\eqref{entropy} consists of two parts. 
The first term in \eqref{entropy} represents a classical part, that is, the Shannon entropy 
\begin{align}
\label{def:S_cl}
    S^{(1)}_{cl}=-\sum_{k=0}^Np_k\log p_k
\end{align}
for the probability distribution $\{p_k\}$.
The second term in \eqref{entropy}, 
\begin{align}
\label{def:S_q}
    S_{q}=\sum_{k=0}^N p_{k}\, S_{A_k}(\rho_k),
\end{align}
is called a quantum part which is the average of the $k$-th sector entanglement entropy $S_{A_k}(\rho_k)$ for each sector with the probability distribution $\{p_k\}$. 
The target space R\'enyi entropy for subregion $A$ is also defined as
\begin{align}
    S^{(n)}:=\frac{\log\tr_A\rho_A^n}{1-n}=\frac{\log\sum_{k=0}^Np_k^n\tr_{A}\rho_{k,A}^n}{1-n}.
    \label{renyi-entropy}
\end{align}
The limit $\lim_{n\rightarrow 1}S^{(n)}$ agrees with the entanglement entropy \eqref{entropy}. 

In \cite{Sugishita:2021vih}, it is shown that, if $\rho_N$ is a pure state whose wave function is given by the Slater determinant, the entanglement entropy \eqref{entropy} and also the R\'enyi entropy \eqref{renyi-entropy} follow simple formulae, and have the following upper bound
\begin{align}
\label{up-bound}
    S^{(n)}\leq N \log 2,
\end{align}
independently of the R\'enyi parameter $n$ and choices of the subregion.
However, the derivation of the bound \eqref{up-bound} cannot be applied to general mixed states $\rho_N$. 
Thus, it is not sure whether the R\'enyi entropy for thermal states of $N$ fermions follows the bound \eqref{up-bound}.
We will discuss that the bound \eqref{up-bound} does not hold for general thermal states.

\subsection{Probability for finite temperature systems}
We consider the probability $p_k$ of finding $k$ particles in region $A$ for thermal state \eqref{th-den-mat}.
By definition \eqref{probability}, $p_k$ is given by
\begin{align}
    p_k&=\mathrm{tr}\left(\Pi_k\rho_N\Pi_k\right)=\mathrm{tr}\left(\rho_N\Pi_k\right)
    \nn
    &=\binom{N}{k}\mathrm{tr}(\rho_NP^-\Pi_A^{(N,k)}P^-)
    =\binom{N}{k}\mathrm{tr}(\rho_N\Pi_A^{(N,k)}),
    \label{pk:trace}
\end{align}
where $\Pi_A^{(N,k)}:=\Pi_A^{\otimes k}\otimes\Pi_{\bar{A}}^{\otimes(N-k)}$, and we have used $\rho_N=P^{-}\rho_NP^{-}$ because $\rho_N$ is anti-symmetric as \eqref{th-den-mat}.
In the energy eigenstate basis \eqref{def:I-basis},  \eqref{pk:trace} is computed as 
\begin{align}
     p_k&=\binom{N}{k}\sum_{I}\braket{I|\rho_N\Pi_A^{(N,k)}|I}
    =\frac{\binom{N}{k}}{Z_N}\sum_{I}e^{-\beta(E_{n_1}+\cdots+E_{n_N})}\braket{I|\Pi_A^{(N,k)}|I}\nn
    &=\frac{\binom{N}{k}}{(N!)^2 Z_N}\sum_{n_1,\cdots,n_N}\sum_{\sigma,\sigma' \in S_N}(-)^{\sigma\sigma'}\prod_{i=1}^{k}e^{-\beta E_{n_i}}\braket{n_{\sigma_i}|\Pi_A|n_{{\sigma'_i}}}\prod_{j=k+1}^{N}e^{-\beta E_{n_j}}\braket{n_{\sigma_{j}}|\Pi_{\bar{A}}|n_{\sigma'_{j}}}.
\end{align}
It is useful to introduce the weighted overlap matrices \begin{align}
    Y_{m,n}:=e^{-\frac{\beta}{2}E_m}\braket{m|\Pi_A|n}e^{-\frac{\beta}{2}E_n},
    \quad
    \bar{Y}_{m,n}:=e^{-\frac{\beta}{2}E_m}\braket{m|\Pi_{\bar{A}}|n}e^{-\frac{\beta}{2}E_n}.
    \label{def_of_y}
\end{align}
The probability $p_k$ is then expressed by the wedge product as
\begin{align}
    p_k&=\frac{\binom{N}{k}}{(N!)^2 Z_N}\sum_{n_1,\cdots,n_N}\sum_{\sigma,\sigma' \in S_N}(-)^{\sigma\sigma'}\prod_{i=1}^{k}Y_{n_{\sigma_i},n_{\sigma'_i}}\prod_{j=k+1}^{N}\bar{Y}_{n_{\sigma_j},n_{\sigma'_j}}\nn
    &=\frac{\binom{N}{k}}{Z_N}\tr(\underbrace{Y\wedge\cdots \wedge Y}_{k}\wedge\underbrace{\bar{Y}\cdots\wedge\bar{Y}}_{N-k}),\label{probabilitywedge}
\end{align}
where the formula for the trace of wedge products of operators is given in appendix~\ref{app:wedge}.
\section{R\'enyi entropy for $N$ fermions at finite temperature}\label{sec:Renyi_T}

The target space (R\'enyi) entanglement entropy is defined by \eqref{entropy} and \eqref{renyi-entropy}.
Our aim is to compute it for $N$ fermions at temperature $1/\beta$.
As we will see in the next subsection, it is easy for a single particle case $(N=1)$ to compute the entropies  \eqref{entropy} and \eqref{renyi-entropy}.
For multiple fermions, it is also easy to compute the classical part \eqref{def:S_cl} (at least numerically) because probabilities $\{p_k\}$ can be computed by \eqref{probabilitywedge}.

However, for the following reason, it is difficult to compute the quantum part \eqref{def:S_q} for multiple fermions.
When we would like to perform numerical computations, we have to introduce a cutoff making the dimension of Hilbert spaces finite. 
Let the single particle Hilbert space $\mathcal{H}$ be cut off so that the dimension is $d_{\mathrm{max}}<\infty$. 
Then, the dimension of the Hilbert space $\mathcal{H}_N$ for $N$ fermions is $\binom{d_{\mathrm{max}}}{N}$ which is combinatorially large. 
Thus, since the reduced density matrix $\rho_{k,A}$ in \eqref{def:S_Ak} also has combinatorially large size, it is difficult to diagonalize it and to numerically compute $S( \rho_{k})$ in the quantum part \eqref{def:S_q}. 
Similarly, it is difficult to compute the R\'enyi entropy $S^{(n)}$ with general parameters $n$.
Nevertheless, for integer $n$, we can show that $S^{(n)}$ reduces to a combination of computations of matrices with size $d_{\mathrm{max}}$.
More precisely, $S^{(n)}$ with integers $n$ can be written by using the weighted overlap matrices $Y, \bar{Y}$ like the probability $p_k$ in \eqref{probabilitywedge}. 
For simplicity, we will investigate the case $n=2$, that is, the 2nd R\'enyi entropy
\begin{align}
    S^{(2)}=-\log\left(\sum_{k=0}^{N}p_{k}^2\tr_{A_k}\rho_{k,A}^2\right).
    \label{2ndrenyi}
\end{align}
The formula for $S^{(2)}$ at finite temperature is given by \eqref{Formula_of_Renyi} which is in fact written by the weighted overlap matrices \eqref{def_of_y}. 

In addition, the situation becomes simple for large $N$. In that case, we may use the grand canonical ensemble, and can compute the entanglement entropy as explained in subsec.~\ref{subsec:largeN_EE}.

\subsection{Entanglement entropy for a single particle at finite temperature}
As a first step, 
we consider the single particle ($N=1$) case. 
In this case, we have the two sectors $k=0$ and $k=1$.
The projected density matrix on each sector defined in \eqref{densitymatrixofeachk} is
\begin{align}
\label{rho0rho1}
    \rho_0=\frac{1}{p_0}\Pi_{\bar{A}}\rho\Pi_{\bar{A}},\; \rho_1=\frac{1}{p_1}\Pi_{A}\rho\Pi_{A}.
\end{align}
Then, the reduced density matrices on $A$ are
\begin{align}
    \rho_{0,A}=\tr_{\bar{A}}\rho_0=1,\quad
    \rho_{1,A}=\tr_{\bar{A}}\rho_1=\rho_1.
\end{align}

We are interested in thermal state $\rho=\frac{1}{Z_1}e^{-\beta H_1}$. 
For the thermal state, the probabilities $p_k$ are expressed by the weighted overlap matrices $Y, \bar{Y}$ in \eqref{def_of_y}
as \eqref{probabilitywedge}.
For $N=1$, they are given by 
\begin{align}
    p_1={1}/{Z_1}\tr Y=\tr W,\;
    p_0=1-\tr W,
\end{align}
where we have define
$W={Y}/{Z_1}$. 

The $n$-th R\'enyi entropy \eqref{renyi-entropy} is also computed as follows. 
First, note that $\tr_A (\rho_{0,A}^n)=1$. Next, 
since $\rho_1^n=\frac{1}{p_1^n}\Pi_A (\rho \Pi_A)^n$, we obtain
\begin{align}
    &\tr_{A}(\rho_{1,A}^n)=\tr(\rho_1^n)\underset{\text{\eqref{def_of_y}}}{=}\frac{\tr (Y^n)}{Z_1^np_1^n}=\frac{\tr (Y^n)}{(\tr Y)^n}=\frac{\tr (W^n)}{(\tr W)^n}.
\end{align}
Thus, from \eqref{renyi-entropy} and \eqref{entropy}, the $n$-th R\'enyi entropy and the entanglement entropy are given as follows
\begin{align}
    S^{(n)}&=\frac{\log\left((1-\tr W)^n+\tr (W^n)\right)}{1-n},\\
    S^{(1)}&=-p_0\log p_0-p_1\log p_1-p_1\tr\left(\frac{W}{\tr W}\log\frac{W}{\tr{W}}\right)\nn
    &=-(1-\tr W)\log(1-\tr W)-\tr(W\log W).
\end{align}
In particular, the 2nd R\'enyi entropy for $N=1$ at finite temperature takes the form
\begin{align}
    S^{(2)}=-\log\left((1-\tr W)^2+\tr (W^2)\right).
\end{align}
\subsection{Reduced density matrix}\label{subsec:partial}
As written in the beginning of this section, we will investigate a formula of the 2nd R\'enyi entropy \eqref{2ndrenyi} for multiple fermions at finite temperature. 
To obtain it, in this subsection, we will consider the reduced density matrix.

We would like to compute the reduced density matrix $p_k\rho_{k,A}=\tr_{\bar{A}}(\Pi_k\rho_N\Pi_k)$ (see \eqref{densitymatrixofeachk}, \eqref{def:rho_k}). 
For this purpose, it is convenient to introduce the projection of $e^{-\beta H_1}$ on $A,\bar{A}$ as
\begin{align}
    \tilde{Y}:=\Pi_{A}e^{-\beta H_1}\Pi_{A},\quad 
    \bar{\tilde{Y}}:=\Pi_{\bar{A}}e^{-\beta H_1}\Pi_{\bar{A}}.
\end{align}
Since the projection $\Pi_k$ is given by \eqref{projection}, 
$\Pi_k\rho_N\Pi_k$ is written as
\begin{align}
   \frac{\binom{N}{k}^2}{Z_N}
    (\underbrace{\Pi_A\wedge\cdots \wedge \Pi_A}_{k}\wedge\underbrace{\Pi_{\bar{A}}\cdots\wedge\Pi_{\bar{A}}}_{N-k})
    \left(e^{-\beta H_1}\wedge \cdots \wedge e^{-\beta H_1}\right)
    (\underbrace{\Pi_A\wedge\cdots \wedge \Pi_A}_{k}\wedge\underbrace{\Pi_{\bar{A}}\cdots\wedge\Pi_{\bar{A}}}_{N-k}).
    \label{wedge-expand}
\end{align}
Because the products of wedge products of operators can be computed as 
\eqref{prod-wedge}, $\Pi_k\rho_N\Pi_k$ can be written as wedge products of operators $e^{-\beta H_1}$ sandwiched between $\Pi_A$ or $\Pi_{\bar{A}}$, i.e., $\tilde{Y}, \bar{\tilde{Y}}$ and also $\Pi_A e^{-\beta H_1}\Pi_{\bar{A}}$, $\Pi_{\bar{A}} e^{-\beta H_1}\Pi_A$.
After taking the trace over $\bar{A}$ of \eqref{wedge-expand}, $e^{-\beta H_1}$ with $\Pi_{\bar{A}}$ are connected, and $\tr_{\bar{A}}(\Pi_k\rho_N\Pi_k)$ can be  written by using operators
\begin{align}
    \tilde{Y}^{(m)}&:=\Pi_A e^{-\beta H}(\Pi_{\bar{A}}e^{-\beta H})^{m-1}\Pi_A,
\end{align}
and also several traces with forms
\begin{align}
    \tr(\bar{\tilde{Y}}\wedge\cdots \wedge \bar{\tilde{Y}}).
\end{align}
We will write these types of traces as
\begin{align}
    \tr_{j}A&:=\tr(\underbrace{A\wedge\cdots \wedge A}_{j}).
\end{align}
Noting also that $\tr_{\bar{A}}(\Pi_k\rho_N\Pi_k)$ should be an operator
on $\bigwedge\nolimits^{\!k} (\Pi_A \mathcal{H})$,
we can find that the reduced density matrix $p_k\rho_{k,A}$ is expanded as
\begin{align}
    p_k\rho_{k,A}
    =\tr_{\bar{A}}(\Pi_k\rho_N\Pi_k)
    &=\frac{1}{Z_N}\sum_{\{m\}}C_{\{m\}}(\tr_{N-\sum_{i}m_{i}}\bar{\tilde{Y}})\tilde{Y}^{(m_1)}\wedge\cdots\wedge\tilde{Y}^{(m_k)}
    \label{Reduced-density-matrix},
\end{align}
where $\{m\}$ represent sets of $k$ integers as $\{m\}=(m_1,\cdots,m_k)$ $(1\leq m_1\leq m_2\leq\cdots\leq m_k\leq N-k+1)$,  
and $C_{\{m\}}$ are numerical constants. 

We will determine the coefficients $C_{\{m\}}$ below, by computing the matrix elements of the both sides of \eqref{Reduced-density-matrix} in the position basis.

First, we can compute the matrix elements of the left hand side of \eqref{Reduced-density-matrix} as follows.
By definition (see \cite{Sugishita:2021vih}), we have
\begin{align}
\label{y-p_k-rho_k}
\bra{\vec{y}}p_k \rho_{k,A}\ket{\vec{y'}}
   =\binom{N}{k}\int_{\bar{A}}d^{N-k}z
    \bra{\vec{y},\vec{z}}\rho_N\ket{\vec{y'},\vec{z}},
\end{align}
where $\vec{y}$, $\vec{y'}$ represent $k$-component position vectors in $A$, and 
$\vec{z}$, $\vec{z'}$ do $(N-k)$-component vectors in $\bar{A}$.
In the following, we will also use this notation, that is, $y$ denotes coordinates of $A$, and $z$ does coordinates of $\bar{A}$.
Since $\rho_N$ takes the form \eqref{th-den-mat}, 
the matrix elements in the position basis are given by
\begin{align}
    \bra{x_1,\dots,x_N}\rho_N\ket{x'_1,\dots,x'_N}
    &=\frac{1}{(N!)^2 Z_N}\sum_{\sigma, \sigma'\in S_N}(-)^{\sigma\sigma'}\prod_{i=1}^N\bra{x_{\sigma(i)}}e^{-\beta H_1}\ket{x'_{\sigma'(i)}}
    \nn
    &=\frac{1}{N! Z_N}\sum_{\sigma\in  S_N}(-)^{\sigma}\prod_{i=1}^N\bra{x_{i}}e^{-\beta H_1}\ket{x'_{\sigma(i)}}.
    \label{posi-prod}
\end{align}
Thus, \eqref{y-p_k-rho_k} becomes
\begin{align}
\label{determine-c1}
   \bra{\vec{y}}p_k \rho_{k,A}\ket{\vec{y'}}
    &=\frac{\binom{N}{k}}{N! Z_N}\int_{\bar{A}}d^{N-k}z\sum_{\sigma \in S_N}(-)^{\sigma}\prod_{i=1}^N\bra{x_{i}}e^{-\beta H_1}\ket{x'_{\sigma(i)}},
\end{align}
where we have set $(\vec{y}, \vec{z})=(x_1,\dots, x_N)$ and $(\vec{y'}, \vec{z})=(x'_1,\dots, x'_N)$.

On the other hand, the matrix elements of the right hand side of \eqref{Reduced-density-matrix} are given by
\begin{align}
\frac{1}{Z_N}\sum_{\{m\}}C_{\{m\}}(\tr_{N-\sum_{i}m_{i}}\bar{\tilde{Y}})\frac{1}{(k!)^2}\sum_{\tau,\tau'\in S_k}(-)^{\tau,\tau'}\prod_{i=1}^k\bra{y_{\tau(i)}}\tilde{Y}^{(m_i)}\ket{y'_{\tau'(i)}},
\label{determine-c2}
\end{align}
because we have 
\begin{align}
    \bra{\vec{y}}\tilde{Y}^{(m_1)}\wedge\cdots\wedge\tilde{Y}^{(m_k)}\ket{\vec{y'}}
    =\frac{1}{(k!)^2}\sum_{\tau,\tau'\in S_k}(-)^{\tau,\tau'}\prod_{i=1}^k\bra{y_{\tau(i)}}\tilde{Y}^{(m_i)}\ket{y'_{\tau'(i)}}.
\end{align}

Accordingly, \eqref{determine-c1} should be the same as \eqref{determine-c2}. Thus, $C_{\{m\}}$ should satisfy
\begin{align}
    &\frac{\binom{N}{k}}{N! Z_N}\int_{\bar{A}}d^{N-k}z\sum_{\sigma \in S_N}(-)^{\sigma}\prod_{i=1}^N\bra{x_{i}}e^{-\beta H_1}\ket{x'_{\sigma(i)}}
    \nn
    &=\frac{1}{Z_N}\sum_{\{m\}}C_{\{m\}}(\tr_{N-\sum_{i}m_{i}}\bar{\tilde{Y}})\frac{1}{(k!)^2}\sum_{\tau,\tau'\in S_k}(-)^{\tau,\tau'}\prod_{i=1}^k\bra{y_{\tau(i)}}\tilde{Y}^{(m_i)}\ket{y'_{\tau'(i)}}.\label{determine-c}
\end{align}
From this equation, we show below that $C_{\{m\}}$ is determined as
\begin{align}
    C_{\{m\}}&=(-1)^{\sum_{i}m_i-k}\frac{k!}{r_1!\cdots r_k!}
=k!\left(\prod_{i}^k \frac{(-1)^{m_i-1}}{r_i!}\right),
\end{align}
where $r_j$ is the number of $m_i$ satisfying $m_i=j$ in $\{m\}=(m_1, \cdots, m_k)$. For example, for $\{m\}=(1,1,2,3,3)$, $r_1=2, r_2=1,r_3=2, r_4=r_5=0$.

The left hand side of \eqref{determine-c} can be written using a determinant as
\begin{align}
    \frac{\binom{N}{k}}{N! Z_N}\int_{\bar{A}}d^{N-k}z
    \left|\begin{array}{ccc|ccc}
    (e^{-\beta H_1})_{y_1,y_1'} & \cdots & (e^{-\beta H_1})_{y_1,y_k'} & (e^{-\beta H_1})_{y_1,z_1} & \cdots & (e^{-\beta H_1})_{y_1,z_{N-k}}\\
    \vdots &\ddots &\vdots &\vdots &\ddots &\vdots \\
    (e^{-\beta H_1})_{y_k,y_1'} & \cdots & (e^{-\beta H_1})_{y_k,y_k'} & (e^{-\beta H_1})_{y_k,z_1} & \cdots & (e^{-\beta H_1})_{y_k,z_{N-k}}\\ \hline
    (e^{-\beta H_1})_{z_1,y_1'} & \cdots & (e^{-\beta H_1})_{z_1,y_k'} & (e^{-\beta H_1})_{z_1,z_1} & \cdots & (e^{-\beta H_1})_{z_1,z_{N-k}}\\
    \vdots &\ddots &\vdots &\vdots &\ddots &\vdots \\
    (e^{-\beta H_1})_{z_{N-k},y_1'} & \cdots & (e^{-\beta H_1})_{z_{N-k},y_k'} & (e^{-\beta H_1})_{z_{N-k},z_1} & \cdots & (e^{-\beta H_1})_{z_{N-k},z_{N-k}}
    \end{array}\right|.
    \label{lhs1}
\end{align}
In the cofactor expansion of the above determinant, there is a term taking the following form:
\begin{align}
&(e^{-\beta H_1})_{y_1,z_1}(e^{-\beta H_1})_{z_1,z_2}\cdots (e^{-\beta H_1})_{z_{m_1-1},y'_1}(-1)^{m_1-1}
\nn
&\times(e^{-\beta H_1})_{y_2,z_{m_1}}(e^{-\beta H_1})_{z_{m_1},z_{m_1+1}}\cdots (e^{-\beta H_1})_{z_{m_1+m_2-2},y'_2}(-1)^{m_2-1}
\nn
&\times\cdots
\nn
&\times(e^{-\beta H_1})_{y_{k},z_{\sum_{i=1}^{k-1}m_i-k+2}}(e^{-\beta H_1})_{z_{\sum_{i=1}^{k-1}m_i-k+2},z_{\sum_{i=1}^{k-1}m_i-k+3}}\cdots (e^{-\beta H_1})_{z_{\sum_{i=1}^{k}m_i-k},y_k}(-1)^{m_k-1}
\nn
&\times\left((e^{-\beta H_1})_{z_{\sum_{i=1}^{k}m_i-k+1},z_{\sum_{i=1}^{k}m_i-k+1}}\cdots (e^{-\beta H_1})_{z_{N-k},z_{N-k}}\right).\label{lhs}
\end{align}
Performing the $z$-integral $\frac{\binom{N}{k}}{N! Z_N}\int_{\bar{A}}d^{N-k}z$ of it, we obtain
\begin{align}
\frac{\binom{N}{k}}{N!Z_{N}}(-1)^{\sum_{i}m_{i}-k}
\tilde{Y}_{y_1,y_1'}^{(m_1)}\cdots\tilde{Y}_{y_k,y_k'}^{(m_k)}
\left(\tr\bar{\tilde{Y}}\right)^{N-\sum_{i}m_i}.
\end{align}
Other terms in the cofactor expansion also becomes the same term after the $z$-integral, and the number of the terms is  $\frac{(N-k)!}{(N-\sum_{i}m_i)!}$ which is a number of ways to ordering $\sum_{i}m_i-k$ elements from $z_1, \cdots, z_{N-k}$.
Thus, in the left hand side of \eqref{determine-c}, there is a term
\begin{align}
\frac{1}{k! (N-\sum_{i}m_i)!Z_{N}}(-1)^{\sum_{i}m_{i}-k}\tilde{Y}_{y_1,y_1'}^{(m_1)}\cdots\tilde{Y}_{y_k,y_k'}^{(m_k)}
\left(\tr\bar{\tilde{Y}}\right)^{N-\sum_{i}m_i}.
\label{lhs-c}
\end{align}

On the other hand, in the right hand side of \eqref{determine-c}, we have the following terms including $\tilde{Y}_{y_1,y_1'}^{(m_1)}\cdots\tilde{Y}_{y_k,y_k'}^{(m_k)}$; 
\begin{align}
\frac{1}{Z_N}C_{\{m\}}(\tr_{N-\sum_{i}m_i}\bar{\tilde{Y}})\frac{1}{(k!)^2}r_1!\cdots r_k!\tilde{Y}^{(m_1)}_{y_1,y_1'}\cdots\tilde{Y}^{(m_k)}_{y_k,y_k'}.\label{rhs-c}
\end{align}
Since 
$\tr_{N-\sum_{i}m_i}\bar{\tilde{Y}}=\frac{1}{(N-\sum_{i}m_i)!}\left((\tr\bar{\tilde{Y}})^{N-\sum_{i}m_i}+\cdots\right)$ (see \eqref{tr-wedge_same}), the same type of terms as \eqref{lhs-c} appear in \eqref{rhs-c} as 
\begin{align}
\frac{r_1!\cdots r_k!}{(k!)^2 (N-\sum_{i}m_i)!Z_N}C_{\{m\}}\tilde{Y}^{(m_1)}_{y_1,y_1'}\cdots\tilde{Y}^{(m_k)}_{y_k,y_k'}\left(\tr\bar{\tilde{Y}}\right)^{N-\sum_{i}m_i}.
\label{rhs-c-2}
\end{align}
By comparing \eqref{lhs-c} with \eqref{rhs-c-2}, the coefficient $C_{\{m\}}$ is fixed as 
\begin{align}
C_{\{m\}}
&=(-1)^{\sum_{i}m_i-k}\frac{k!}{r_1!\cdots r_k!}
=k!\left(\prod_{i}^k \frac{(-1)^{m_i-1}}{r_i!}\right).
\label{coefficient-c}
\end{align}

To summarize, the reduced density matrix on $k$-th sector is given by the formula
\begin{align}
    p_k\rho_{k,A}
    &=\frac{1}{Z_N}\sum_{\{m\}}C_{\{m\}}(\tr_{N-\sum_{i}m_{i}}\bar{\tilde{Y}})\tilde{Y}^{(m_1)}\wedge\cdots\wedge\tilde{Y}^{(m_k)}
    \label{Reduced-density-matrix1},
\end{align}
with coefficients \eqref{coefficient-c}.

\subsection{Second R\'enyi entropy with multi fermions}
Since we have obtained the reduced density matrices \eqref{Reduced-density-matrix1}, 
it is straightforward to compute the 2nd R\'enyi entropy
\begin{align}
    S^{(2)}=-\log(\sum_{k=0}^{N}\tr_{A}(p_{k}\rho_{k,A})^2).
\end{align}
Using the expression \eqref{Reduced-density-matrix1} and  the product formula \eqref{prod-wedge},   $\tr_{A}(p_{k}\rho_{k,A})^2$ is computed as
\begin{align}
    \tr_A(p_k\rho_{k,A})^2&=\frac{1}{Z_N^2}\sum_{\{m\}}\sum_{\{n\}}C_{\{m\}}C_{\{n\}}(\tr_{N-\sum_{i}m_{i}}\bar{\tilde{Y}})(\tr_{N-\sum_{i}n_{i}}\bar{\tilde{Y}})\nn
    &\qquad\qquad\qquad
    \tr\left((\tilde{Y}^{(m_1)}\wedge\cdots\wedge\tilde{Y}^{(m_k)})(\tilde{Y}^{(n_1)}\wedge\cdots\wedge\tilde{Y}^{(n_k)})\right)\\
    &=\frac{1}{Z_N^2}\sum_{\{m\}}\sum_{\{n\}}C_{\{m\}}C_{\{n\}}(\tr_{N-\sum_{i}m_{i}}\bar{\tilde{Y}})(\tr_{N-\sum_{i}n_{i}}\bar{\tilde{Y}})\nn
    &\qquad\qquad\qquad
    \frac{1}{k!}\sum_{\sigma\in S_k}\tr(\tilde{Y}^{(m_1)}\tilde{Y}^{(n_{\sigma_1})}\wedge\cdots\wedge\tilde{Y}^{(m_k)}\tilde{Y}^{(n_{\sigma_k})}).
\end{align}
In this expression, 
we can replace $\tilde{Y},\bar{\tilde{Y}}$ by the weighted overlap matrices $Y,\bar{Y}$ defined in \eqref{def_of_y} because we have
\begin{align}
    &\tr(\bar{\tilde{Y}}^n)=\tr(\bar{Y}^n),
    \qquad 
     \tr_j \bar{\tilde{Y}} =\tr_j \bar{Y}, \\
   & \tr(\tilde{Y}^{(m_1)}\cdots \tilde{Y}^{(m_j)}) =\tr(Y\bar{Y}^{m_1-1}\cdots Y\bar{Y}^{m_j-1})
    =\tr(Y^{(m_1)}\cdots Y^{(m_j)}),
\end{align}
where we have defined $Y^{(m)}:=Y\bar{Y}^{m-1}$. 

Therefore, the 2nd R\'enyi entropy for $N$ fermions at finite temperature is given by
\begin{align}
    S^{(2)}=-\log[\sum_{k=0}^N\frac{1}{Z_N^2\, k!}\sum_{\{m\}, \{n\}}
    &C_{\{m\}}C_{\{n\}}(\tr_{N-\sum_{i}m_{i}}\bar{Y})(\tr_{N-\sum_{i}n_{i}}\bar{Y})
    \nn
    &\times\sum_{\sigma\in S_k}\tr(Y^{(m_1)}Y^{(n_{\sigma_1})}\wedge\cdots\wedge Y^{(m_k)}Y^{(n_{\sigma_k})})].
    \label{Formula_of_Renyi}
\end{align}
We emphasize that \eqref{Formula_of_Renyi} is written in terms of the weighted overlap matrices $Y$ and $\bar{Y}$ whose sizes are independent of the particle number $N$, and the combinatorial complexity is reduced to the sums over $\{m\}, \{n\}$ and $\sigma \in S_k$.

\subsection{Low temperature approximation}
In this section we introduce low-temperature approximation of the 2nd Re\`nyi entropy. 
We do not use the formula \eqref{Formula_of_Renyi} for low temperature in our numerical computations because if we use it for large $\beta$ we need to perform highly accurate computations as explained in section~\ref{subsec:numerical}.
We therefore use an approximation around the ground states which is valid at low temperature. 

The density matrix of $N$ fermions at finite temperature is written as 
\begin{align}
    \rho_{N}=\frac{1}{Z_N}\sum_{I}e^{-\beta E_{I}}\ket{I}\bra{I}.
\end{align}
At low temperature (large $\beta)$, the ground states are dominant in the sum. 
Supposing that the ground states are degenerated with degeneracy $d_0$, we represent them by $\ket{1}, \ket{2}, \cdots, \ket{d_0}$. 
Then, the density matrix becomes
\begin{align}
    \rho_N\to
    \dfrac{1}{d_0}\left(\ket{1}\bra{1}+\ket{2}\bra{2}+\cdots+\ket{d_0}\bra{d_0}\right)
    \label{GSDM}
\end{align}
in the limit $\beta\to\infty$. 
Note that the density matrix even for the ground states is a mixed state as \eqref{GSDM} when the states are degenerated. 

At low temperature, it is sufficient to consider the perturbation around the ground state \eqref{GSDM}.
We now consider a perturbation up to the first excited states.
We suppose that the degeneracy of the ground states with energy $E_0$ is $d_0$ and that of the first excited states with energy $E_1$ is $d_1$.\footnote{In appendix A, we summarize the degeneracy of $N$ fermion system on one-dimensional circle.} 
Then, the reduced density matrix is approximated as
\begin{align}
 &\rho=\frac{1}{\mathcal{N}}(\rho_0+\delta\rho)\qquad \text{with} \quad
    \rho_0=e^{-\beta E_0}\sum_{i=1}^{d_0}\underbrace{\ket{a_i}\bra{a_i}}_{=\rho_{a_i}},\quad\delta \rho=e^{-\beta E_1}\sum_{j=1}^{d_1}\underbrace{\ket{b_j}\bra{b_j}}_{=\rho{_{b_{j}}}},
\end{align}
where $\mathcal{N}$ is the normalization factor given by
\begin{align}
    \mathcal{N}=d_0e^{-\beta E_0}+d_1e^{-\beta E_1}.
\end{align}
We will consider low temperature such as  $e^{-\beta (E_1-E_0)} \ll 1$, and keep only the linear order of $e^{-\beta (E_1-E_0)}$.
Then the approximation of the 2nd R\'enyi entropy is given by
\begin{align}
    S^{(2)}_A
    &=-\log\left(\dfrac{\sum_{i,i'=1}^{d_0}\sum_{k}\tr(\rho_{a_i,k,A}\rho_{a_{i'},k,A})+2e^{-\beta(E_1-E_0)}\sum_{i=1}^{d_0}\sum_{j=1}^{d_1}\sum_{k}\tr(\rho_{a_i,k,A}\rho_{b_j,k,A})}{d_0^2+2d_0d_1e^{-\beta(E_1-E_0)}}\right),
    \label{low_temp_Renyi}
\end{align}
where $\rho_{a_i,k,A}$ and  $\rho_{b_j,k,A}$ are the $k$-th sector non-normalized reduced density matrices on $A$ given by
\begin{align}
  \rho_{a_i,k,A}:=\Pi_k \ket{a_i}\bra{a_i} \Pi_k, \quad
   \rho_{b_j,k,A}:=\Pi_k \ket{b_j}\bra{b_j} \Pi_k.
\end{align}

Therefore, if we obtain a general formula of $\tr(\rho_{a,k,A}\rho_{b,k,A})$ for $N$-fermion states $\ket{a}, \ket{b}$, we can compute \eqref{low_temp_Renyi}. 
Since the $N$-fermion states $\ket{a}, \ket{b}$ are labeled by $N$ distinct one-body states, we represent them  by $\{n_{1},n_{2},\cdots,n_{N}\}$ for $\ket{a}$ and  $\{m_{1},m_{2},\cdots,m_{N}\}$ for $\ket{b}$.
Then the $N$-body wave functions are given by the Slater determinants as
\begin{align}
    &\psi_{a}(\vx)=\braket{\vx|a}=\frac{1}{\sqrt{N!}}\det(\chi_{n_{k}}(x_l)),\\
    &\psi_{b}(\vx)=\braket{\vx|b}=\frac{1}{\sqrt{N!}}\det(\chi_{m_{k}}(x_l))),
\end{align}
where $\chi_k(x)$ are single-body wave functions normalized as
\begin{align}
    \int_{M}dx\,  \chi_k^*(x)\chi_l(x)=\delta_{kl}.
\end{align}
We also introduce the overlap matrices $X,\bar{X}$ on subregions $A,\bar{A}$ as 
\begin{align}
\label{def:omX}
    X_{kl}:=\int_{A}dy\,\chi_k^*(y)\chi_l(y),\quad \bar{X}_{kl}:=\int_{\bar{A}}dz\,\chi_k^*(z)\chi_l(z).
\end{align}

Matrix elements of non-normalized reduced density matrix $\rho_{{a},k,A}=\Pi_k \ket{a}\bra{a} \Pi_k$ are given by
\begin{align}
    &\braket{\vec{y}|\rho_{{a},k,A}|\vec{y'}}
    =\binom{N}{k}\int_{\bar{A}}d^{N-k}z\;\psi_{a}(\vec{y},\vec{z})\psi_{a}^{*}(\vec{y'},\vec{z})\nn
    &=\frac{1}{N!}\binom{N}{k}\sum_{\sigma,\sigma'\in S_N}(-)^{\sigma\sigma'}\int d^{N-k}z\: \chi_{n_{{\sigma{(1)}}}}(y_1)\cdots\chi_{n{_{{\sigma{(k)}}}}}(y_{k})\chi_{n_{{\sigma(k+1)}}}(z_1)\cdots\chi_{n_{{\sigma(N)}}}(z_{N-k})\notag\nn
    &\quad\quad\quad\quad\quad\quad\quad\quad\quad\quad\quad\quad\quad\times\chi^{*}_{n_{{\sigma'(1)}}}(y'_{1})\cdots\chi^{*}_{n_{{\sigma'(k)}}}(y'_{{k}})\chi^{*}_{n_{{\sigma'(k+1)}}}(z_1)\cdots\chi^{*}_{{n_{{\sigma'(N)}}}}(z_{N-k})\nn
    &
    =\frac{1}{N!}\binom{N}{k}\sum_{\sigma,\sigma'\in S_N}(-)^{\sigma\sigma'}\chi_{n_{{\sigma(1)}}}(y_1)\cdots\chi_{n_{{\sigma(k)}}}(y_{k})\chi^{*}_{n_{{\sigma'(1)}}}(y'_{1})\cdots\chi^{*}_{n_{{\sigma'(k)}}}(y'_{{k}})\notag\\
    &\quad\quad\quad\quad\quad\quad\quad\quad\quad\times\bar{X}_{n_{{\sigma'(k+1)}}n_{i_{\sigma(k+1)}}}\cdots\bar{X}_{n_{{\sigma'(N)}}n_{{\sigma(N)}}}.
\end{align}
Then, $\tr(\rho_{a,k,A}\rho_{b,k,A})$ is computed as 
\begin{align}
    &\tr(\rho_{a,k,A}\rho_{b,k,A})
    =
    \binom{N}{k}^2
    \int_A d^ky d^ky' \int_{\bar{A}}d^{N-k}z d^{N-k}z'
    \psi_{a}(\vec{y},\vec{z})\psi^\ast_{a}(\vec{y'},\vec{z})\psi_{b}(\vec{y'},\vec{z'})\psi^\ast_{b}(\vec{y},\vec{z'})
    \nn
    &=\frac{{\binom{N}{k}}^2}{{N!}^2}
    \sum_{\sigma,\sigma',\tau,\tau'\in S_N}(-)^{\sigma,\sigma',\tau,\tau'}\,
    X_{m_{\tau'(1)}n_{\sigma(1)}}\cdots X_{m_{\tau'(k)}n_{\sigma(k)}}
    \times\bar{X}_{n_{\sigma'(k+1)}n_{\sigma(k+1)}}\cdots \bar{X}_{n_{\sigma'(N)}n_{\sigma(N)}}\nn
    &\qquad\qquad\qquad\qquad\qquad
    \times X_{n_{\sigma'(1)}m_{\tau(1)}}\cdots X_{n_{\sigma'(k)}m_{\tau(k)}}
    \times\bar{X}_{m_{\tau'(k+1)}m_{\tau(k+1)}}\cdots \bar{X}_{m_{\tau'(N)}m_{\tau(N)}}.
    \label{trrho2-perm}
\end{align}

We can rewrite the above expression by using minor determinants as follows. 
Let $F_{N,k}(n)$ be the set of all subsets of $k$ ordered different elements taken from $(n_{1},n_{2},\cdots,n_{N})$.
For example, when $N=3$, $k=2$, we have
$F_{N=3,k=2}(n)=\{(n_1,n_2), (n_1,n_3), (n_2,n_3)\}$. 
We use $I_1, I_2$ to represent elements of $F_{N,k}(n)$, and also $\bar{I}_1$ and $\bar{I}_2$ for the complements of $I_1$ and $I_2$ respectively. In the above example with $N=3$, $k=2$, if $I_1=(n_1,n_2)$, the complement is $\bar{I}_1=(n_3)$.
Similarly, we represent the set of subsets of $(m_{1},m_{2},\cdots,m_{N})$ by $F_{N,k}(m)$. 
The elements are denoted by $I_3, I_4$, and the complements are by $\bar{I}_3, \bar{I}_4$. 
We then define minor determinants $\det(X_{I_i,I_j})$ of $k \times k$ submatrix of $X$ associated with sets $I_i, I_j$.
For example, when $I_1=(n_{i_1}, \cdots, n_{i_k})$ and $I_4=(m_{j_1}, \cdots, m_{j_k})$, the minor determinant $\det(X_{I_4,I_1})$ is
\begin{align}
   \det(X_{I_4,I_1})= \sum_{\sigma \in S_k}(-)^\sigma X_{m_{j_1}, n_{i_{\sigma(1)}}} \cdots X_{m_{j_k}, n_{i_{\sigma(k)}}}. 
\end{align}
We also define minor determinants $\det(\bar{X}_{\bar{I}_i,\bar{I}_j})$ associated with sets $\bar{I}_i,\bar{I}_j$.
In addition we define the sign $\mathrm{sgn}(I_1)$ as the sign of permutation from $(n_{1},n_{2},\cdots,n_{N})$ to $I_1\cup \bar{I}_1$, which is a set combining $I_1$ and $ \bar{I}_1$ without changing the order.
For example, when $N=3$ and $k=2$,  if $I_1=(n_1, n_3)$, we have $I_1\cup\bar{I_1}$ is $(n_1,n_3,n_2)$ and $\mathrm{sgn}(I_1)$ is the sign of the permutation from $(n_1,n_3,n_2)$ to $(n_1,n_2,n_3)$, i.e., $\mathrm{sgn}(I_1)=-1$. 
We also define the sign for $I_2, I_3, I_4$ in a similar way, and represent the product of sign as
\begin{align}
    (-)^{I_1 I_2 I_3 I_4}:=\prod_{i=1}^4 \mathrm{sgn}(I_i).
\end{align}

Using these minor determinants and sign, \eqref{trrho2-perm} can be rewritten as
\begin{align}
    &\tr(\rho_{a,k,A}\rho_{b,k,A})
    =\frac{{\binom{N}{k}}^2}{{N!}^2}
    \sum_{I_1,I_2 \in F_{N,k}(n)}\,\sum_{I_3,I_4\in F_{N,k}(m)}(-)^{I_1 I_2 I_3 I_4}\,
    k!\det(X_{I_4,I_1})
    \times(N-k)!\det(\bar{X}_{\bar{I}_2\bar{I}_1})
    \nn
   &\qquad\qquad\qquad\qquad\qquad\qquad\qquad\qquad\qquad\qquad
    \times k!\det(X_{I_2,I_3})
    \times(N-k)!\det(\bar{X}_{\bar{I}_4,\bar{I}_3})
    \nn
    &=\sum_{I_1,I_2\in F_{N,k}(n)}\,\sum_{I_3,I_4 \in  F_{N,k}(m)}(-)^{I_1 I_2 I_3 I_4}\, \det(X_{I_4,I_1})\det(\bar{X}_{\bar{I}_2\bar{I}_1})\det(X_{I_2,I_3})\det(\bar{X}_{\bar{I}_4,\bar{I}_3}).
    \label{trrho2-I}
\end{align}

Therefore, the 2nd R\'enyi entropy at low temperature can be computed by \eqref{low_temp_Renyi} with applying the formula  \eqref{trrho2-I} for $\tr(\rho_{a_i,A,k}\rho_{b_j,A,k})$.

\subsection{Large $N$ and the grand canonical ensemble}\label{subsec:largeN_EE}
A difficulty of the computations of entropy in the canonical ensemble is due to the condition that the particle number $N$ is fixed. 
However, for large $N$, the canonical ensemble with fixed $N$ can be approximated well by the grand canonical ensemble (see, e.g., \cite{Dean:2016sug} for the proof). 
Thus, we may use the grand canonical ensemble for large $N$, and the computations are easier than the canonical ensemble as explained below. 
For the grand canonical ensemble, it is convenient to use the second quantized picture (or a field theory description). 
In the second quantized picture, the entropy is just a base space entanglement.\footnote{See \cite{Mazenc:2019ety, Das:2020jhy,  Sugishita:2021vih} for the equivalence between the target space entanglement entropy in the first quantization and the base space entanglement in the second quantization. } 
The base space entanglement for non-interacting non-relativistic fermions at zero temperature is investigated, e.g., in \cite{Das:1995vj, Klich:2004pb, klich2008scaling, Calabrese:2011zzb, Calabrese:2011vh, Song:2011gv, Mintchev:2022xqh, Mintchev:2022yuo}.

The density matrix for the grand canonical ensemble at temperature $1/\beta$ with the chemical potential $\mu$ is given as
\begin{align}
    \rho(\beta, \mu)=&e^{\Omega(\beta,\mu)}\sum_{N=0}^\infty \sum_{I} e^{-\beta E_I -\beta \mu N}\ket{I} \bra{I} 
    \label{gc-density}
    \\
    &\text{with} \quad \Omega(\beta,\mu) = - \log\left[\sum_{N=0}^\infty \sum_{I} e^{-\beta E_I -\beta \mu N}\right],
\end{align}
where $\ket{I}$ are energy eigenstates with energy $E_I$ for $N$-particle states.
For non-interacting fermions, the grand potential $\Omega$ is simplified as 
\begin{align}
   \Omega(\beta,\mu) = - \sum_{n}\log\left[1+ e^{-\beta E_n -\beta\mu}\right] ,
\end{align}
where $n$ runs over all single-body energy eigenstates.

Let us work in the second quantized picture.
Suppose that $c_n, c_n^\dagger$ are the annihilation and creation operators of the single-body energy eigenstates $\ket{n}$. They  satisfy $\{c_m, c_n\}=0$ and $\{c_m, c_n^\dagger\}=\delta_{m,n}$. 
Then, the density matrix \eqref{gc-density} is written as 
\begin{align}
\label{rho_grand}
    \rho(\beta, \mu)=&e^{\Omega(\beta,\mu)} e^{-\sum_n \beta( E_n +\mu)c_n^\dagger c_n}.
\end{align}
It is easy to compute the entanglement entropy for this density matrix (see, e.g.,  \cite{peschel2003calculation, Casini:2009sr, Song:2011gv}) because it is Gaussian.

We introduce the creation operators at $x$ as 
\begin{align}
    c^\dagger (x):= \sum_n \chi_n(x) c_n^\dagger
\end{align}
where $\chi_n(x)$ are the single-body wave functions for state $\ket{n}$, and also the two-point correlation function as
\begin{align}
    G(x; x'):=\tr (c^\dagger (x) c (x')\rho ).
\end{align}
Then, any $(k+\ell)$-point  correlation functions, 
\begin{align}
    G_{k,\ell}(x_1, \cdots, x_k; x'_1, \cdots, x'_\ell)
    :=\tr (c^\dagger(x_k)\cdots c^\dagger(x_1) c(y_1) \cdots c(y_\ell)\rho ),
\end{align}
satisfy the following  Wick’s contraction rule:
\begin{align}
\label{eq:Wick}
   G_{k,\ell}(x_1, \cdots, x_k; x'_1, \cdots, x'_\ell)
    =\delta_{k,\ell}\sum_{\sigma\in S_k}(-)^\sigma G(x_{\sigma(1)};x'_1)\cdots G(x_{\sigma(k)};x'_k).
\end{align}
For a system satisfying Wick’s contraction rule, the reduced density matrix for subregion $A$ can be computed from the restricted two-point function $G_A(x;x')$ which is $G(x;x')$ restricted on $A$ as $x,x' \in A$ \cite{peschel2003calculation, Casini:2009sr, Song:2011gv}. 
The R\'enyi entropy is then given by
\begin{align}
    S^{(n)}=\frac{1}{1-n}\tr_A \log \left[G_A^n+(1_A-G_A)^n\right],
\end{align}
where $G_A^n(x;x')=\int_A dy\, G_A^{n-1}(x;y)G_A (y;x')$ and $\tr_A G_A^n= \int_A dy\, G_A^{n}(y;y)$.

For the density matrix \eqref{rho_grand}, the two-point function $G$ is given by
\begin{align}
    G(x;x')=\sum_m \bar{n}_m\chi_m(x) \chi^\ast_m(x') 
\end{align}
with 
\begin{align}
    \bar{n}_m:=\frac{1}{1+e^{\beta( E_m+\mu)}}
\end{align}
which is the average number of particles in state $\ket{m}$.
Then, we have 
\begin{align}
    \tr_A G_A^n= \tr Y_G^n
\end{align}
for any $n$ where $Y_G$ is a matrix with elements,
\begin{align}
\label{def:Y_G}
    (Y_G)_{m,n}=\sqrt{\bar{n}_m}X_{m,n} \sqrt{\bar{n}_n}\,.
\end{align}
Note that $X_{m,n}$ is the overlap matrix on $A$ defined by \eqref{def:omX}.
Thus, the R\'enyi entropy is given by
\begin{align}
\label{Sn_YG}
    S^{(n)}=\frac{1}{1-n}\tr \log \left[Y_G^n+(1-Y_G)^n\right]=\frac{1}{1-n} \log \det \left[Y_G^n+(1-Y_G)^n\right].
\end{align}
In particular,  the von Neumann entropy $(n=1)$ is given by
\begin{align}
\label{ee-grand}
    S^{(1)}=-\tr\left[Y_G \log Y_G +(1-Y_G)\log (1-Y_G)\right].
\end{align}

The R\'enyi entropy \eqref{Sn_YG} is a function of inverse temperature $\beta$ and chemical potential $\mu$. 
When we compare it  with the result for fixed particle number $N$, we will fix $\mu$ so that the average number of particles is $N$. 
The condition is given by
\begin{align}
\label{cond-chemic}
    N= \braket{\hat{N}}=\frac{1}{\beta}\partial_\mu \Omega(\beta,\mu).
\end{align}
This condition can also be written as
\begin{align}
    N=\sum_m \bar{n}_m.
\end{align}

We also note that the thermal entropy of the grand canonical ensemble, \begin{align}
    S^{(1)}_\mathrm{th}
    &=-\tr \rho \log \rho
    =-\Omega 
    +\beta \braket{\hat{E}}
    +\beta \mu \braket{\hat{N}}
    \\
    &=-\sum_m \left[
    \bar{n}_m \log \bar{n}_m+(1-\bar{n}_m)\log (1-\bar{n}_m)
    \right],
\end{align}
agrees with \eqref{ee-grand} when we take the subregion $A$ as the entire region because for the entire region we have
\begin{align}
    (Y_G)_{m,n}=\bar{n}_m\delta_{m,n}\,.
\end{align}
\section{Fermions on circle}\label{sec:circle}
We have obtained formulae of the 2nd R\'enyi entropy for $N$ fermions at finite temperature. 
We will apply them to a concrete system, which is non-relativistic free fermions on a one-dimensional circle.

\subsection{Setup}
We consider non-interacting $N$ fermions on a one-dimensional circle with length $L$. 
The energy spectrum of a single particle on the circle is labeled by integers as $E_n=\frac{1}{2m}\left(\frac{2\pi n}{L}\right)^2$ $(-\infty<n<\infty)$.
We summarize the spectrum for the ground states and the first excited ones for $N$-particle case in appendix \ref{app:spec}.

In our numerical computations, we cannot deal with the full Hilbert space with infinite dimensions, and have to introduce a cutoff $n_\mathrm{max}$ so that the labels $n$ are in the range $-n_\mathrm{max}\leq n \leq n_\mathrm{max}$.
We will choose a sufficiently large $n_\mathrm{max}$ so that the high temperature behavior of the thermal partition is reproduced in a precise order.
With the cutoff, the dimension of the effective Hilbert space of $N$ fermions is $\binom{2n_\mathrm{max}+1}{N}$, which is combinatorially large.
Thus, it is still difficult to directly diagonal the reduced density matrix if we take a sufficiently large cutoff.
Our formula \eqref{Formula_of_Renyi} avoids the obstruction to diagonalizing such combinatorially large size matrices because the right-hand side of \eqref{Formula_of_Renyi} can be obtained by computing several traces of $(2n_\mathrm{max}+1) \times (2n_\mathrm{max}+1)$ matrices $Y, \bar{Y}$.

We define a dimensionless inverse temperature $\bar{\beta}$ as
\begin{align}
    \bar{\beta}:=\frac{1}{2m}\left(\frac{2\pi }{L}\right)^2\beta.
\end{align}
Then, the Boltzmann factor is given by $e^{-\beta E_n}=e^{-\bar{\beta} n^2}$.

For $N=1$, the thermal partition function $Z_1(\bar{\beta})$ is given by
\begin{align}
    Z_1(\bar{\beta})=\sum_{n=-\infty}^{\infty}e^{-\bar{\beta}n^2}=\theta_3(e^{-\bar{\beta}}),
\end{align}
where $\theta_3$ is the Jacobi theta function, $\theta_3(q):=\sum_{n=-\infty}^{\infty} q^{n^2}$.
The partition function $Z_N(\bar{\beta})$ for $N$ fermions can be computed from $Z_1(\bar{\beta})$ by using the formula \eqref{ZN-det}.

We take a subregion $A$ as an interval with length $r L$ in the circle with the whole length $L$. 
The target space entanglement entropy for this subregion can be computed by using the weighted overlap matrix $Y_{m,n}, \bar{Y}_{m,n}$ in \eqref{def_of_y}. 
Let $\ket{n}$ be single-particle energy eigenstates.
We then have
\begin{align}
    \bra{m}\Pi_A\ket{n}=\frac{\sin[\pi(m-n)r]}{\pi(m-n)}.
\end{align}
Thus, $Y_{m,n}$ for interval $A$ is
\begin{align}
    Y_{m,n}=e^{-\frac{\bar{\beta}}{2}(m^2+n^2)}\frac{\sin[\pi(m-n)r]}{\pi(m-n)},\label{Yt}
\end{align}
and $\bar{Y}$, which satisfies  $Y+\bar{Y}=\Pi_Ae^{-\beta H}\Pi_A+\Pi_{\bar{A}}e^{-\beta H}\Pi_{\bar{A}}=e^{-\beta H}$, is
\begin{align}
    \bar{Y}_{m,n}=e^{-\bar{\beta}m^2}
    \delta_{m,n}-Y_{m,n}.\label{bYt}
\end{align}
Since the overlap matrices $Y, \bar{Y}$ are given, we can compute the R\'enyi entropy from the formula \eqref{Formula_of_Renyi}.
Note also that $Y$ and $\bar{Y}$ cannot be diagonalized simultaneously. 
This is the reason why it is difficult to compute general R\'enyi entropies.

\subsection{Numerical results}\label{subsec:numerical}
We numerically evaluate the 2nd R\'enyi entropy for $N=1, \cdots, 9$.\footnote{The results for large $N$ using the grand canonical ensemble are given in section~\ref{sec:num-largeN}.}
We consider high temperature\footnote{We regard $\bar{\beta}$ as high temperature if $\beta (E_\mathrm{1ES}-E_\mathrm{GS}) \ll 1$, where $E_\mathrm{GS}, E_\mathrm{1ES}$ are energy of the ground and first excited states (see appendix~\ref{app:spec}). Since $\beta (E_\mathrm{1ES}-E_\mathrm{GS}) \sim \bar{\beta}N$, we can regard $\bar{\beta}=0.001$ as high temperature for $N\sim \mathcal{O}(10)$.} $\bar{\beta}=0.001$, middle one $\bar{\beta}=0.01$ and low one  $\bar{\beta}=1$.
As we mentioned above, we need to introduce the cutoff $n_{\mathrm{max}}$ to perform numerical computations. 
We take $n_{\mathrm{max}}$ so that the partition function at high temperature $\bar{\beta}=0.001$ is computed with good accuracy.\footnote{At low temperature, we can take small $n_{\mathrm{max}}$ because high energy states are suppressed.} 
We write the partition function with cutoff as $Z_N(\bar{\beta};n_{\mathrm{max}})$.
Then, for $N=9$, we have 
\begin{align}
    \frac{Z_{N=9}(\bar{\beta}=0.001;n_{\mathrm{max}}=90)}{Z_{N=9}(\bar{\beta}=0.001;n_{\mathrm{max}}=\infty)}=1+\mathcal{O}(10^{-4}).
\end{align}
Thus, it is reasonable to take $n_{\mathrm{max}}=90$ for $N \leq 9$. We have to  take larger $n_{\mathrm{max}}$ if we increase $N$ (see subsec.~\ref{sec:num-largeN}).

At high and middle temperature, we use the formula \eqref{Formula_of_Renyi} to compute the 2nd R\'enyi entropy. 
However, at low temperature $\bar{\beta}=1$, it is not efficient to use the formula. 
In fact, for large $\bar{\beta}$, we need to perform highly accurate computations like the sign problem. 
To illustrate this problem, let us consider the partition function $Z_N(\beta)$ using the formula \eqref{ZN-det}. 
For large $\beta$, the dominant term of $Z_1(\beta)$ is $e^{-\beta E^{(1)}_{GS}}$ where $E^{(1)}_{GS}$ is the lowest energy for the single particle. 
Since the right-hand side of \eqref{ZN-det} contains a term $Z_1(\beta)^N$, it seems that the right-hand side of \eqref{ZN-det}  have a term  $(e^{-\beta E^{(1)}_{GS}})^N$. However, we know that the dominant term of $Z_N(\beta)$ for large $\beta$ is $e^{-\beta E^{(N)}_{GS}}$ where $E^{(N)}_{GS}$ is the lowest energy for $N$ fermions. 
Thus, $(e^{-\beta E^{(1)}_{GS}})^N$ is canceled out by other terms in \eqref{ZN-det}. 
To precisely see the cancellation in the numerical computation of the right-hand side of \eqref{ZN-det}, we need high precision computations because $e^{-\beta E^{(N)}_{GS}}$ is very small compared to $(e^{-\beta E^{(1)}_{GS}})^N$ for large $\beta$.
For fermions on the circle, we have $\beta E^{(1)}_{GS}=0$ and $\beta E^{(N)}_{GS}\sim \bar{\beta}N^3/12$.
For example, if $N=9$ and $\bar{\beta}=1$, we have  $e^{-\beta E^{(N)}_{GS}}\sim 8.8 \times 10^{-27}$, which is much less than $(e^{-\beta E^{(1)}_{GS}})^N=1$, and thus, to see the cancellation, we have to perform numerical computations with 27-digits accuracy if we use the formula \eqref{ZN-det}.
A situation is worse if we consider the 2nd R\'enyi thermal entropy which involves $Z_{N}(2\beta)$, and we need 54-digits accuracy.
A similar problem happens, if we use \eqref{Formula_of_Renyi} for large $\beta$. 
Thus, we instead use the approximation \eqref{low_temp_Renyi} at low temperature $\bar{\beta}\geq1$. 

\begin{figure}[H]
\vspace{-0.01\columnwidth}
\centering
\includegraphics[width=10cm]{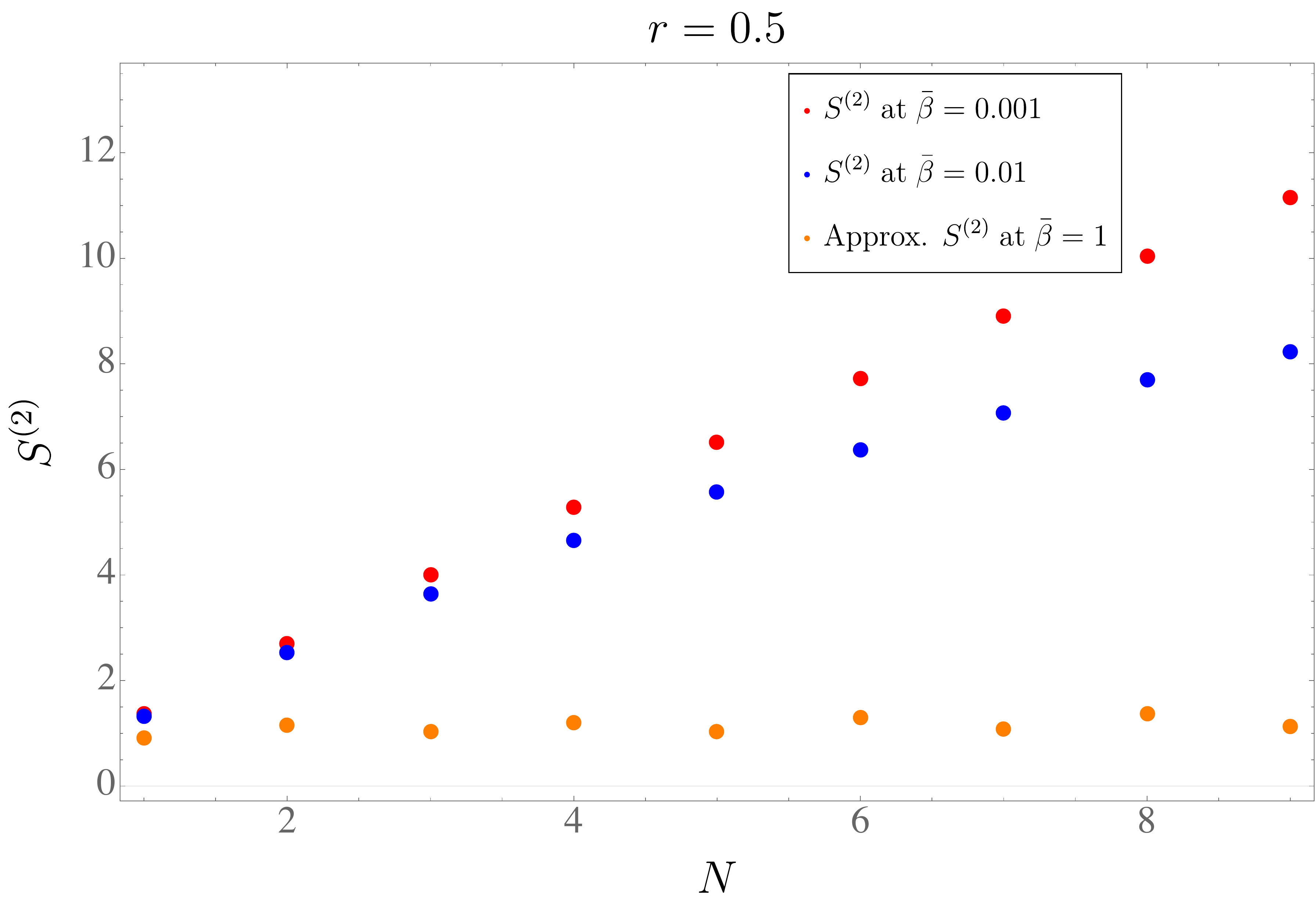}
\vspace{-0.8em}
\caption{$N$-dependence of the 2nd R\'enyi entropy for the half region ($r=0.5$) in the circle. Red and blue points represent $S^{(2)}$ at $\bar\beta=0.001$ and $0.01$ respectively. 
Orange points are the approximation \eqref{low_temp_Renyi}  at $\bar\beta=1$.}
\label{N_Renyi_r=0.5}
\vspace{-0.01\columnwidth}
\end{figure}

First, we show the $N$-dependence of the 2nd R\'enyi entropy in Fig.~\ref{N_Renyi_r=0.5} for half region $r=0.5$.
We also show the results with other values of $r$ (small region $r=0.1$ and large region $r=0.9$) in Fig.~\ref{N_Renyi_r=0.1,0.9} in appendix~\ref{app:plots}. 
The plots indicate that the 2nd R\'enyi entropy for thermal state does not follow the upper bound \eqref{up-bound} which holds for pure states.
The plots also suggest that the 2nd R\'enyi entropy generally increases with $N$. 
However, we will see in subsec.~\ref{sec:num-largeN} that this increase will not continue if we consider the grand canonical ensemble as in subsec.~\ref{subsec:largeN_EE}.
Although it is difficult to compute directly the 2nd R\'enyi entropy for the canonical ensemble with fixed particle number $N$ for large $N \gg 10$, the equivalence of the canonical and grand canonical ensemble for large $N$ suggests that the 2nd R\'enyi entropy with fixed temperature $\beta$ does not grow in linear in $N$ for large $N$.
This result is natural for the following reason. 
If we increase $N$ with fixed temperature, the system is effectively reduced to the ground states since the energy gap between the ground states and the first excited ones is proportional to $N$ (see appendix.~\ref{app:spec}). 
Since we know that the R\'enyi entropy for the ground state is proportional to $\log N$ for large $N$ \cite{Sugishita:2021vih}, we expect that the 2nd R\'enyi with fixed temperature is also.

\begin{figure}[H]
\begin{center}
\begin{minipage}[b]{0.8\columnwidth}
\includegraphics[width=\columnwidth]{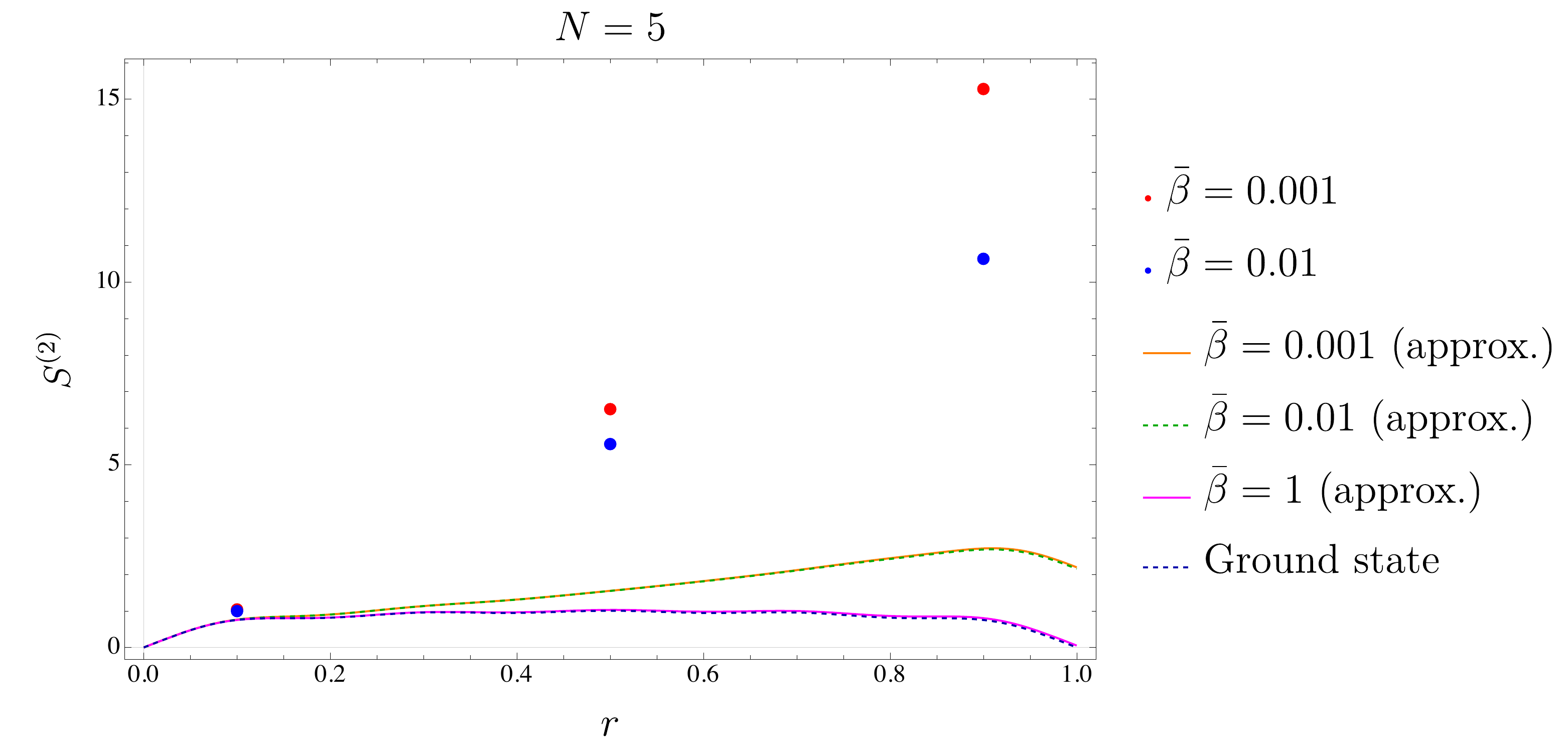}
\end{minipage}
\vspace{0.02\columnwidth}
\\
\begin{minipage}[b]{0.8\columnwidth}
\includegraphics[width=\columnwidth]{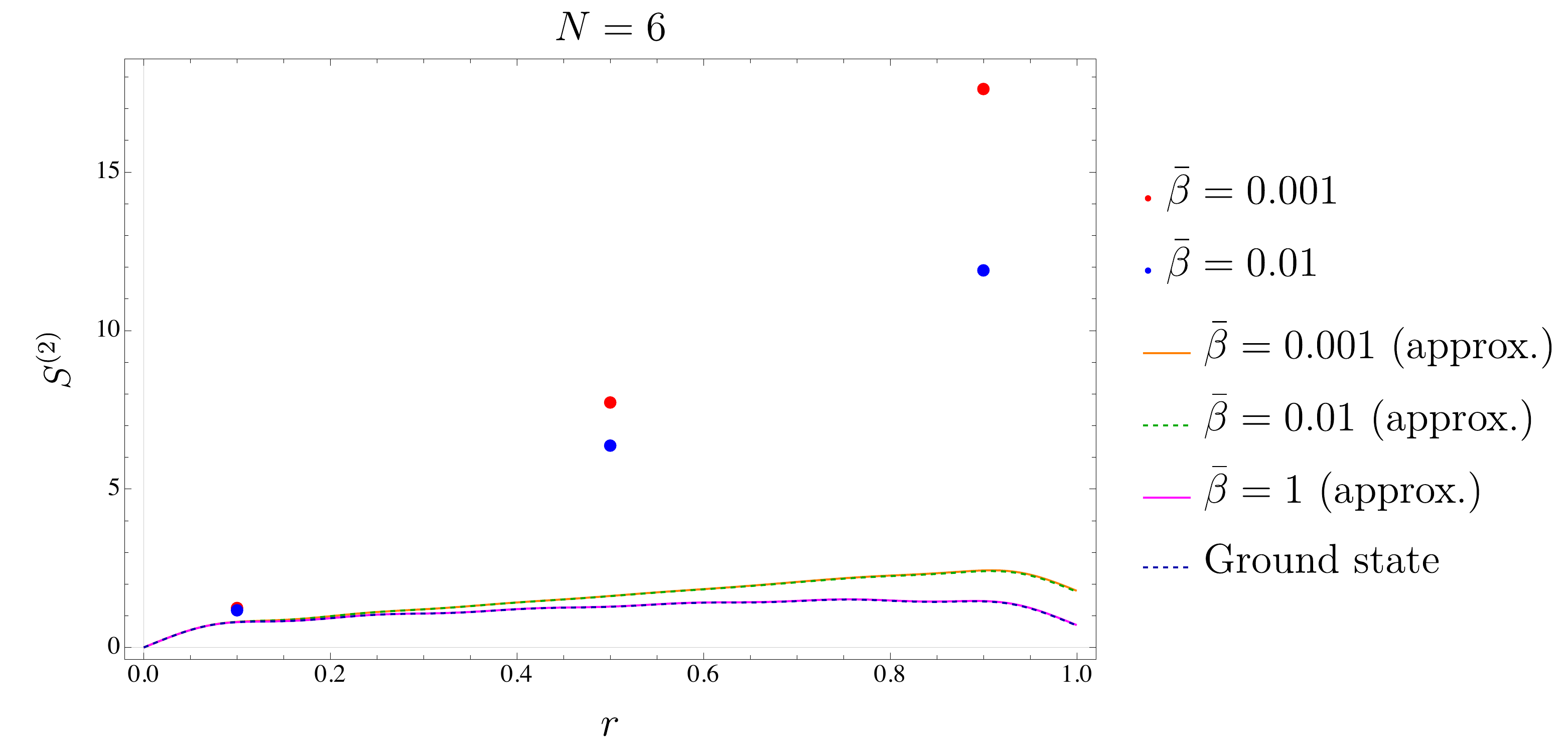}
\end{minipage}
\vspace{-0.8em}
\caption{$r$-dependence of the 2nd R\'enyi entropy for $N=5, 6$ at $\bar{\beta}=0.001, 0.01, 1$. We also show the results of the ground state (dashed blue curve). 
For low temperature $\bar{\beta}=1$, the approximation \eqref{low_temp_Renyi} is used without using the formula \eqref{Formula_of_Renyi}, and the result is almost the same as that of the ground state as expected. 
For comparison, we also show the results of the approximation \eqref{low_temp_Renyi} for $\bar{\beta}=0.001, 0.01$ although the approximation is not valid for the high temperature. In fact, the approximation differs from the results of \eqref{Formula_of_Renyi} (blue and red points) except for small $r$.}
\label{N=5_beta=1_r}
\end{center}
\vspace{-0.02\columnwidth}
\end{figure}

Next, we show $r$-dependence of the 2nd R\'enyi entropy for $N=5,6$ in Fig.~\ref{N=5_beta=1_r}. 
We calculate it by using \eqref{Formula_of_Renyi} at $\bar{\beta}=0.001, 0.01$ and \eqref{low_temp_Renyi} at $\bar{\beta}=1$. 
Since the states are mixed except for the ground states for odd $N$, we do not have $S^{(2)}_A=S^{(2)}_{\bar{A}}$. 
That is, the $r$-dependence is not symmetric for the exchange $r \leftrightarrow 1-r$.\footnote{For the ground state with even $N$, the R\'enyi entropy $S^{(n)}$ approaches  $\log 2$ in the limit $r\to 1$, while it vanishes in the limit $r\to 0$. 
This $\log 2$ is the residual entropy for the degeneracy of the ground states.} 

\begin{figure}[H]
\centering
\includegraphics[width=10cm]{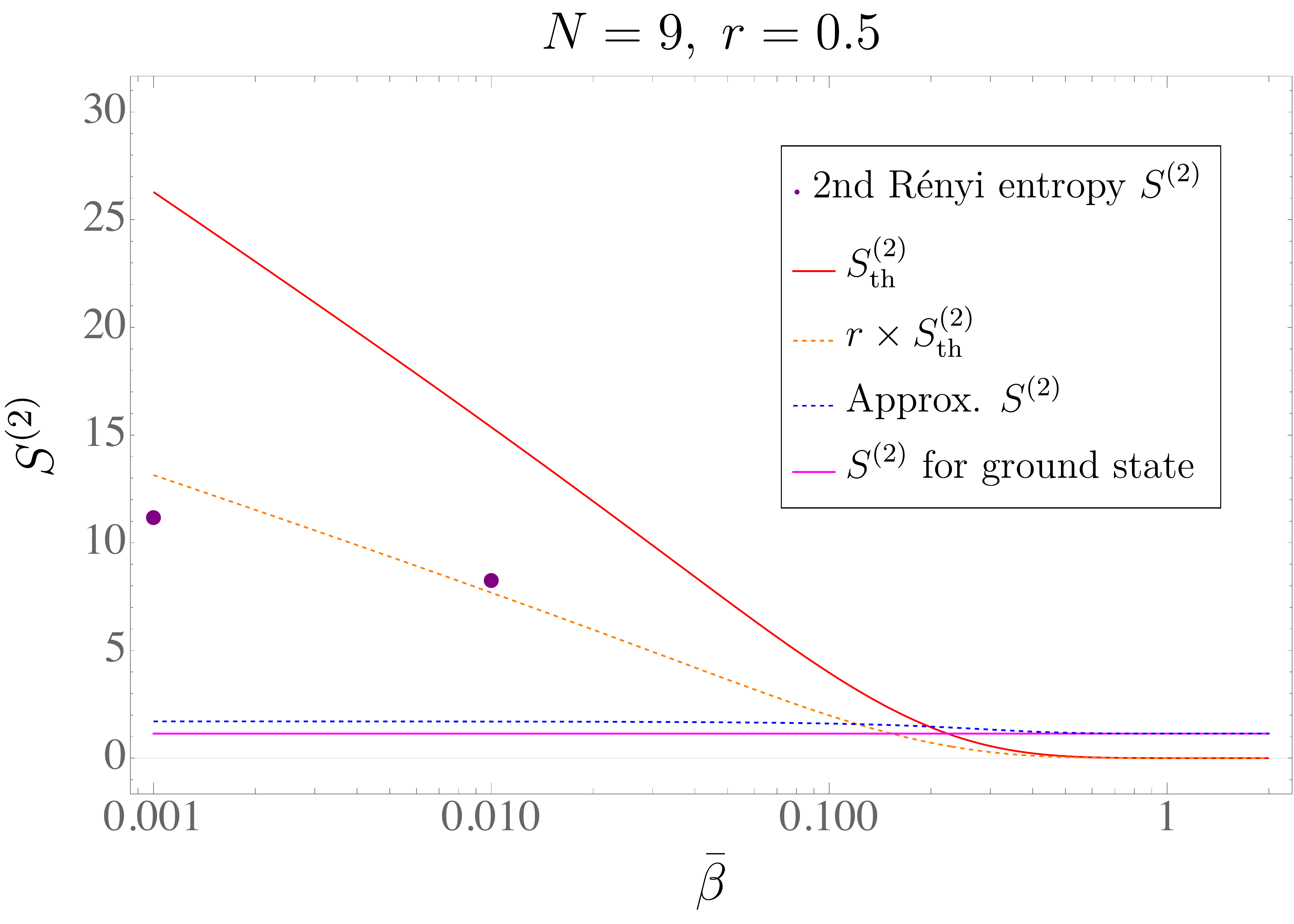}
\vspace{-0.8em}
\caption{$\bar{\beta}$-dependence of the 2nd R\'enyi entropy for $N=9$ with $r=0.5$. The purple points represent the 2nd R\'enyi entropy computed via \eqref{Formula_of_Renyi}. 
The red curve is the thermal 2nd R\'enyi entropy $S^{(2)}_{\mathrm{th}}$ in \eqref{n-renyi-thermal}. 
The orange dashed curve is $r\times S^{(2)}_{\mathrm{th}}$. 
The blue dashed curve is computed by the approximation \eqref{low_temp_Renyi}. The magenta curve is $S^{(2)}$ for the ground state.
This result implies $S^{(2)}\sim r\times S^{(2)}_{\mathrm{th}}$ except for low temperature.
}
\label{S_th_and_Renyi_N=9_r=0.5}
\end{figure}

We next consider the $\bar{\beta}$-dependence of the 2nd R\'enyi entropy.
Fig.~\ref{S_th_and_Renyi_N=9_r=0.5} represents the $\bar{\beta}$-dependence for $N=9$ and $r=0.5$. 
We also present similar plots with other parameters ($N=8,\;9$ and $r=0.1,\;0.5,\;0.9$) in Fig.~\ref{S_th_and_Renyi} in appendix~\ref{app:plots}. 
The figures imply that $S^{(2)}\sim r\times S^{(2)}_{\mathrm{th}}$ holds qualitatively except for low temperature. 
It means that most of the (R\'enyi) entanglement entropy is given by a portion of the thermal (R\'enyi) entropy in the subregion.
We can naively interpret that the deviation from $r\times S^{(2)}_{\mathrm{th}}$ is related to quantum entanglement between the subregion $A$  and $\bar{A}$. 
We will also argue the deviation in  subsec.~\ref{sec:SSA}.

\begin{figure}[H]
\centering
\includegraphics[width=10cm]{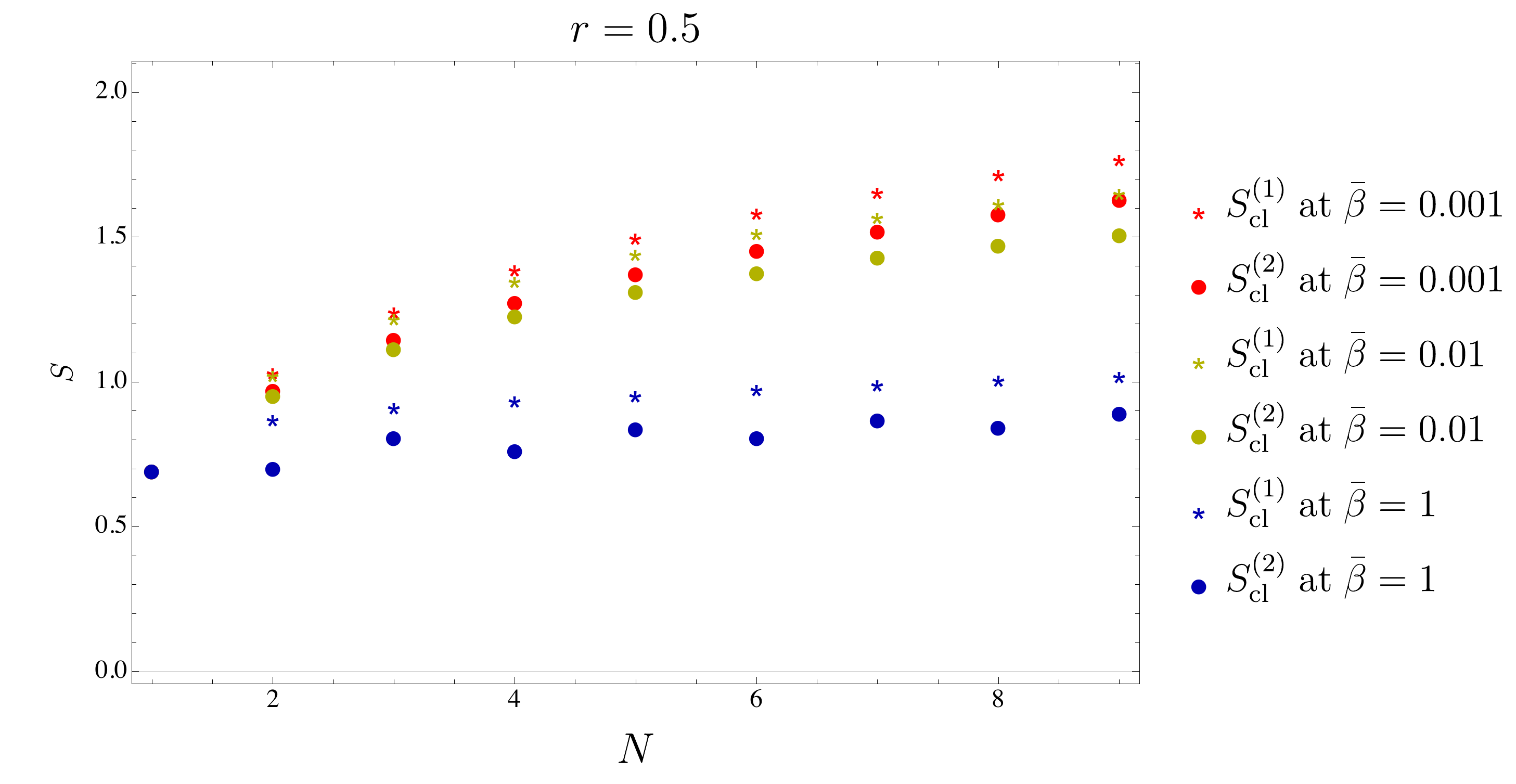}
\vspace{-0.8em}
\caption{$N$-dependence of $S^{(1)}_{\mathrm{cl}}$ and $S^{(2)}_{\mathrm{cl}}$ for the half subregion ($r=0.5$).}
\label{r=0.5_Renyi_and_entangle_classical}
\end{figure}

As in \eqref{entropy}, the entanglement entropy has the classical part \eqref{def:S_cl} associated with the probability $\{p_k\}$ for finding $k$ particles in the subregion $A$, which is the Shannon entropy $S^{(1)}_{\mathrm{cl}}=\sum_{k=0}^{N}\left(-p_k\log p_k\right)$.
We can similarly define the classical part of the 2nd R\'enyi entropy as $S^{(2)}_{\mathrm{cl}}=-\log\left(\sum_{k=0}^{N}p_k^2\right)$.
The probabilities $p_k$ can be computed from \eqref{probabilitywedge} with  \eqref{Yt} and \eqref{bYt}. 
We present the results of classical parts $S^{(1)}_{\mathrm{cl}}, S^{(2)}_{\mathrm{cl}}$ for half subregion ($r=0.5$) in  Fig.~\ref{r=0.5_Renyi_and_entangle_classical}.
Similar figures with other subregions $(r=0.1, 0.9)$ are also shown in Fig.~\ref{Renyi_minus_S_classical} in appendix~\ref{app:plots}.

\begin{figure}[H]
\centering
\includegraphics[width=12cm]{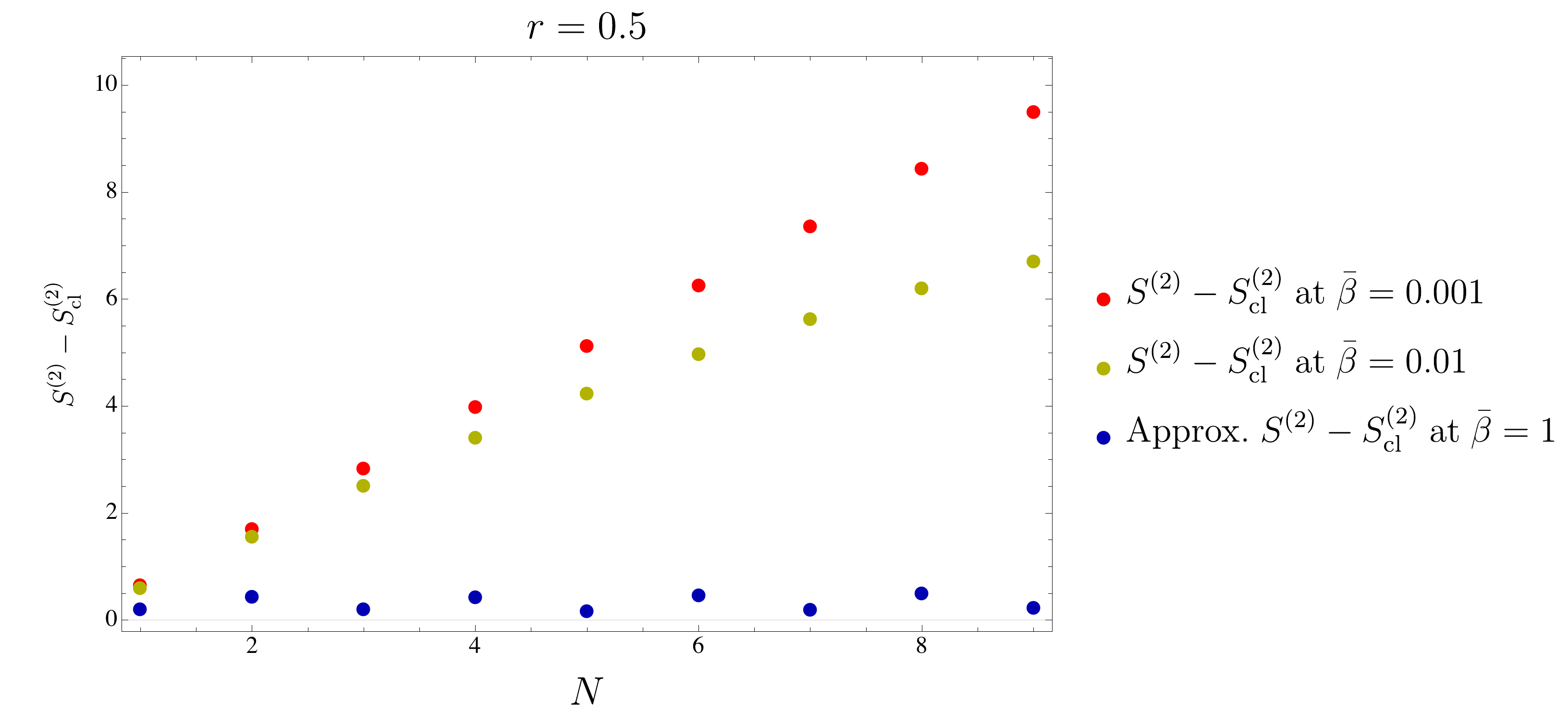}
\vspace{-0.8em}
\caption{Differences between the 2nd R\'enyi entropy and the classical part for the half subregion ($r=0.5$).}
\label{r=0.5_Renyi_minus_S_classical}
\end{figure}

We can also compute the difference $S^{(2)}-S^{(2)}_{\mathrm{cl}}$ which can be interpreted as the quantum part of the 2nd R\'enyi entropy.
Fig.~\ref{r=0.5_Renyi_minus_S_classical} is the result for the half region $r=0.5$. The results for other regions $(r=0.1, 0.9)$ are also shown in  Fig.~\ref{Renyi_minus_S_classical} in appendix~\ref{app:plots}. 
These results indicate that the difference increases with increasing temperature and $N$.
Thus, it means that quantum effects are important at high temperature.

\subsection{Large $N$ results}\label{sec:num-largeN}

\begin{figure}[H]
\centering
\includegraphics[width=11cm]{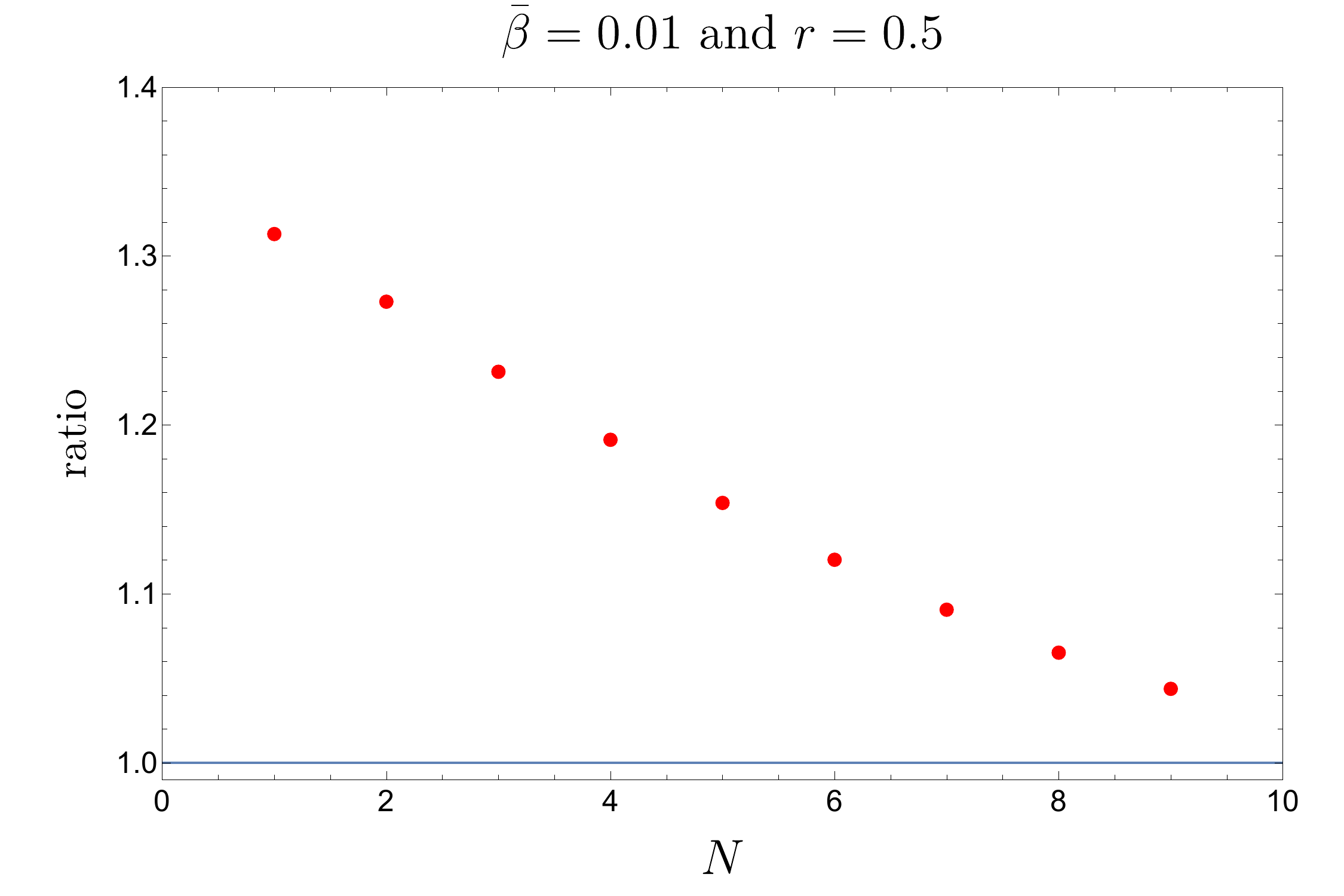}
\vspace{-0.8em}
\caption{The ratio (red points), $S^{(2)}\text{ (fixed $N$)}$ over $S^{(2)}\text{ (grand canonical)}$, for the half subregion at $\bar{\beta}=0.01$.
It indicates that the ratio approaches 1 as $N$ increases.}
\label{Ratio_GC_and_exact_beta=0.01_r=0.5}
\end{figure}

For large particle numbers $N$, we expect that the density matrix \eqref{rhoN-sumI} with fixed $N$ can be replaced by that for the grand canonical ensemble \eqref{gc-density} with an appropriate chemical potential satisfying \eqref{cond-chemic}. 
As shown in subsec.~\ref{subsec:largeN_EE}, it is easy to compute the (R\'enyi) entanglement entropy for the grand canonical ensemble.
Fig.~\ref{Ratio_GC_and_exact_beta=0.01_r=0.5} shows the plot of the ratio of the 2nd R\'enyi entropies for fixed $N$ to the grand canonical with $\bar{\beta}=0.01, r=0.5$ (we also show similar plots with other parameters in Fig.~\ref{Ratio_GC_and_exact_beta_other_plot} in appendix~\ref{app:plots}).
It indicates that the grand canonical ensemble is a good approximation of the fixed-$N$ ensemble for sufficiently large $N$.

Thus, we may investigate large $N$ behaviors by using the grand canonical ensemble. 
In the ensemble, the R\'enyi entropy is computed by eq.~\eqref{Sn_YG} with matrix $Y_G$ in \eqref{def:Y_G}. 
Unlike the case for fixed $N$, 
the computational cost is small, and the results for $N=1, \cdots, 100$ are obtained relatively easily. 
The result for the 2nd R\'enyi entropy for the half region $r=0.5$ is shown in Fig.~\ref{GC_Renyi_r=0.5} (see also Fig.~\ref{GC_renyi_entangle_other} with other $r$ in appendix~\ref{app:plots}).
In the computation, we need to introduce a cutoff $n_\mathrm{max}$ as in the fixed-$N$ case. 
We take $n_\mathrm{max}=150$ at high temperature $\bar{\beta}=0.001, 0.01$ because the thermal partition function with $n_\mathrm{max}$ seems to almost converge for $n_\mathrm{max}>100$.
At low temperature, smaller $n_\mathrm{max}$ gives us a good accuracy and we take $n_\mathrm{max}=75$ for $\bar{\beta}=1$.
In Fig.~\ref{GC_Renyi_r=0.5}, we also show the asymptotic expression of the R\'enyi entropy for the ground state explained below.

In \cite{Sugishita:2021vih}, the target space entanglement  for the ground state of the same system ($N$ free fermions on the circle) was investigated. 
It was shown that the target space R\'enyi entropy for a single interval at the ground state takes the following asymptotic form at large $N$ (where $N$ is assumed to be odd numbers):\footnote{For even $N$, \eqref{Renyi-asympt} cannot be applied. This is due to the difference of the degeneracy of the ground states between even and odd $N$. (see appendix~\ref{app:spec}).
In fact, in Fig.~\ref{GC_Renyi_r=0.5}, the low temperature result ($\bar{\beta}=1$) agrees well with the asymptotic form \eqref{Renyi-asympt} for odd $N$, while it slightly deviates from \eqref{Renyi-asympt} for even $N$.} 
\begin{align}
\label{Renyi-asympt}
    S^{(n)}_A \sim 
    \frac{1}{6}\left(1+\frac{1}{n}\right) \log  [2N \sin(\pi r)]
+\Upsilon_n,
\end{align}
where $\Upsilon_n$ is a constant given by
\begin{align}
  \Upsilon_n:=\frac{n
   }{i(1-n)}
   \int^{\infty}_{-\infty}dw [\tanh(\pi n w)-\tanh(\pi w)]
   \log \frac{\Gamma\left(\frac{1}{2}+i w\right)}{\Gamma\left(\frac{1}{2}-i w\right)}.
   \label{def:Upn}
\end{align}
In particular, for $n=1$ and $n=2$, we have $\Upsilon_1\sim 0.495$ and $\Upsilon_2\sim 0.404$.
\begin{figure}[H]
\centering
\includegraphics[width=12cm]{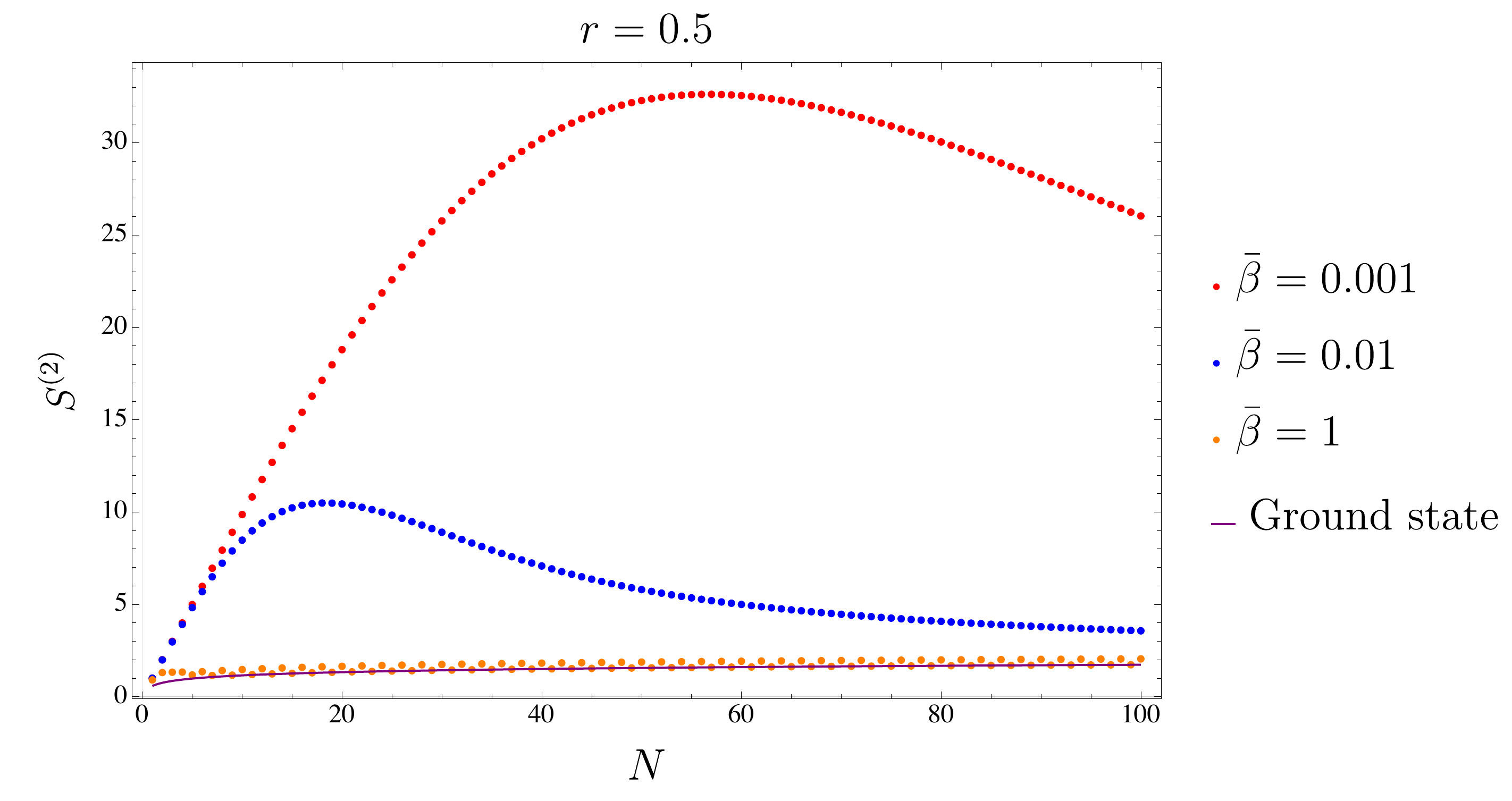}
\vspace{-0.8em}
\caption{2nd R\'enyi entropy for the half region $(r=0.5)$ for the grand canonical ensemble. The red, blue and orange points represent $S^{(2)}$ at $\bar\beta=0.001,\;0.01,\;\text{and}\;1$ respectively. The purple curve represents the asymptotic form \eqref{Renyi-asympt} of the 2nd R\'enyi entropy for the ground state.}
\label{GC_Renyi_r=0.5}
\end{figure}

\begin{figure}[H]
\centering
\includegraphics[width=12cm]{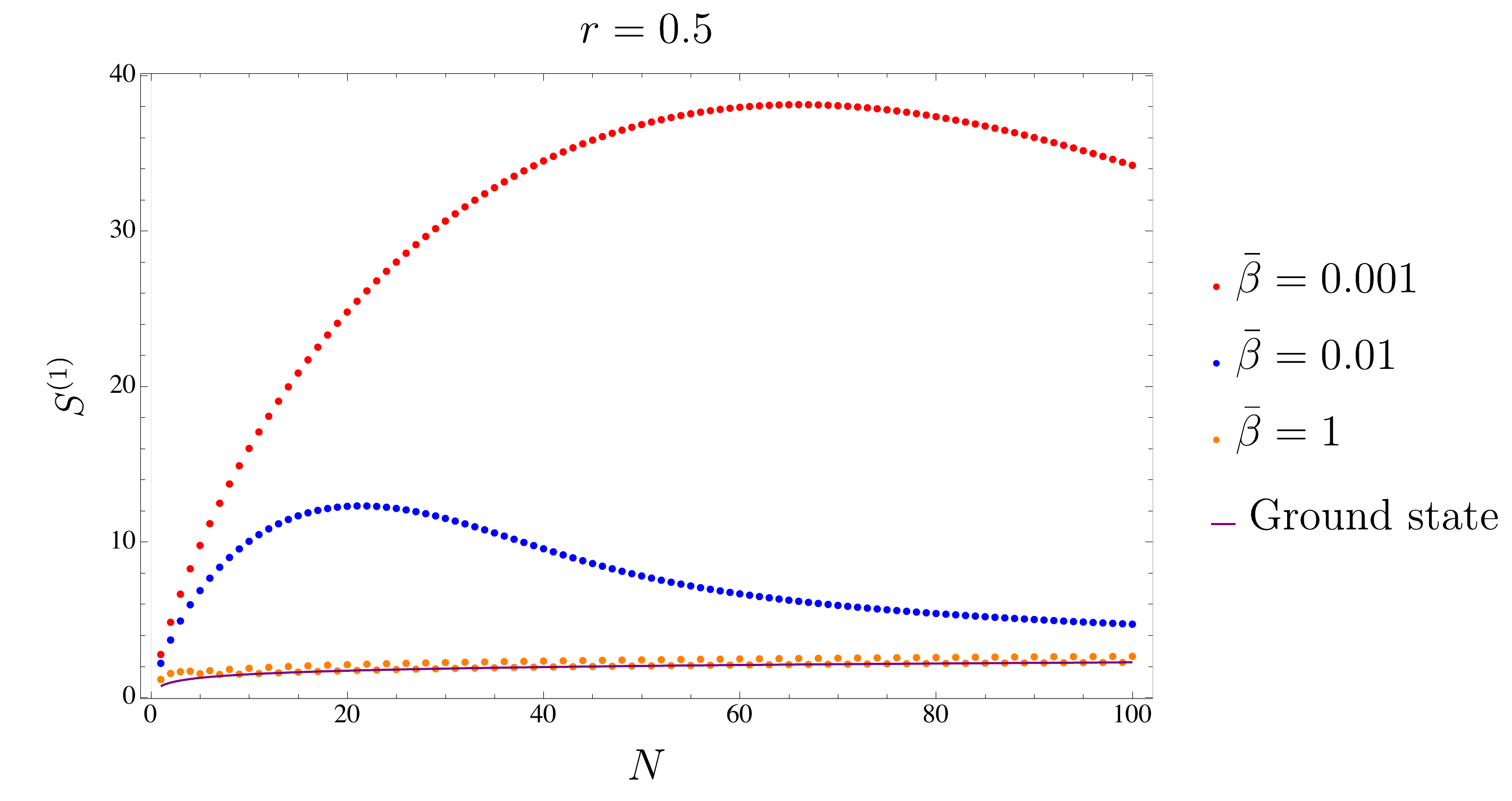}
\vspace{-0.8em}
\caption{Entanglement entropy for the half region $(r=0.5)$ for the grand canonical ensemble. The red, blue and orange points represent $S^{(1)}$ at $\bar\beta=0.001,\;0.01$, and $1$ respectively. The purple curve represents  the asymptotic form \eqref{Renyi-asympt} of the entanglement  entropy for the ground state.}
\label{GC_Entanglement_r=0.5}
\end{figure}

We can also compute the entanglement entropy $(n=1)$ for the grand canonical ensemble by \eqref{ee-grand}.
Fig.~\ref{GC_Entanglement_r=0.5} is the result for the half region (see also Fig.~\ref{GC_renyi_entangle_other} with other $r$ in appendix~\ref{app:plots}).

Figs.~\ref{GC_Renyi_r=0.5}, \ref{GC_Entanglement_r=0.5} indicates that the (R\'enyi) entanglement entropy in the large $N$ limit with fixed $\beta$ is reduced to that for the ground state. 
This is due to the fact that temperature is effectively small in the large $N$ limit because contributions from  excited states could be neglected as argued in the previous subsection.
If we take a different large $N$ limit such that $\bar{\beta} N$ is fixed, contributions from  excited states may remain, and the entropies could grow more than $\mathcal{O}(\log N)$.
However, it is difficult to perform numerical computations with the limit because, the larger $\bar{\beta}$ is, the larger cutoff $n_\mathrm{max}$ we have to take.

\subsection{Entropy inequality}\label{sec:SSA}
In subsection \ref{subsec:numerical}, we have seen that the 2nd R\'enyi entropy $S^{(2)}_A$ for interval $A$ is qualitatively the same as a portion of the thermal entropy in the subregion as $S^{(2)}_A \sim r S^{(2)}_{th}$. 
It indicates that we have $S^{(2)}_A + S^{(2)}_{\bar{A}}\sim  S^{(2)}_{th}$. 

On the other hand, it is known that the von Neumann entropy follows the Araki-Lieb inequality \cite{Araki:1970ba}:
\begin{align}
    |S^{(1)}_A-S^{(1)}_{\bar{A}}| \leq S^{(1)}_{A\cup \bar{A}}\leq S^{(1)}_A+S^{(1)}_{\bar{A}}.
\end{align}
If the total system is a thermal state, we have $S^{(1)}_{A\cup \bar{A}}=S^{(1)}_{th}$. 
Thus, the inequalities are written as
\begin{align}
    |S_A^{(1)}-S_{\bar{A}}^{(1)}| \leq S_{th}^{(1)}\leq S_A^{(1)}+S_{\bar{A}}^{(1)}.
\end{align}
Hence, for von Neumann entropy, we have $S^{(1)}_A + S^{(1)}_{\bar{A}} \geq   S^{(1)}_{th}$.

We would like to check whether a similar inequality holds or not for the 2nd R\'enyi entropy.
For general systems, the R\'enyi entropies except for $n=1$ do not follow such an inequality because they are generally not subadditive. 
Nevertheless, it is known that the 2nd R\'enyi for Gaussian states satisfies the strong subadditivity \cite{Adesso:2012ni}.
Thus, we should have $S^{(2)}_A+S^{(2)}_{\bar{A}}- S^{(2)}_{th} \geq 0$ for Gaussian states.
Here we check whether the target space 2nd  R\'enyi entropy satisfies the following inequalities or not:
\begin{align}
    \delta_{\pm} S^{(2)}\geq 0,
    \label{a}
\end{align}
where
\begin{align}
\delta_{+} S^{(2)}:=S^{(2)}_A+S^{(2)}_{\bar{A}}-S^{(2)}_{th}, \quad
    \delta_{-} S^{(2)}:=S^{(2)}_{th}-|S^{(2)}_A-S^{(2)}_{\bar{A}}|.
\end{align}

\begin{figure}[htbp]
\centering
\includegraphics[width=10cm]{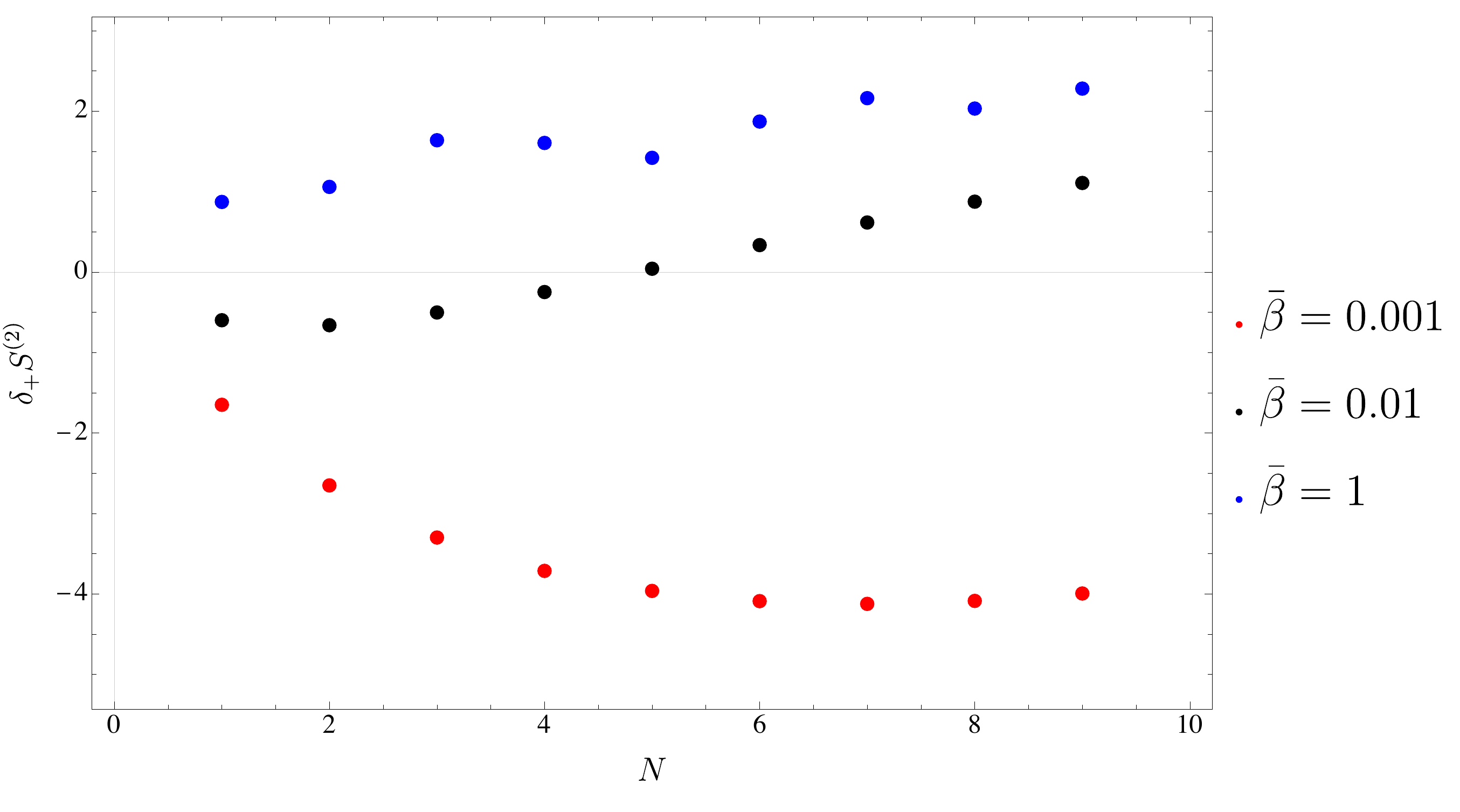}
\vspace{-0.8em}
\caption{$N$-dependence of $\delta _{+}S^{(2)}$ at $\bar{\beta}=0.001,0.01,1$ with $r=0.5$, that is $\delta _{+}S^{(2)}=2S_{S(r=0.5)}^{(2)}-S_{th}^{(2)}$. 
It shows that $\delta_{+} S^{(2)}\geq0$ does not hold at high temperature.}
\label{N_delta_s2}
\end{figure}

\begin{figure}[H]
\centering
\includegraphics[width=10cm]{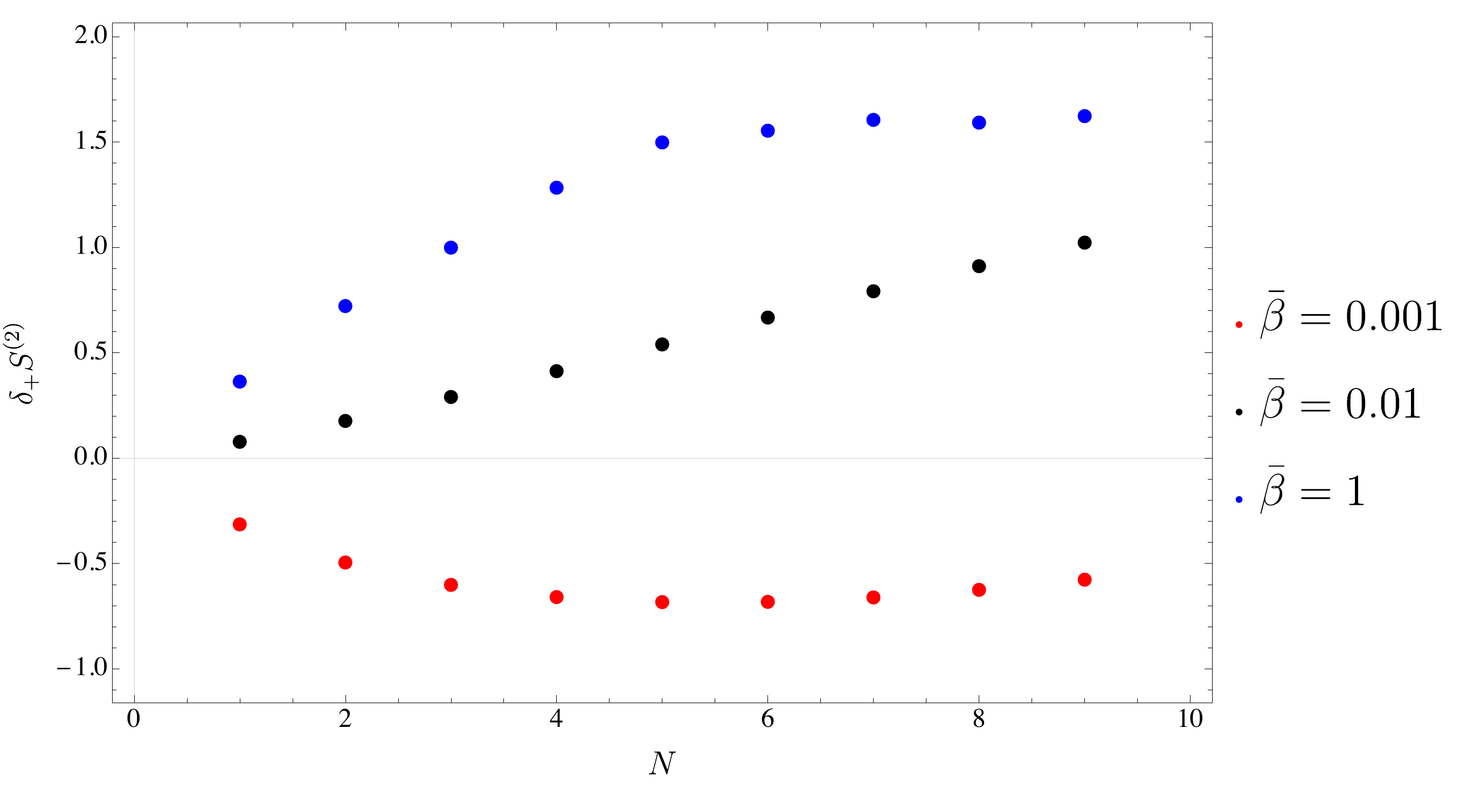}
\vspace{-0.8em}
\caption{$N$-dependence of $\delta _{+}S^{(2)}$ at $\bar{\beta}=0.001,0.01,1$ with $r=0.1$, that is  $\delta_{+}S^{(2)}=S^{(2)}_{A(r=0.1)}+S^{(2)}_{\bar{A}(r=0.9)}-S^{(2)}_{th}$.  It shows that $\delta_{+} S^{(2)}\geq0$ does not hold at high temperature as similar to Fig.~\ref{N_delta_s2}.}
\label{N_delta_plus_s2_0109}
\end{figure}

We show the $N$-dependence of $\delta_{+}S^{(2)}$ for the case where subregion $A$ is the half region $r=0.5$ at temperature $\bar{\beta}= 0.001, 0.01, 1$ in Fig.~\ref{N_delta_s2}, and also the case where $A$ is a small interval $r=0.1$ in Fig.~\ref{N_delta_plus_s2_0109}.
The figures show that the inequality $\delta_{+}S^{(2)}\geq 0$ does not hold at high temperature. 


\begin{figure}[H]
\centering
\includegraphics[width=10cm]{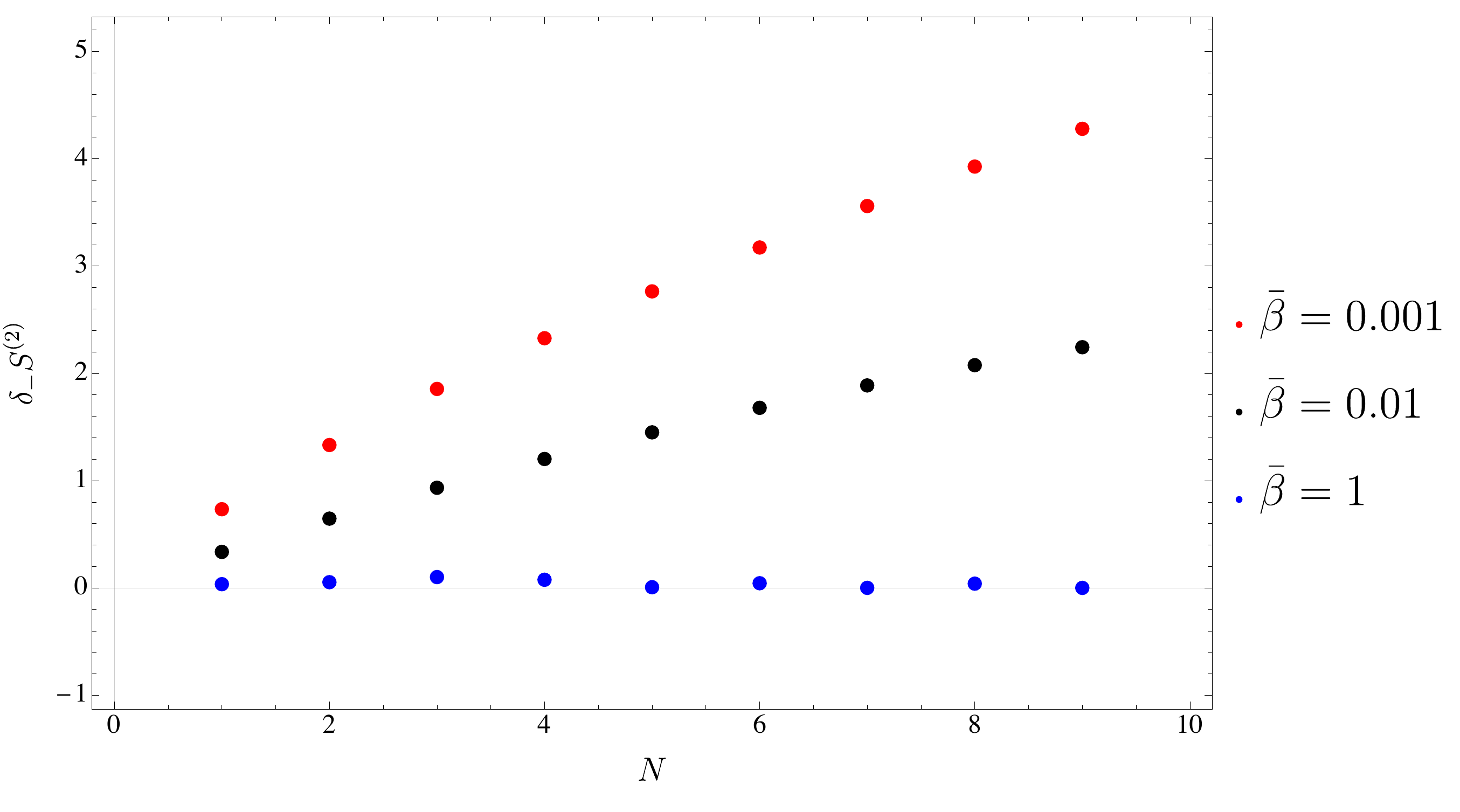}
\vspace{-0.8em}
\caption{$N$-dependence of $\delta_{-} S^{(2)}$ at $\bar{\beta}=0.001,0.01,1$ with $r=0.1$, that is,  $\delta_{-}S^{(2)}=S^{(2)}_{th}-|S^{(2)}_{A(r=0.1)}-S^{(2)}_{\bar{A}(r=0.9)}|$.  All of the values are greater than zero, and it indicates that  $\delta_{-} S^{(2)}\geq0$ holds.}
\label{N_delta_minus_s2_0109}
\end{figure}

We also show in Fig.~\ref{N_delta_minus_s2_0109} the $N$-dependence of $\delta_{-}S^{(2)}$ at temperature $\bar{\beta}=0.001, \;0.01, \;1$ for $r=0.1$, that is, $\delta_{-}S^{(2)}=S_{th}^{(2)}-|S_{A(r=0.1)}^{(2)}-S_{\bar{A}(r=0.9)}^{(2)}|$.
It indicates that $\delta_{-}S^{(2)}\geq 0$ holds in this system.

$\bar{\beta}$-dependence of the $\delta_{+} S^{(2)}$ is shown in Fig.~\ref{Araki-Lieb_N=9} for $N=9, r=0.5$ (and also in Fig.~\ref{Araki-Lieb_N=8} in appendix~\ref{app:plots} for $N=8, r=0.5$). 
We also draw a continuous curve using the approximation  \eqref{low_temp_Renyi} although the approximation is not valid for small $\bar{\beta}$.
They also indicate that $\delta_{+} S^{(2)}>0$ holds at low temperature while it is violated at high temperature.
\begin{figure}[htbp]
\centering
\includegraphics[width=11cm]{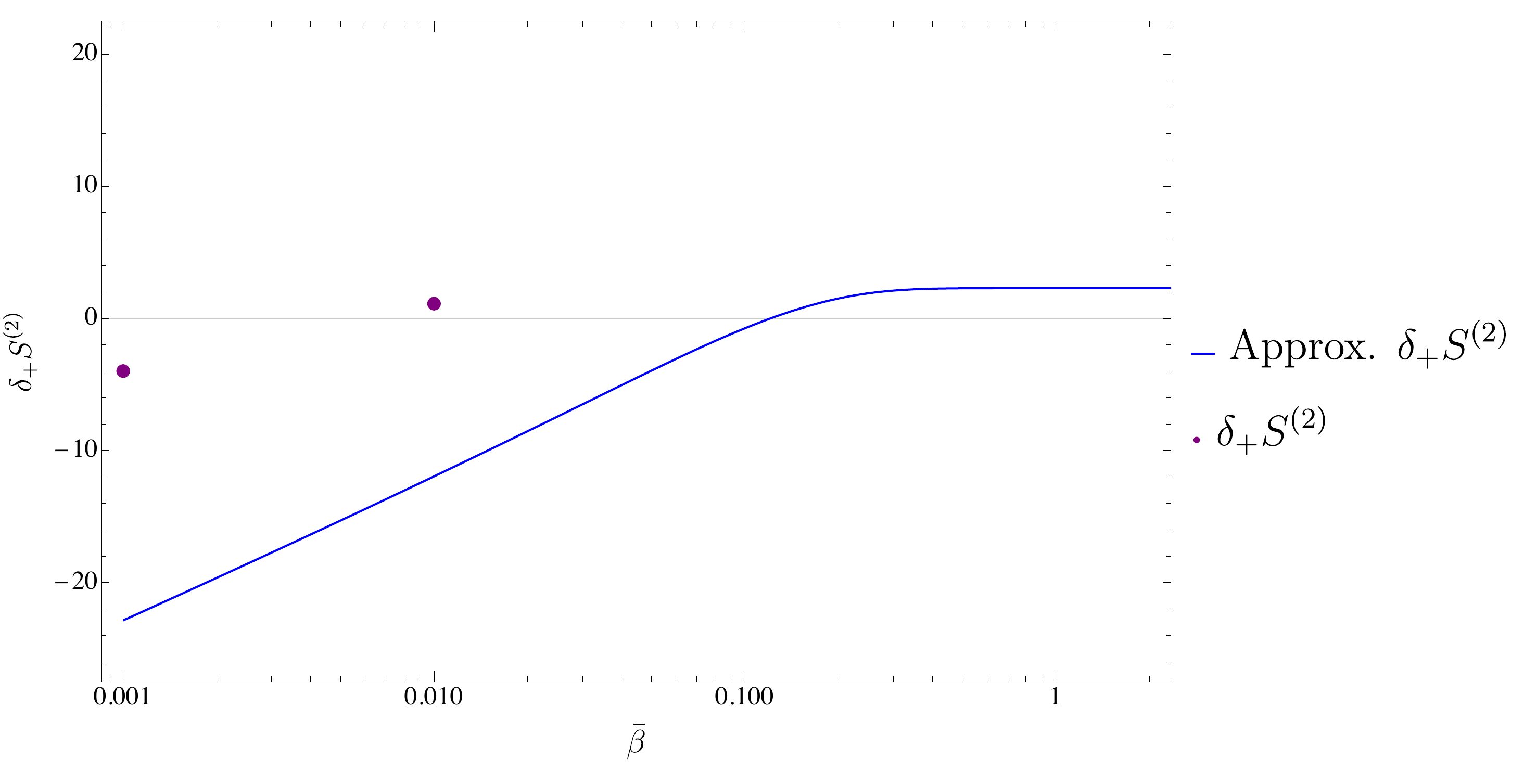}
\vspace{-0.8em}
\caption{$\bar{\beta}$-dependence of $\delta_{+} S^{(2)}$ with $r=0.5$, that is $\delta_{+}S^{(2)}=2S_{A(r=0.5)}^{(2)}-S_{th}^{(2)}$ for $N=9$.
The purple points are computed by \eqref{Formula_of_Renyi}.
The blue curve is the plot of \eqref{low_temp_Renyi} although the approximation is not valid for small $\bar{\beta}$.}
\label{Araki-Lieb_N=9}
\end{figure}

\begin{figure}[H]
\centering
\includegraphics[width=10cm]{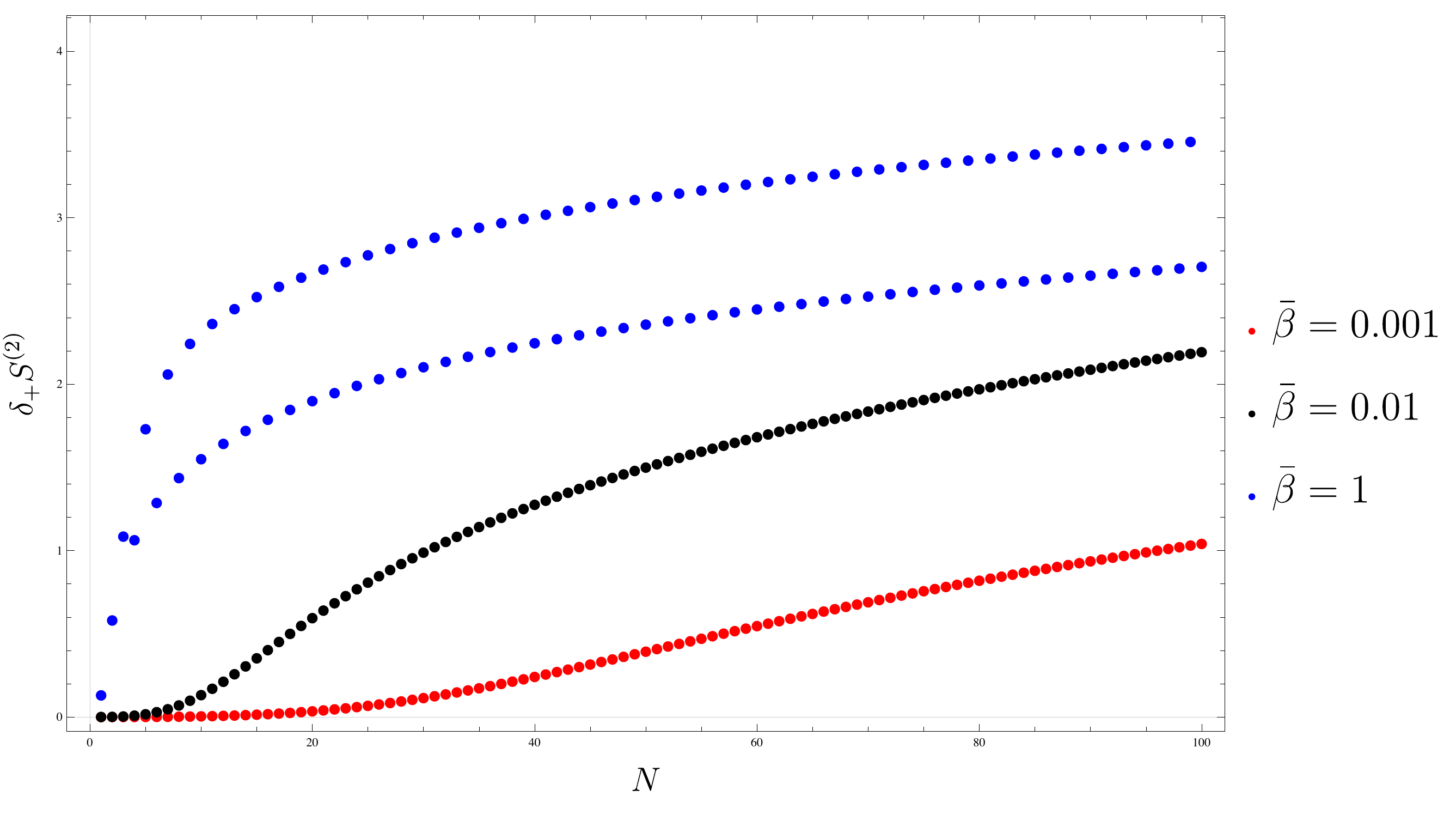}
\vspace{-0.8em}
\caption{$N$-dependence of $\delta_{+} S^{(2)}$ in the grand canonical ensemble with $r=0.5$ for $N=1, \cdots, 100$. All of the points satisfies $\delta_{+}S^{(2)}\geq0$ in contrast to Fig.~\ref{N_delta_s2}.
At low temperature $\bar{\beta}=1$ (blue points), the behaviors for odd $N$ are quite different from those for even $N$. 
This represents the difference of thermal entropy at low temperature arising from that of the degeneracy of the ground states.}
\label{Large_N_delta_plus_s2_0505}
\end{figure}

\begin{figure}[H]
\centering
\includegraphics[width=10cm]{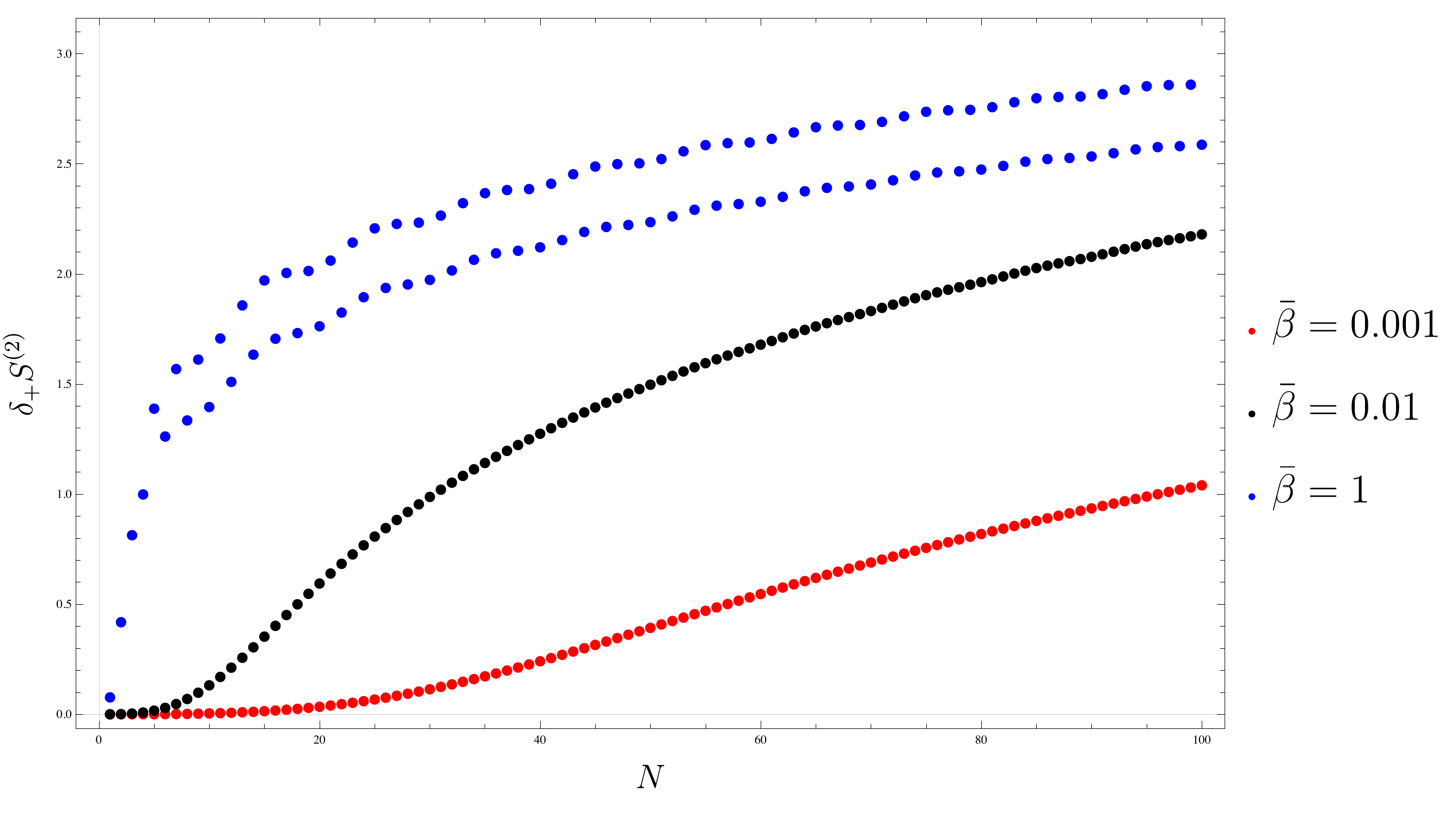}
\vspace{-0.8em}
\caption{$N$-dependence of $\delta_{+} S^{(2)}$ in the grand canonical ensemble with $r=0.1$.  $\delta_{+}S^{(2)}\geq0$ holds in contrast to Fig.~\ref{N_delta_plus_s2_0109}.}
\label{Large_N_delta_plus_s2_0109}
\end{figure}

As in the previous subsection, we may use the grand canonical ensemble for large $N$. 
We can check the inequalities \eqref{a} for the grand canonical ensemble.
Figs.~\ref{Large_N_delta_plus_s2_0505}, \ref{Large_N_delta_plus_s2_0109}, are the results of $\delta_{+} S^{(2)}$ for the subregion $r=0.5, 0.1$. 
Thus, for the grand canonical ensemble, $\delta_{+} S^{(2)}\geq0$ holds in contrast to the above Figs.~\ref{N_delta_s2}, \ref{N_delta_plus_s2_0109} for the fixed-$N$ ensemble.
This is consistent with \cite{Adesso:2012ni} because the grand canonical ensemble is a Gaussian state as the Wick's rule \eqref{eq:Wick} holds.
Therefore, for small $N$, the fixed-$N$ ensemble is quite different from the grand canonical ensemble because the behavior of $\delta_{+} S^{(2)}$ is different.

\begin{figure}[H]
\centering
\includegraphics[width=10cm]{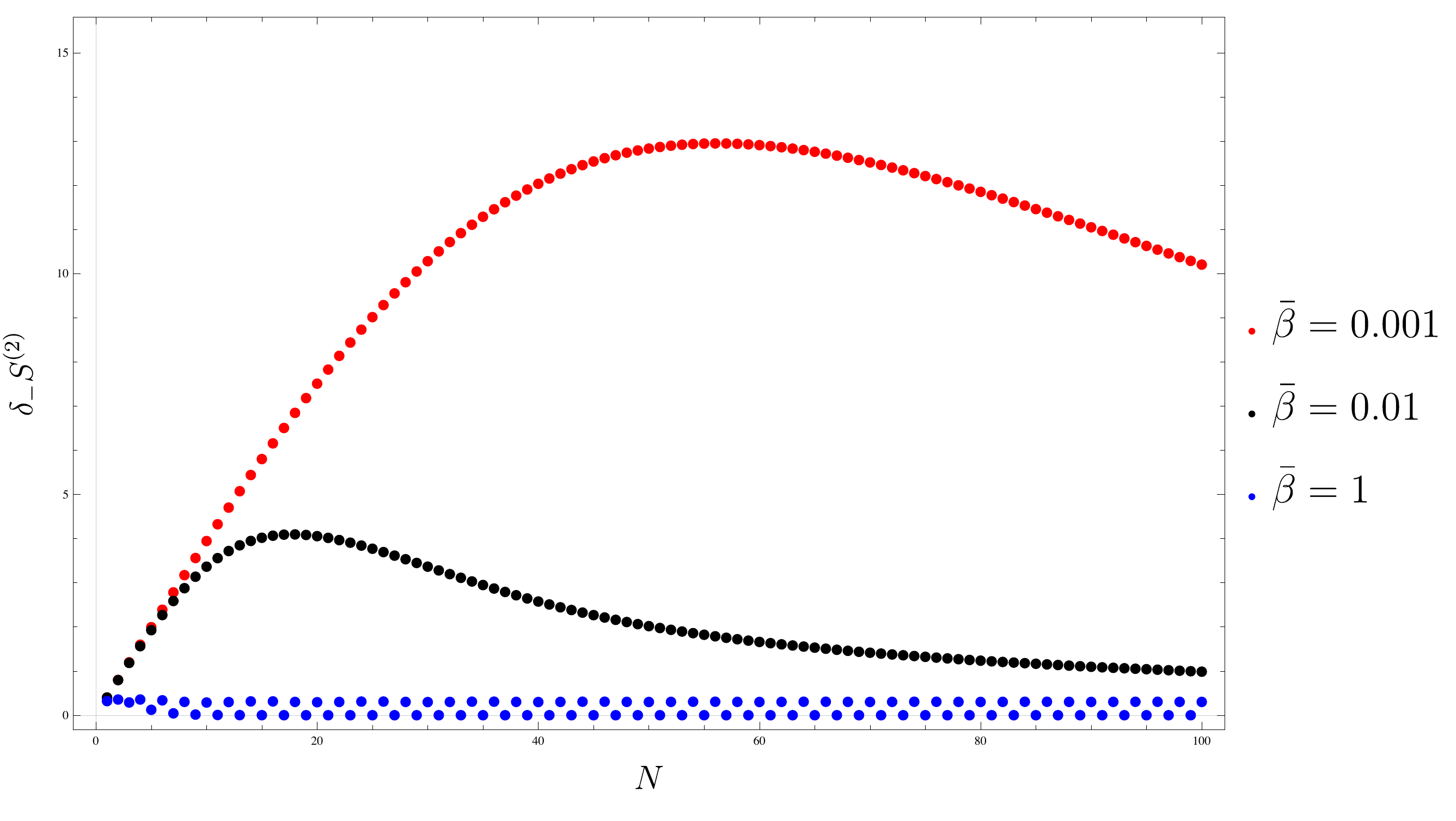}
\vspace{-0.8em}
\caption{$N$-dependence of $\delta_{-} S^{(2)}$ in the grand canonical ensemble with $r=0.1$.   $\delta_{-}S^{(2)}\geq0$ holds as similar to Fig.~\ref{N_delta_minus_s2_0109}.}
\label{Large_N_delta_minus_s2_0109}
\end{figure}

Fig.~\ref{Large_N_delta_minus_s2_0109} is the result for  $\delta_{-} S^{(2)}$ with $r=0.1$. 
It indicates $\delta_{-} S^{(2)}\geq 0$ as similar to Fig.~\ref{N_delta_minus_s2_0109}.

\section{Summary and discussions}

We have investigated the target space R\'enyi entropy for non-interacting $N$ fermions at finite temperature. 
This is an extension of the work \cite{Sugishita:2021vih} for pure states to mixed states. 
The 2nd R\'enyi entropy for thermal states can be computed via the formula \eqref{Formula_of_Renyi} although it is hard to apply to a large $N$ system because of the combinatorial complexity of the formula. 
This difficulty is related to the fact that the fixed-$N$ ensemble is not a Gaussian state.
This is quite different from the grand canonical ensemble.
The grand canonical ensemble for non-interacting fermions is a Gaussian state, and it is relatively easier to compute the entanglement entropy. 
As shown in \cite{Mazenc:2019ety, Das:2020jhy,  Sugishita:2021vih}, the target space entanglement in the first quantized picture agrees with the conventional base space entanglement in the second quantized picture. 
However, it may be not convenient to use the second quantized picture for the fixed-$N$ ensemble since the ensemble is not Gaussian because of the fixed-$N$ condition.
Our formula \eqref{Formula_of_Renyi} is a rare example where an explicit formula of the R\'enyi entropy is available for non-Gaussian states. 

We applied the formula \eqref{Formula_of_Renyi} to a concrete system, $N$ free fermions on one-dimensional circle in section~\ref{sec:circle}.
We numerically computed the 2nd R\'enyi entropy for a single interval subregion. 
The results show that the 2nd R\'enyi entropy for a single interval is qualitatively the same as a portion of the thermal 2nd R\'enyi entropy \eqref{n-renyi-thermal} in the subregion. 
We also used the grand canonical ensemble and compared the result with that for the fixed-$N$ ensemble. 
It seems that they are equivalent for large $N$. 
However, although we expect that the fixed-$N$ ensemble is equivalent to the grand canonical one for large $N$ with fixed temperature, 
it is not sure if the equivalence holds in a different large $N$ limit with $N \beta$ fixed.
In fact, Fig.~\ref{Ratio_GC_and_exact_beta_other_plot} implies that the agreement of the fixed-$N$ and the grand canonical ensemble is worse at higher temperature. 
As we discussed, the large $N$ limit with $\beta$ fixed is somewhat trivial in a sense that the state is reduced to the ground state if the first energy gap grows with $N$ as the model in section~\ref{sec:circle}.
An interesting large $N$ limit in this model is $N \to \infty$ with $N \beta$ fixed. 
In this limit, it is not sure if the fixed-$N$ ensemble can be replaced by the grand canonical one. 
For the ground state, the target space R\'enyi entropy in the large $N$ limit is the same as the base space one for a conformal field theory, i.e., the two-dimensional free compact boson at self dual radius (see \cite{Sugishita:2021vih}). 
It is interesting to see if thermal states also have an agreement with a conformal field theory.

Properties of the target space (R\'enyi) entanglement entropy of $N$ fermions are drastically different between thermal states and a class of pure states considered in \cite{Sugishita:2021vih}.
In that paper, it is shown that the (R\'enyi) entanglement entropy for any pure states whose wave functions are the Slater determinants is bounded by $N \log 2$ independently of the states. 
This represents that only the $N$ states contained in the Slater determinants are involved in the computation of the entropy, and we can effectively regard the dimension of the Hilbert space as $2^N$. 
This fact also makes it easy to numerically compute the entropy. 
In contrast, for thermal states, all of states in the infinite-dimensional Hilbert space are involved.
Therefore, the (R\'enyi) entanglement entropy for thermal states is not bounded from above.
In particular, it diverges in the high temperature limit as the thermal entropy does.

It will be more interesting to investigate the target space entanglement for a one-matrix model dual to a two-dimensional string theory. 
It is also interesting to investigate a time-dependence of the target space entanglement by considering time-dependent states or quantum quench (see \cite{Das:2019qaj}). 
In addition, since the (R\'enyi) entanglement entropy is not a good measure of entanglement for thermal states (more generally for mixed states), it will be important to consider better measures of entanglement such as the target space negativity or relative entropy.

\section*{Acknowledgement}
SS thanks support from JSPS KAKENHI Grant Number 21K13927.
\appendix
\section{Formulae for wedge products of operators}\label{app:wedge}
In section~\ref{sec:QM-fermion}, we have introduced the wedge product of operators $A_1, \cdots, A_N$ as
\begin{align}
\label{def:wedge-op:app}
    A_1 \wedge \cdots \wedge A_N:= P^- (A_1 \otimes \cdots \otimes A_N) P^-, 
\end{align}
whose matrix elements in the basis \eqref{def:I-basis} are
\begin{align}
   (A_1 \wedge \dots \wedge A_N)_{I,I'}
   =\frac{1}{N!} \sum_{\sigma, \sigma' \in S_N}(-)^{\sigma\sigma'}
    (A_1)_{n_{\sigma(1)},n'_{\sigma'(1)}}
    \dots (A_N)_{n_{\sigma(N)},n'_{\sigma'(N)}}.
    \label{wedge-op-elem:app}
\end{align}
In this appendix, we summarize some formulae for the wedge product of operators. 

If we have $A_1=A_2=\cdots=A_N(=A)$, the elements \eqref{wedge-op-elem:app} are written as determinants,  
\begin{align}
    (A \wedge \dots \wedge A)_{I,I'}
    &= \sum_{\sigma\in S_N}(-)^{\sigma}
    A_{n_1,n'_{\sigma(1)}}\cdots A_{n_N,n'_{\sigma(N)}}
    = 
    \begin{vmatrix}
        A_{n_1,n'_1} & A_{n_1,n'_2} & \cdots & A_{n_1,n'_N}\\
        A_{n_2,n'_1} & A_{n_2,n'_2} & \cdots & A_{n_2,n'_N}\\
       \vdots & \vdots & \ddots & \vdots\\
        A_{n_N,n'_1} & A_{n_N,n'_2} & \cdots & A_{n_N,n'_N}
    \end{vmatrix}.
    \label{determinant-2}
\end{align}

We summarize the trace formula of the wedge product  \eqref{def:wedge-op:app}.
The trace is given by
\begin{align}
    \tr(A_1 \wedge \dots \wedge A_N)&:=\sum_{I}
    (A_1 \wedge \dots \wedge A_N)_{I,I}
    \nn
   &= \frac{1}{N!}
   \sum_{I}\sum_{\sigma, \sigma' \in S_N}(-)^{\sigma\sigma'}
    (A_1)_{n_{\sigma(1)},n_{\sigma'(1)}}
    \dots (A_N)_{n_{\sigma(N)},n_{\sigma'(N)}}.
\end{align}
We can replace the sum over index $I=(n_1, \cdots, n_N)$ by $\frac{1}{N!}\sum_{n_1,\cdots,n_N}$ because the summand is symmetric under the permutation of $n_1, \cdots, n_N$.
We thus have 
\begin{align}
    \tr(A_1 \wedge \dots \wedge A_N) &=\frac{1}{(N!)^2}\sum_{n_1,\dots,n_N}\sum_{\sigma, \sigma' \in S_N}(-)^{\sigma\sigma'}
    (A_1)_{n_{\sigma(1)},n_{\sigma'(1)}}
    \dots (A_N)_{n_{\sigma(N)},n_{\sigma'(N)}}
    \nn
    &=
    \frac{1}{(N!)^2}
   \sum_{\sigma, \sigma' \in S_N}\sum_{n_1,\dots,n_N}(-)^{\sigma'}
    (A_{\sigma(1)})_{n_{1},n_{\sigma'(1)}}
    \dots (A_{\sigma(N)})_{n_{N},n_{\sigma'(N)}}.
    \label{trA-permu}
\end{align}
This expression can be simplified furthermore as follows.

Let us focus on the following part 
\begin{align}
     \sum_{\sigma' \in S_N}
     \sum_{n_1,\dots,n_N}
     (-)^{\sigma'}
    (A_{\sigma(1)})_{n_{1},n_{\sigma'(1)}}
    \dots (A_{\sigma(N)})_{n_{N},n_{\sigma'(N)}}
    \label{sum_sigma'}
\end{align}
in \eqref{trA-permu}. 
In the sum over $\sigma' \in S_N$, two permutations have the same contributions if they have the same cycle type.
Since the conjugacy classes of permutation group $S_N$ are classified by the cycle types, 
the sum over $\sigma'$ can be written as a sum over the conjugacy classes of $S_N$. 
In addition, the conjugacy classes are in a one-to-one correspondence with the partitions of integer $N$. 
We use label $\lambda$ for the partitions of integer $N$.
The number of elements in a conjugacy class corresponding to a partition $\lambda$ is given by
\begin{align}
\label{absD}
    \frac{N!}{\prod_{k=1}^N (r_k!\, k^{r_k})},
\end{align}
where $r_k$ are defined for partition $\lambda$ as
\begin{align}
    N=\underbrace{1+1+\cdots+1}_{r_1}+\underbrace{2+2+\cdots+2}_{r_2}+\cdots=\sum_{k=1}^N kr_k.
\end{align}
For instance, for $N=3$, there are the following three partitions, and $r_k$ for each partition is given as follows;
\begin{alignat}{2}
3&=1+1+1\phantom{ab} &\rightarrow r_1=3,\ r_2=0,\ r_3=0,\\
&=1+2\phantom{ab} &\rightarrow r_1=1,\ r_2=1,\ r_3=0,\\
&=3\phantom{ab} &\rightarrow r_1=0,\ r_2=0,\ r_3=1.
\end{alignat}
These $r_k$ denote the numbers of cycles with length $k$ in permutation $\sigma'$ in the conjugacy class corresponding to $\lambda$.
Thus, the signature $(-)^{\sigma'}$ in the class $\lambda$ is determined by $r_k$ as
\begin{align}
\label{sgnD}
    (-1)^{N-\sum_{k=1}^N r_k}.
\end{align}

Furthermore, for each $\sigma'$ in \eqref{sum_sigma'}, if $\sigma'$ is in the class $\lambda$, the indices contract as follows
\begin{align}
 &\tr_{\lambda}(A_1,\dots, A_N)
 \nn
 &:=   \left(\prod_{i_1=1}^{r_1}\tr(A_{i_1})\right)
    \left(\prod_{i_2=1}^{r_2}\tr(A_{r_1+2i_2-1}A_{r_1+2i_2})\right)\cdots
     \left(\tr(\prod_{j_N=1}^N A_{\sum_{j=1}^{N-1}{j r_j}+j_N})\right).
\end{align}
Thus, sum \eqref{sum_sigma'} can be written as
\begin{align}
     \sum_{\lambda\vdash N}D_{\lambda}\tr_{\lambda}(A_{\sigma(1)},\dots, A_{\sigma(N)}),
\end{align}
where the sum of $\lambda$ runs over all the partitions of integer $N$, 
and the coefficient $ D_{\lambda}$ is given by
\begin{align}
    D_{\lambda}:=(-1)^{N-\sum_{k=1}^N r_k}
    \frac{N!}{\prod_{k=1}^N (r_k!\, k^{r_k})},
    \label{D}
\end{align}
whose absolute value $ |D_{\lambda}|$ is \eqref{absD} and sign is \eqref{sgnD}.

Therefore, we obtain the trace formula of the wedge product  \eqref{def:wedge-op:app} as
\begin{align}
    \tr(A_1 \wedge \dots \wedge A_N)= \frac{1}{(N!)^2}
   \sum_{\sigma\in S_N}
   \sum_{\lambda\vdash N}D_{\lambda}\tr_{\lambda}(A_{\sigma(1)},\dots, A_{\sigma(N)}).
   \label{tr-wedge}
\end{align}

If $A_1=\cdots=A_N(=A)$, the trace formula \eqref{tr-wedge} can be written by using determinant as
\begin{align}
     \tr(A \wedge \dots \wedge A)=
\frac{1}{N!}\left|\begin{array}{ccccc}
\tr A & N-1 & 0 & \cdots & \\
\tr A^{2} & \tr A & N-2 & \cdots & \\
\vdots & \vdots & & \ddots & \vdots \\
\tr A^{N-1} & \tr A^{N-2} & & \cdots & 1 \\
\tr A^{N} & \tr A^{N-1} & & \cdots & \tr A
\end{array}\right|.
\label{tr-wedge_same}
\end{align}
For example, if $N=3$ and $A=A_1=A_2=A_3$, we have
\begin{align}
    \tr(A\wedge A\wedge A)=\frac{1}{3!}((\tr A)^3-3\tr A^2\tr A+2\tr A^3).
\end{align}
These coefficients $(1,-3,2)$ can be calculated by (\ref{D}).

We also use 
products of the wedge product of operators. 
It is computed as follows. For example, for $N=2$, we have
\begin{align}
    &((A^{(1)}\wedge A^{(2)})(B^{(1)}\wedge B^{(2)}))_{I,J}=\sum_{I'}(A^{(1)}\wedge A^{(2)})_{I,I'}(B^{(1)}\wedge B^{(2)})_{I',J}.
\end{align} 
Let $I, I', J$ be labeled by set of integers $I=(n_1, n_2)$, $I'=(n'_1, n'_2)$, $J=(m_1, m_2)$ respectively. Then, the product becomes 
\begin{align}
    ((A^{(1)}\wedge A^{(2)})(B^{(1)}\wedge B^{(2)}))_{I,J}
    &=\frac{1}{2!}\sum_{n'_1,n'_2}\sum_{\sigma\in S_2}(-)^\sigma\frac{A^{(1)}_{n_{\sigma(1)},n'_1}A^{(2)}_{n_{\sigma(2)},n'_2}-A^{(1)}_{n_{\sigma(1)},n'_2}A^{(2)}_{n_{\sigma(2)},n'_1}}{2!}
    \nn
    &\qquad\quad \times\sum_{\sigma'\in S_2}(-)^{\sigma'}\frac{B^{(1)}_{n'_1,m_{\sigma'(1)}}B^{(2)}_{n'_2,m_{\sigma'(2)}}-B^{(1)}_{n'_1,m_{\sigma'(2)}}B^{(2)}_{n'_2,m_{\sigma'(1)}}}{2!}
    \nn
    &=\frac{1}{2!}\left[((A^{(1)}B^{(1)})\wedge(A^{(2)}B^{(2)}))_{I,J}+((A^{(1)}B^{(2)})\wedge(A^{(2)}B^{(1)}))_{I,J}\right]\nn
    &=\frac{1}{2!}\sum_{\sigma\in S_2}((A^{(1)}B^{(\sigma(1))})\wedge(A^{(2)}B^{(\sigma(2))}))_{I,J}.
\end{align} 
In a similar way, we can show that the product formula for general $N$ is  given by
\begin{align}
    ((A_1\wedge\cdots\wedge A_N)(B_1\wedge\cdots\wedge B_N))_{I,J}
    &=\frac{1}{N!}\sum_{\sigma\in S_N}\left((A_1B_{\sigma(1)})\wedge\cdots\wedge(A_{N}B_{\sigma(N)})\right)_{I,J}.
    \label{prod-wedge}
\end{align}

\section{Spectrum of fermions on circle}\label{app:spec}
In this appendix, we summarize the spectrum of the model considered in section~\ref{sec:circle}, i.e, fermions on a one-dimensional circle with length $L$.
The energy eigenvalues of a single free particle on the circle are labeled by an integer $n$ as $E_n=\frac{1}{2m}\left(\frac{2\pi n}{L}\right)^2$
$(-\infty<n<\infty)$. 
Thus, the energy eigenstates of $N$-fermion system are labeled by sets of $N$ distinct integers $\{n_1,\cdots,n_N\}$ with energy $E=\frac{1}{2m}\left(\frac{2\pi}{L}\right)^2 \sum_{i=1}^N n_i^2$.

Here we consider the ground and first excited states of the system, and summarize the degeneracies $d_0$ and $d_1$.
\begin{itemize}
\item For $N=1$, the ground state (GS) is unique  $\{0\}$ with $E_{\mathrm{GS}}=0$. 
Thus, $d_0=1$. The first excited states (1ES) are given by $\{1\}$ and $\{-1\}$  and then $d_1=2$ with $E_{\mathrm{1ES}}=\frac{1}{2m}(\frac{2\pi}{L})^2$. 
\item For $N=2$, the GSs are given by $\{0,1\}$ and $\{-1,0\}$ ($d_0=2$) with $E_{\mathrm{GS}}=\frac{1}{2m}(\frac{2\pi}{L})^2$. 
The 1ES is given by $\{-1,1\}$ $(d_0=1)$ with $E_{\mathrm{1ES}}=\frac{1}{m}(\frac{2\pi}{L})^2$. 
\item For odd $N=2k+1\; (k\geq1)$, the GS is unique $(d_0=1)$ as $\{-k,\cdots,0,\cdots,k\}$  with $E_{\mathrm{GS}}=\frac{1}{2m}(\frac{2\pi}{L})^2\cdot\frac{1}{3}k(k+1)(2k+1)$. 
The 1ES is degenerated as $\{-k-1, -k,\cdots,k-1\},\;\{-k+1,\cdots,k,k+1\},\;\{-k-1 ,-k+1,\cdots,k-1,k\}$ and $\{-k,-k+1,\cdots,k-1,k+1\}$ $(d_1=4)$ with $E_{\mathrm{1ES}}=\frac{1}{2m}(\frac{2\pi}{L})^2\frac{1}{3}(2k+1)(k^2+k+3)$.
\item For even $N=2k\; (k\geq2)$, the GS is degenerated as $\{-k,-k+1,\cdots,k-1\}$ and $\{-k+1,-k+2,\cdots,k\}$ $(d_0=2)$ with $E_{\mathrm{GS}}=\frac{1}{2m}(\frac{2\pi}{L})^2\frac{1}{3}k(2k^2+1)$. 
The 1ES is also degenerated as
$\{-k,-k+1,\cdots,k-3, k-2,k\}$ and $\{-k,-k+2,-k+3,\cdots ,k-1,k\}$ $(d_1=2)$ with $E_{\mathrm{1ES}}=\frac{1}{2m}(\frac{2\pi}{L})^2\cdot\frac{1}{3}(2k^3+7k-3)$. 
\end{itemize}
The degeneracies $d_0$ and $d_1$ are summarized in Table~\ref{tab:degen}.
\begin{table}[htbp]
\centering
\begin{tabular}{l|l|l}
       & $d_0$ & $d_1$ \\ \hline\hline 
$N=1$    & 1    & 2    \\ 
$N=2$    & 2    & 1    \\ 
$N=\text{odd}\;(\geq3)$  & 1    & 4    \\
$N=\text{even}\;(\geq4)$ & 2    & 2    \\
\end{tabular}
\caption{Degeneracies for ground and first excited states.}
\label{tab:degen}
\end{table}

For large $N$, the first energy gap $E_{\mathrm{1ES}}-E_{\mathrm{GS}}$ is $\mathcal{O}(N)$.

\section{Other plots}\label{app:plots}
In this appendix, we show some plots similar to those in section~\ref{sec:circle} with different parameters.
\begin{figure}[H]
\begin{center}
  \begin{minipage}[b]{.5\linewidth}
    \centering
    \includegraphics[width=1\linewidth]{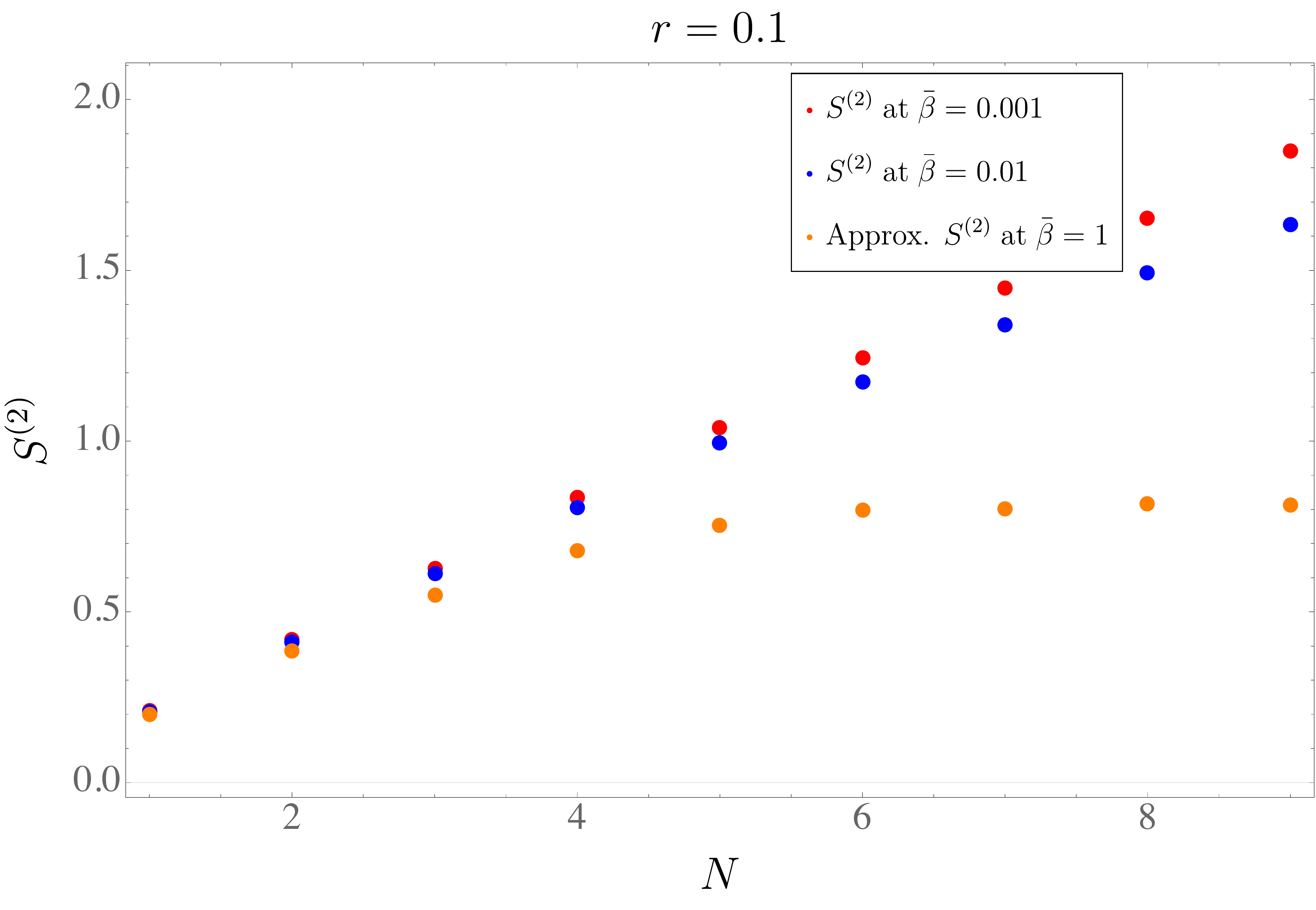}
  \end{minipage}%
  \begin{minipage}[b]{.5\linewidth}
    \centering
    \includegraphics[width=1\linewidth]{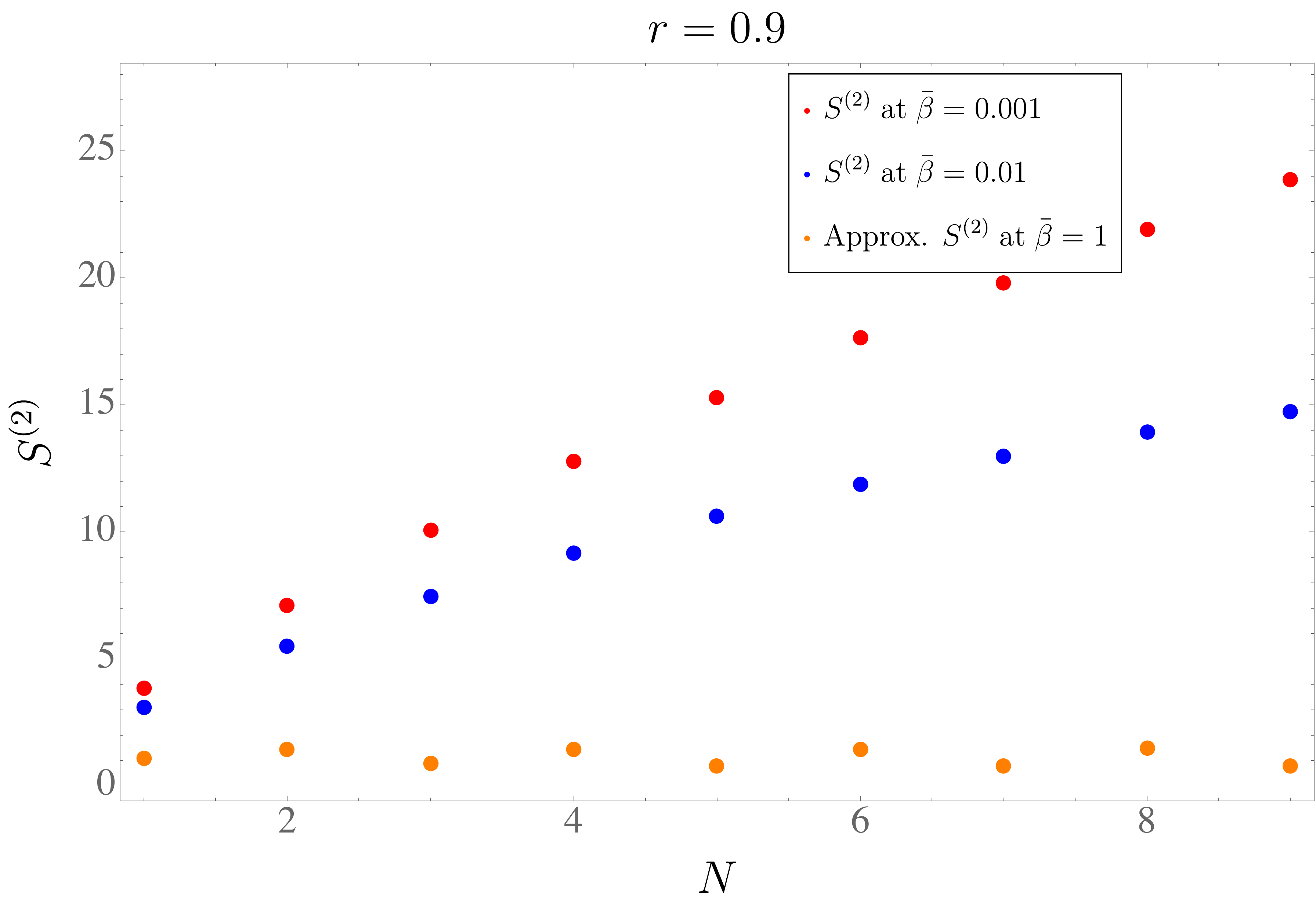}
  \end{minipage}\vspace{-1em}
  \end{center}
  \vspace{-1em}
  \caption{$N$-dependence of the 2nd R\'enyi entropy at $r=0.1,\;0.9$. They are similar to Fig.~\ref{N_Renyi_r=0.5} with $r=0.5$.  The right plot also indicates that the entropy for thermal states can exceed the upper bound $N \log 2$ for pure states \eqref{up-bound}.}
  \label{N_Renyi_r=0.1,0.9}
\end{figure}
\begin{figure}[H]
\begin{tabular}{cc}
\begin{minipage}[t]{0.5\hsize}
\includegraphics[width=8.4cm]{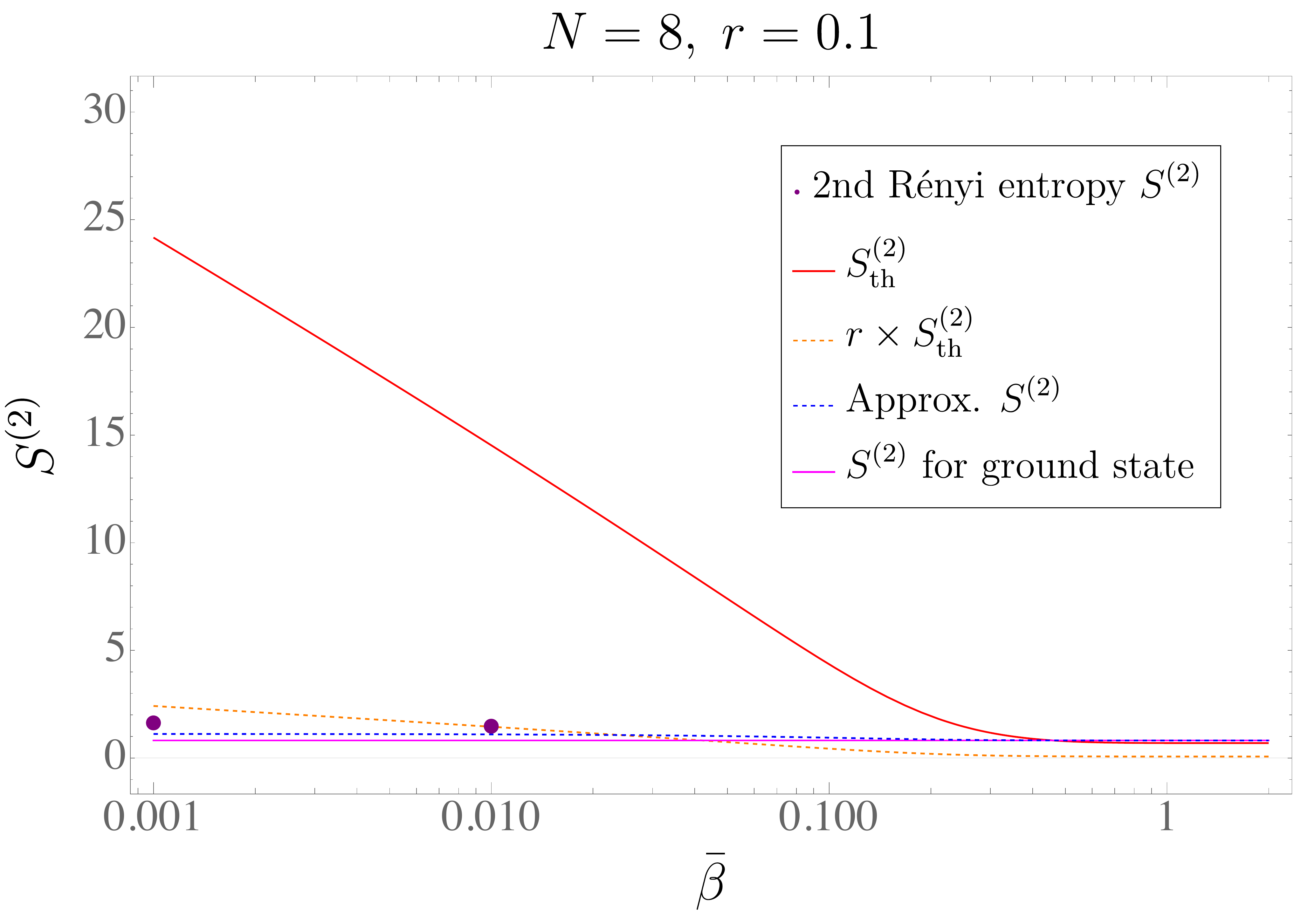}
\end{minipage}
\begin{minipage}[t]{0.5\hsize}
\includegraphics[width=8.4cm]{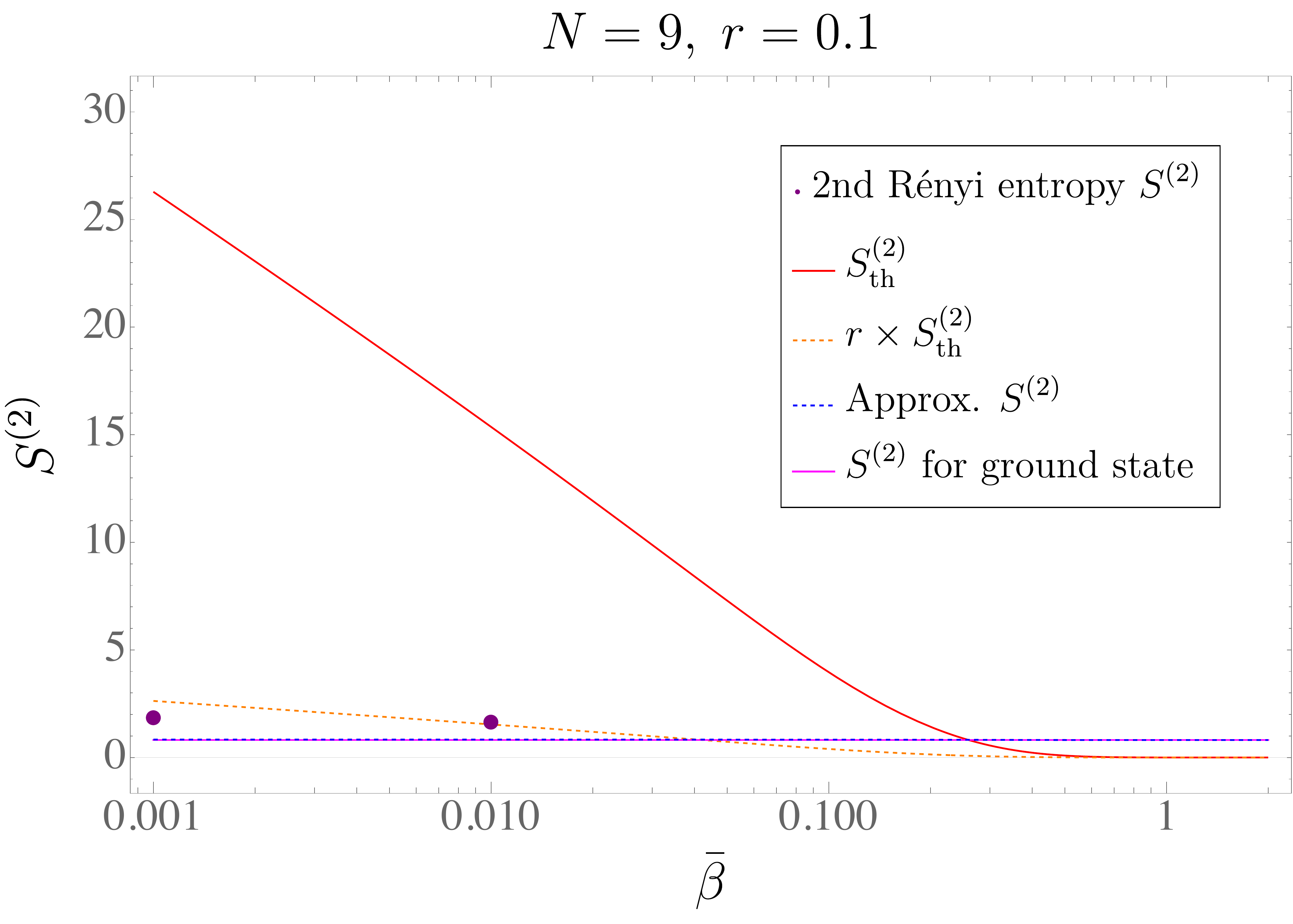}
\end{minipage}
\\
\begin{minipage}[c]{0.5\hsize}
\includegraphics[width=8.4cm]{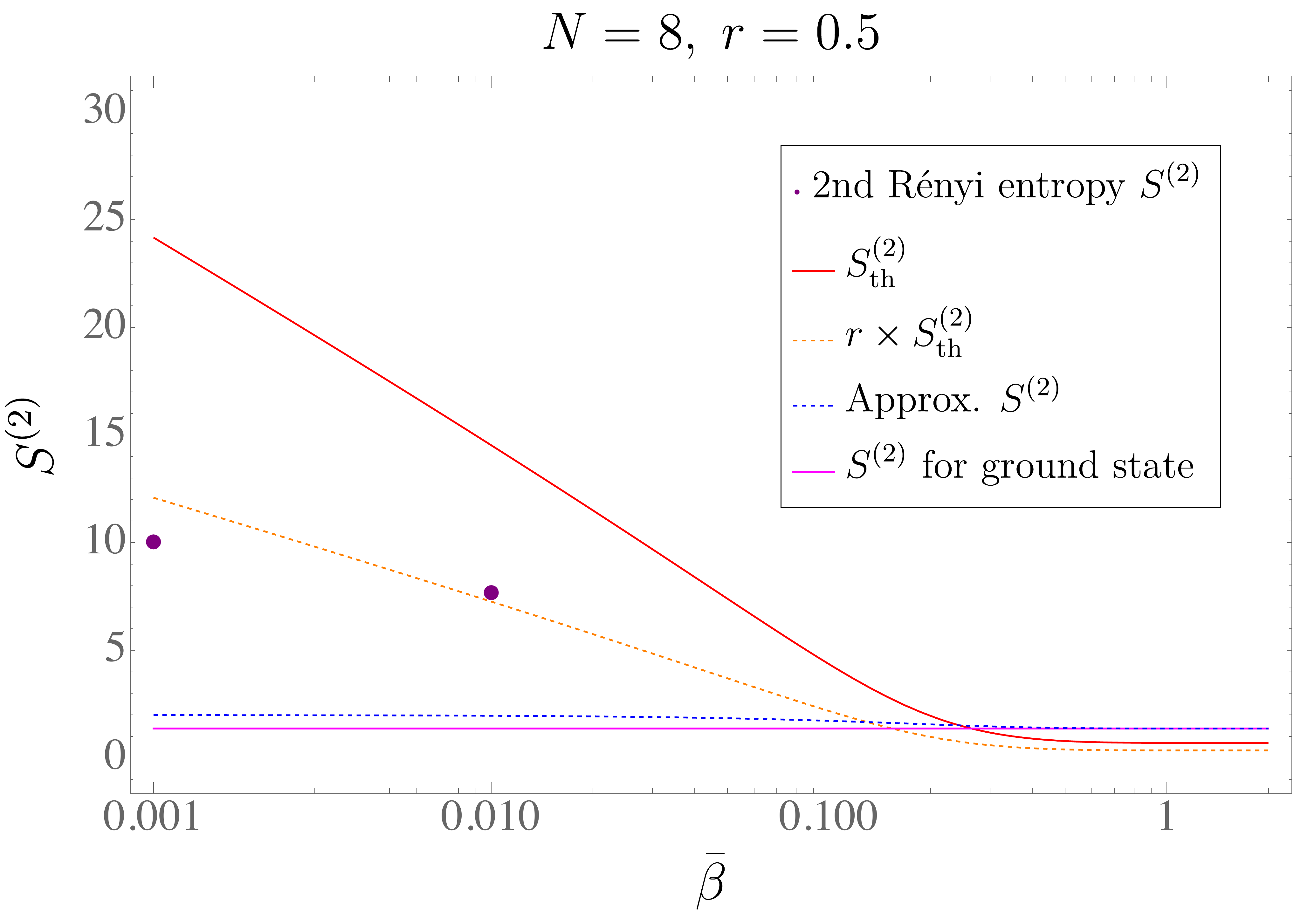}
\end{minipage}
\begin{minipage}[c]{0.5\hsize}
\includegraphics[width=8.4cm]{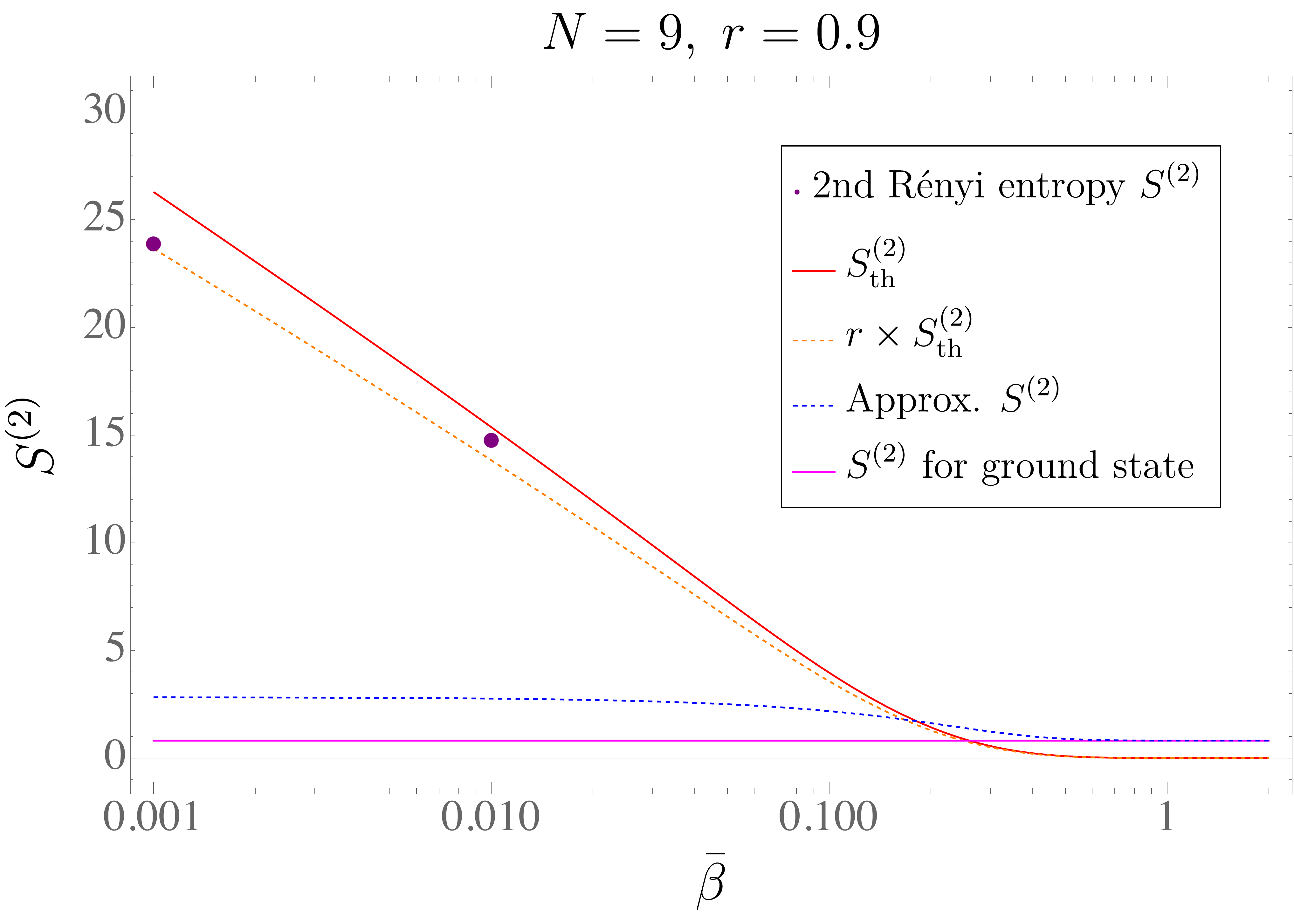}
\end{minipage}
\\
\multicolumn{2}{c}{
\centering
\begin{minipage}[b]{1\hsize}
\includegraphics[width=8.4cm]{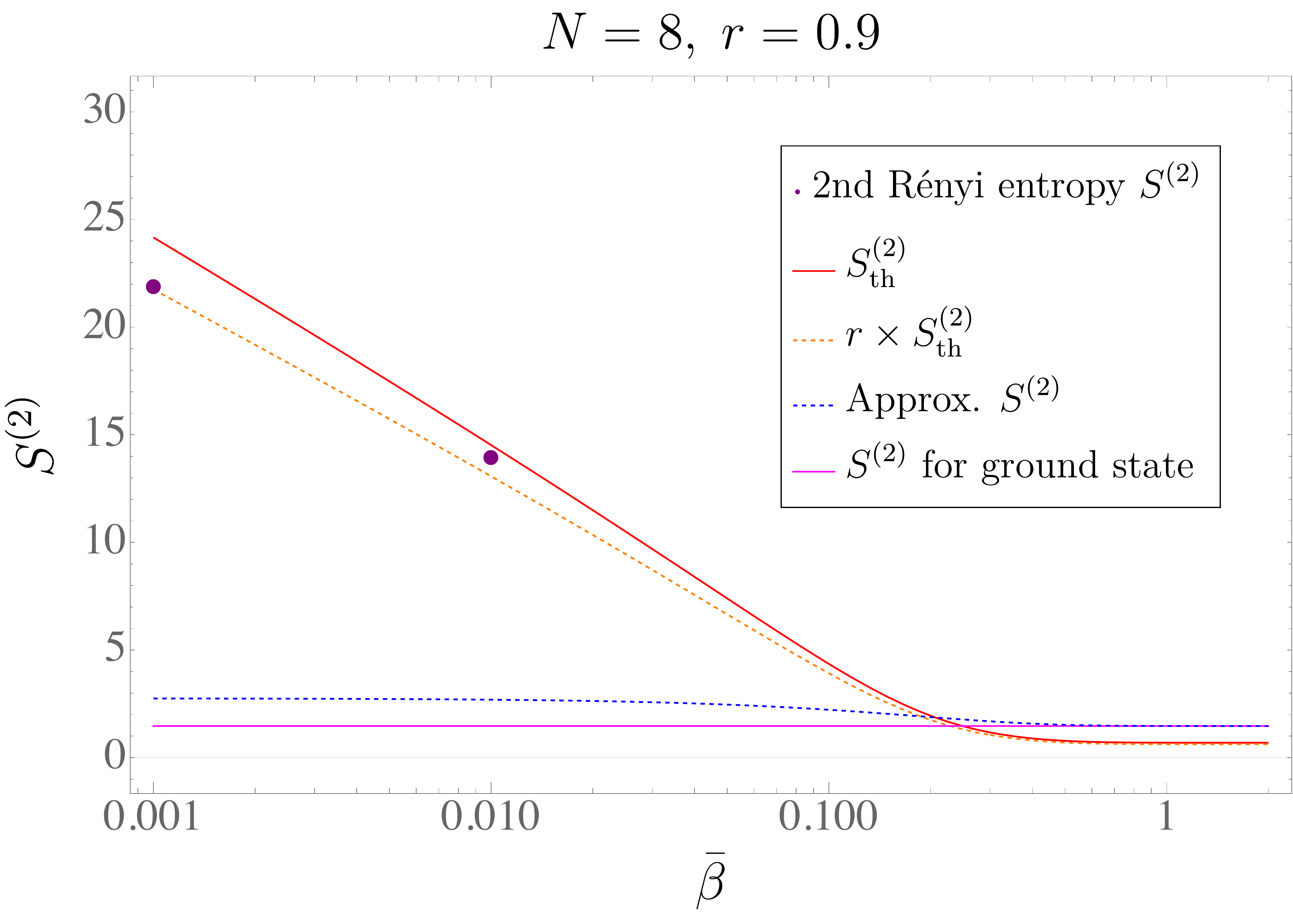}
\end{minipage}
}
\end{tabular}
\vspace{-1em}
\caption{$\bar{\beta}$-dependence of the 2nd R\'enyi entropy for $N=8, 9$ with $r=0.1, 0.5, 0.9$ as similar to Fig.~\ref{S_th_and_Renyi_N=9_r=0.5} with $N=9, r=0.5$.}
\label{S_th_and_Renyi}
\end{figure}
\begin{figure}[H]
\begin{tabular}{cc}
\begin{minipage}[t]{0.5\hsize}
\includegraphics[width=8.6cm]{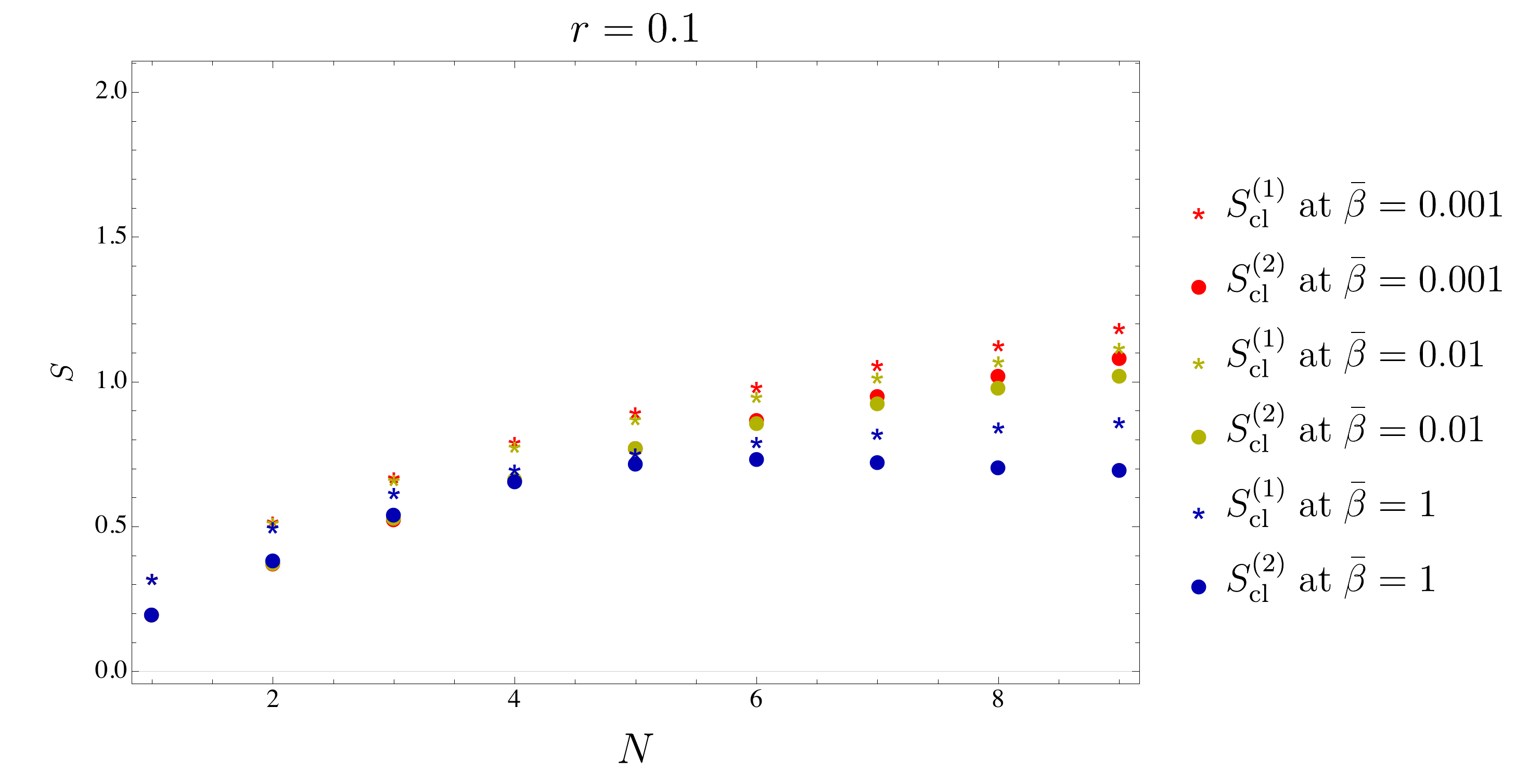}
\end{minipage}
\begin{minipage}[t]{0.5\hsize}
\includegraphics[width=9cm]{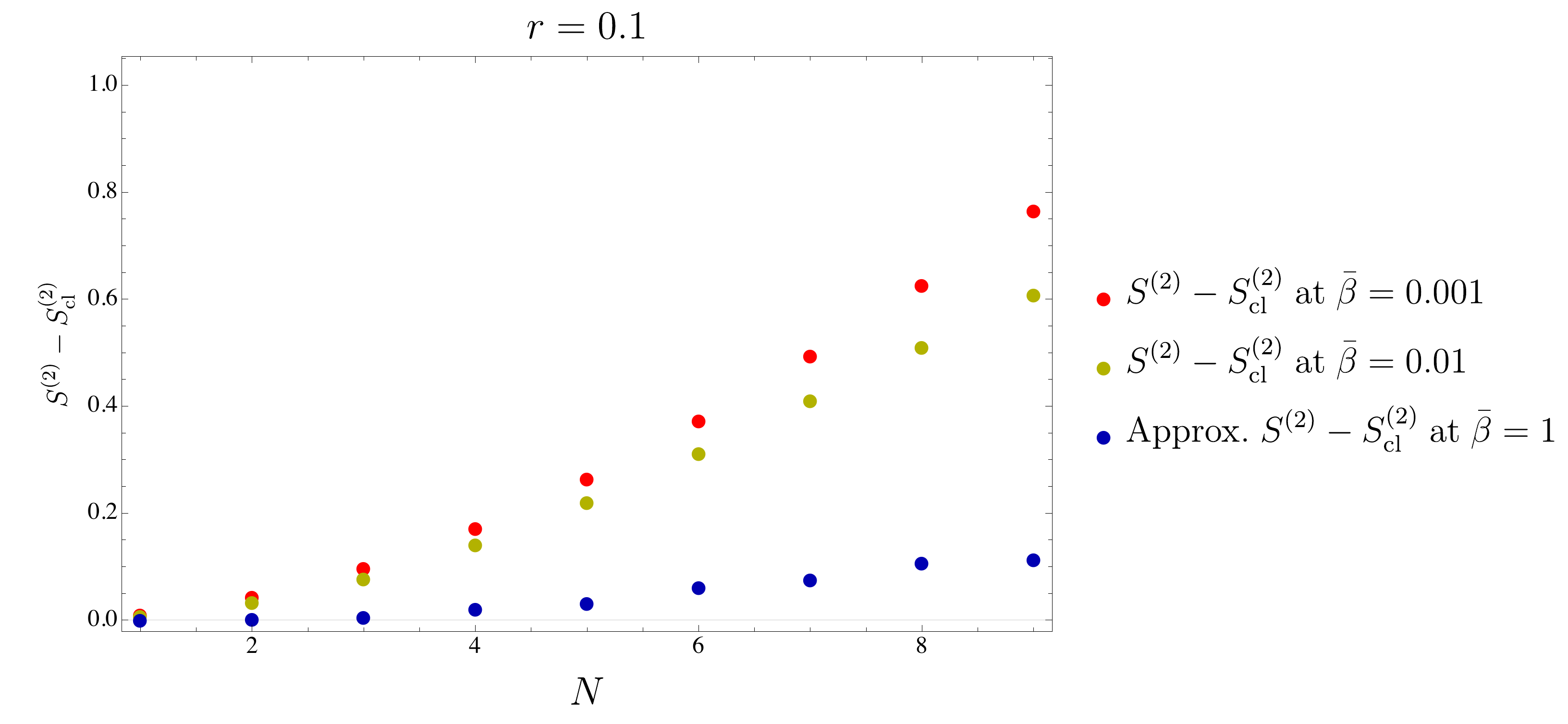}
\end{minipage}
\\
\begin{minipage}[c]{0.5\hsize}
\includegraphics[width=8.6cm]{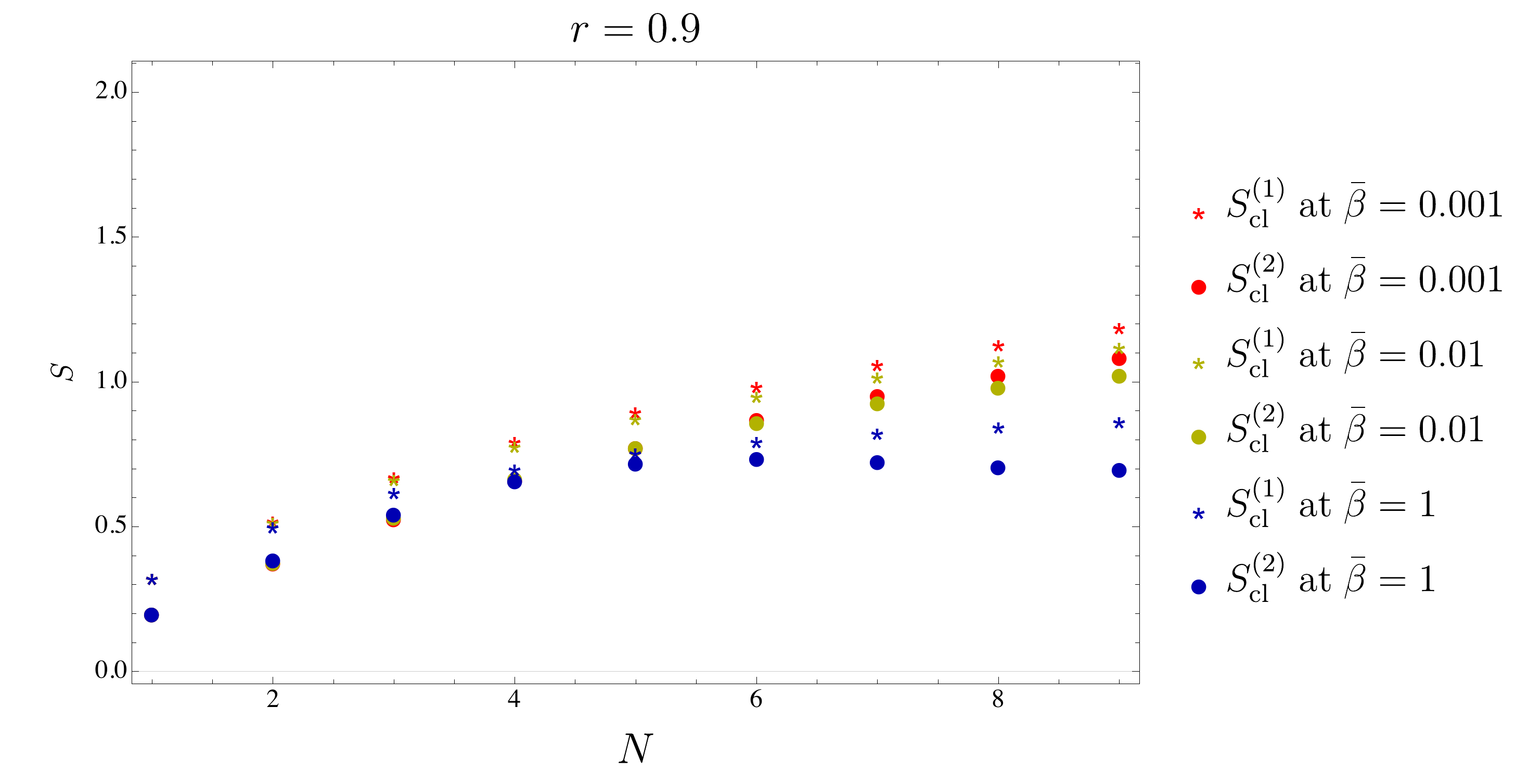}
\end{minipage}
\begin{minipage}[c]{0.5\hsize}
\includegraphics[width=9cm]{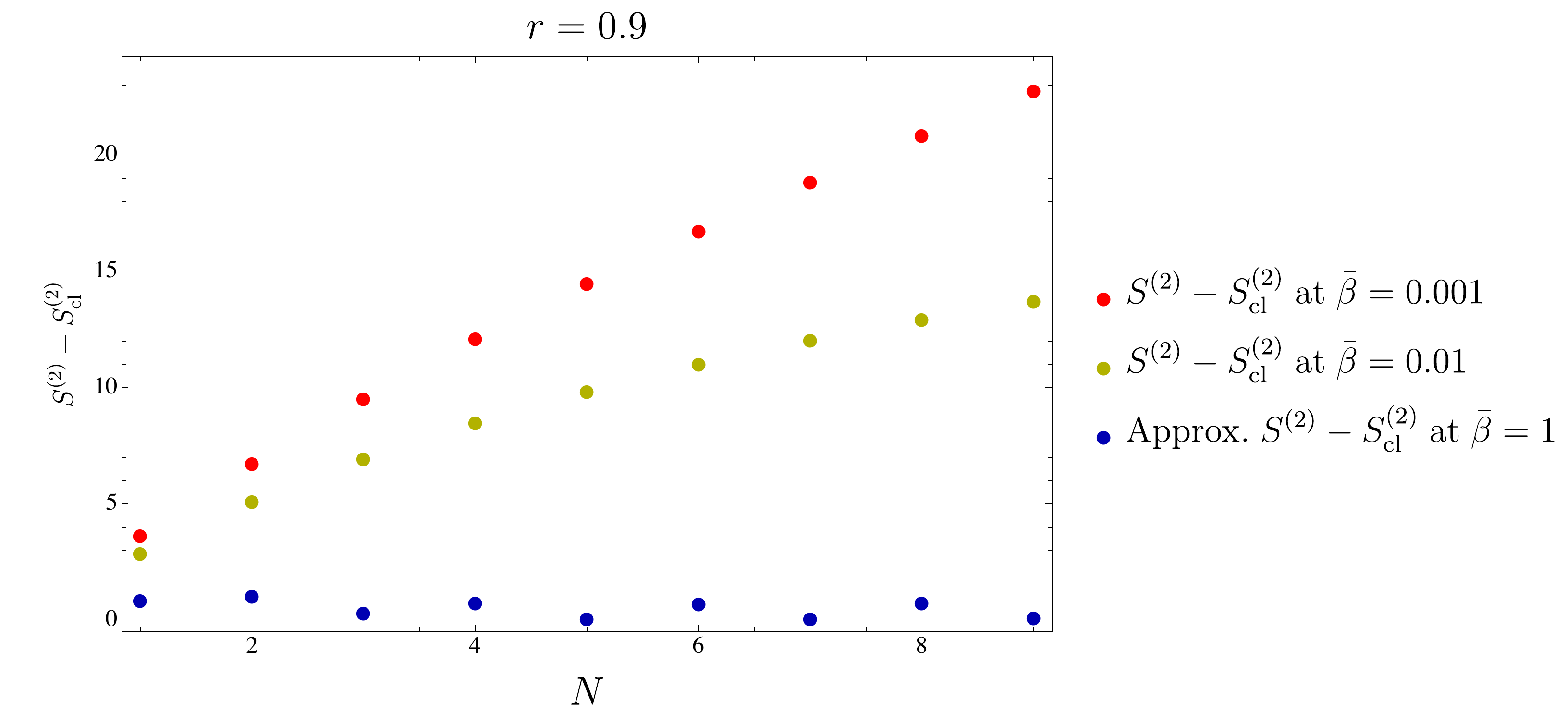}
\end{minipage}
\end{tabular}
\caption{$N$-dependence of the classical term of the 2nd R\'enyi entropy and of the entanglement entropy (left), and the difference between them for the 2nd R\'enyi entropy  (right) as similar to Fig.\ref{r=0.5_Renyi_and_entangle_classical} and Fig.~\ref{r=0.5_Renyi_minus_S_classical}  with $r=0.5$. 
Here we take $r=0.1$ (upper) and $r=0.9$ (lower).
The classical entropies of $r=0.1$ and $r=0.9$ are the same as seen in the left figures because the probability $p_k$ for a region $r=0.1$ is the same as $p_{N-k}$ for the complement region $r=0.9$.}
\label{Renyi_minus_S_classical}
\end{figure}

\begin{figure}[H]
\begin{tabular}{cc}
\begin{minipage}[t]{0.5\hsize}
\includegraphics[width=8.4cm]{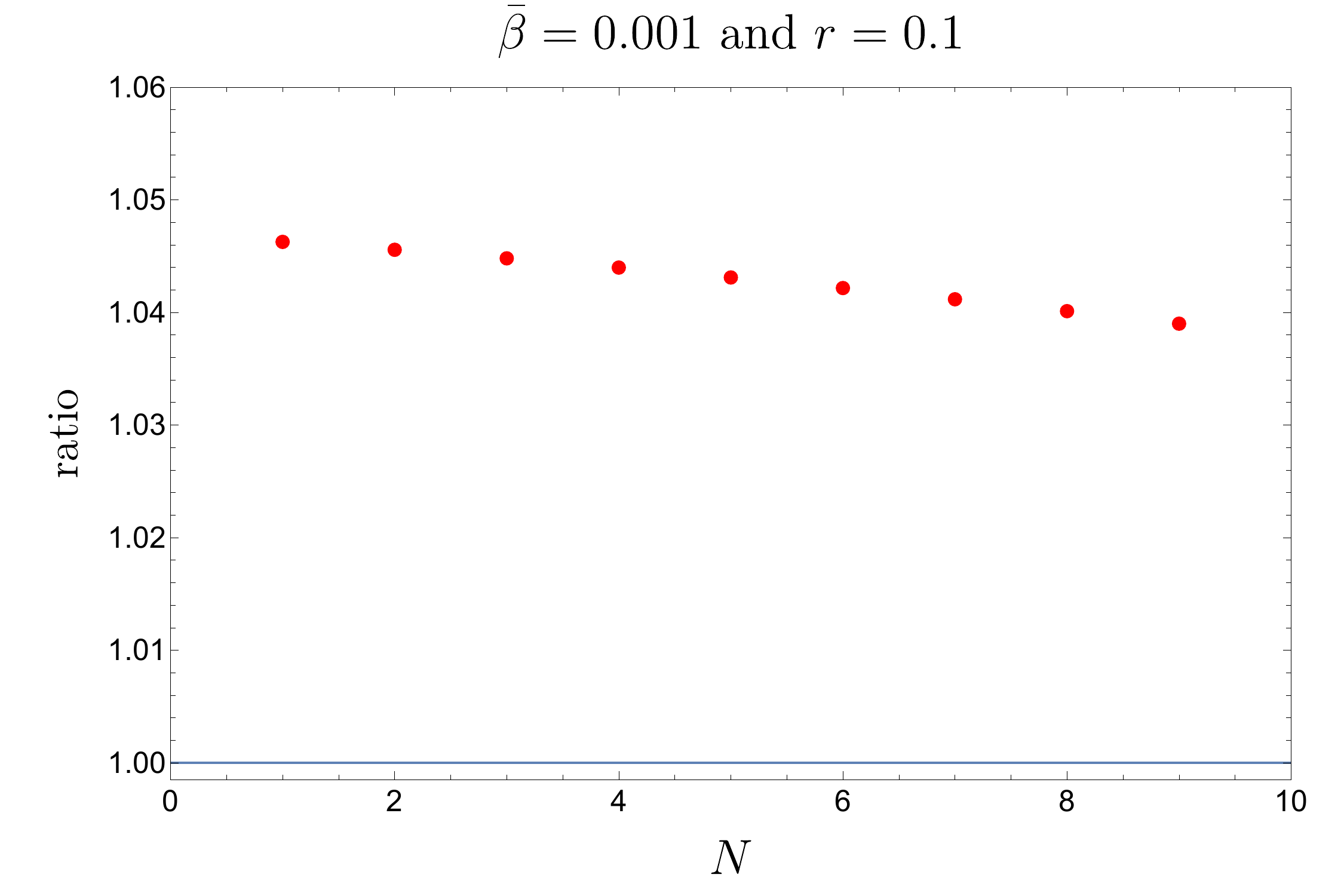}
\end{minipage}
\begin{minipage}[t]{0.5\hsize}
\includegraphics[width=8.4cm]{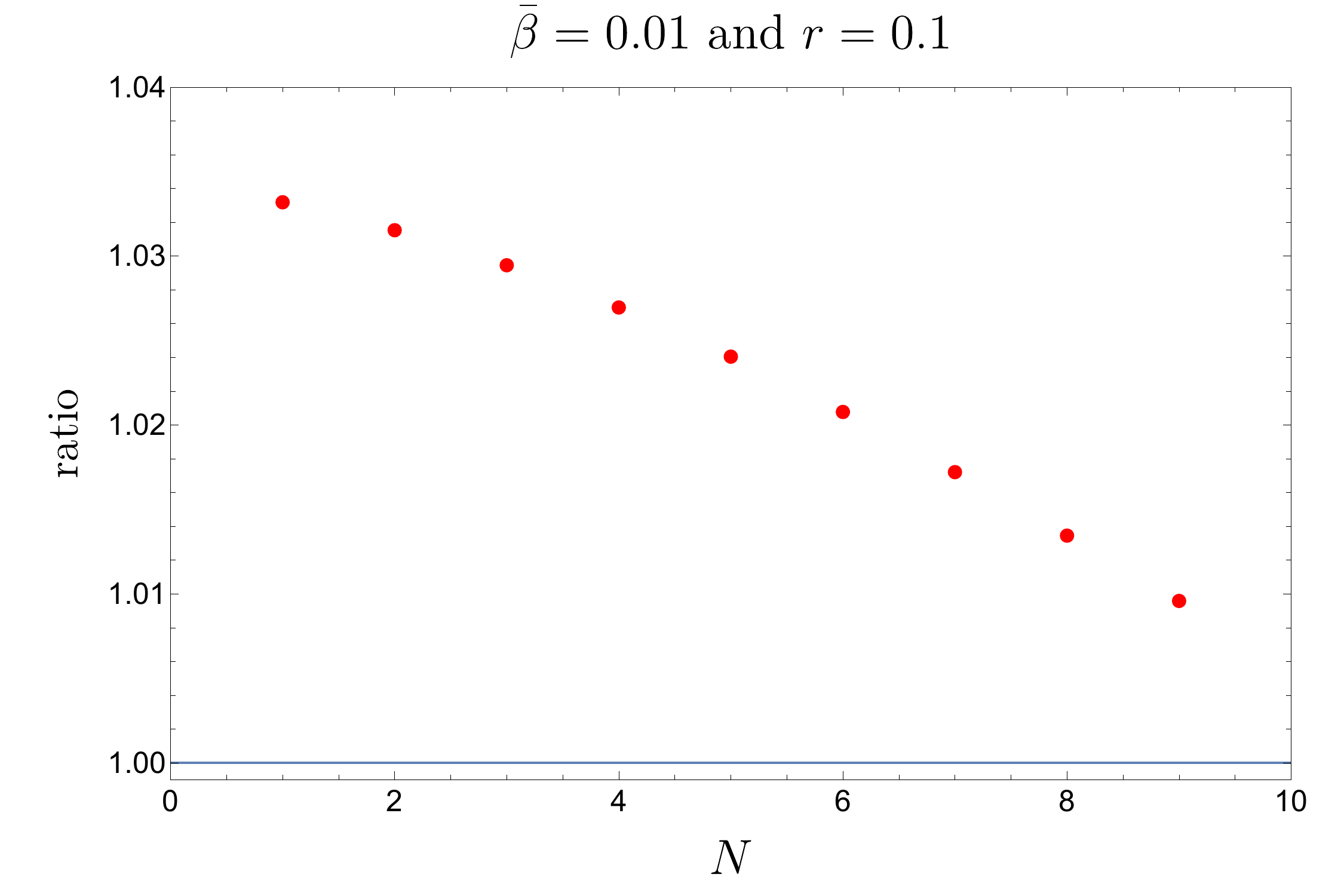}
\end{minipage}
\\
\begin{minipage}[c]{0.5\hsize}
\includegraphics[width=8.4cm]{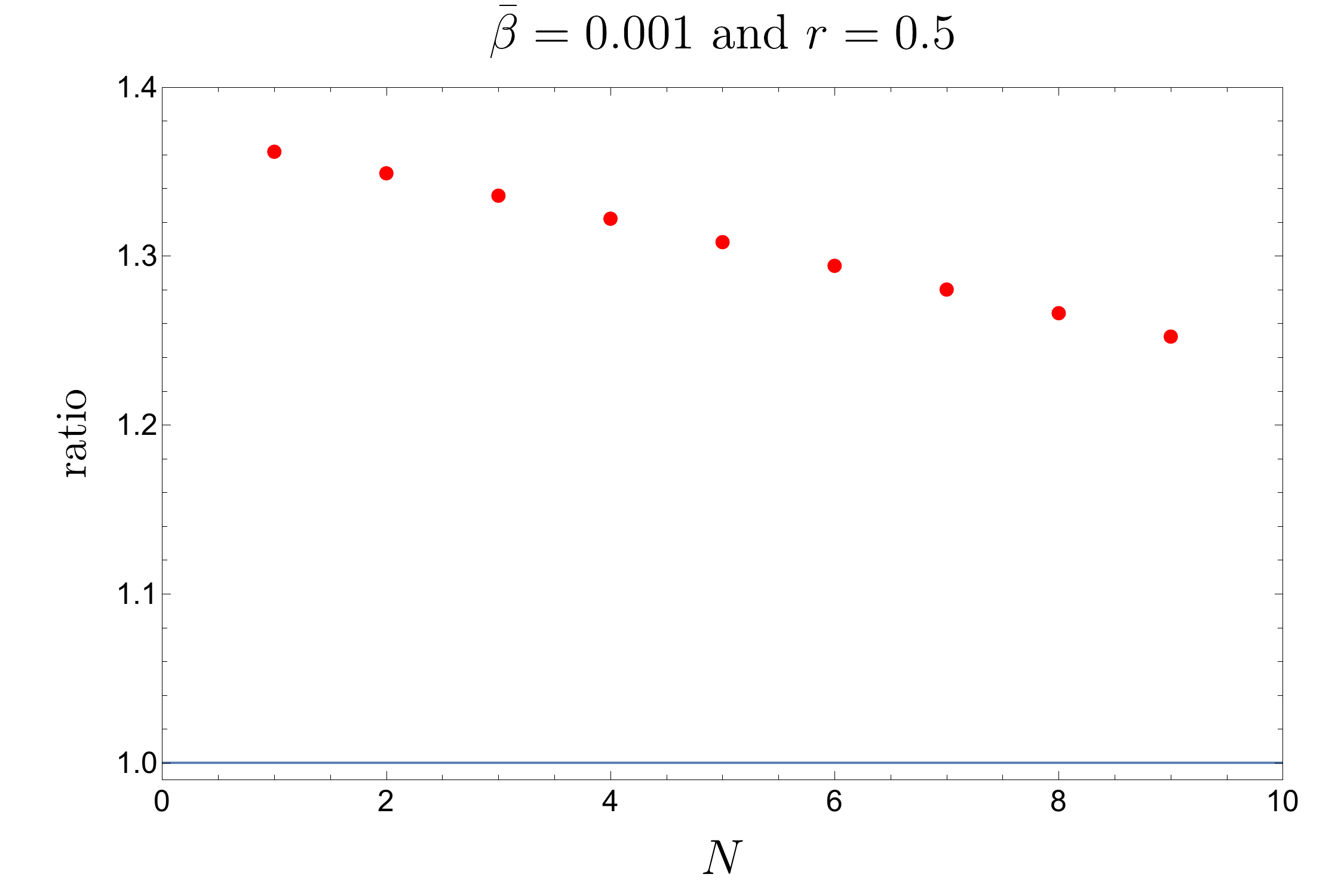}
\end{minipage}
\begin{minipage}[c]{0.5\hsize}
\includegraphics[width=8.4cm]{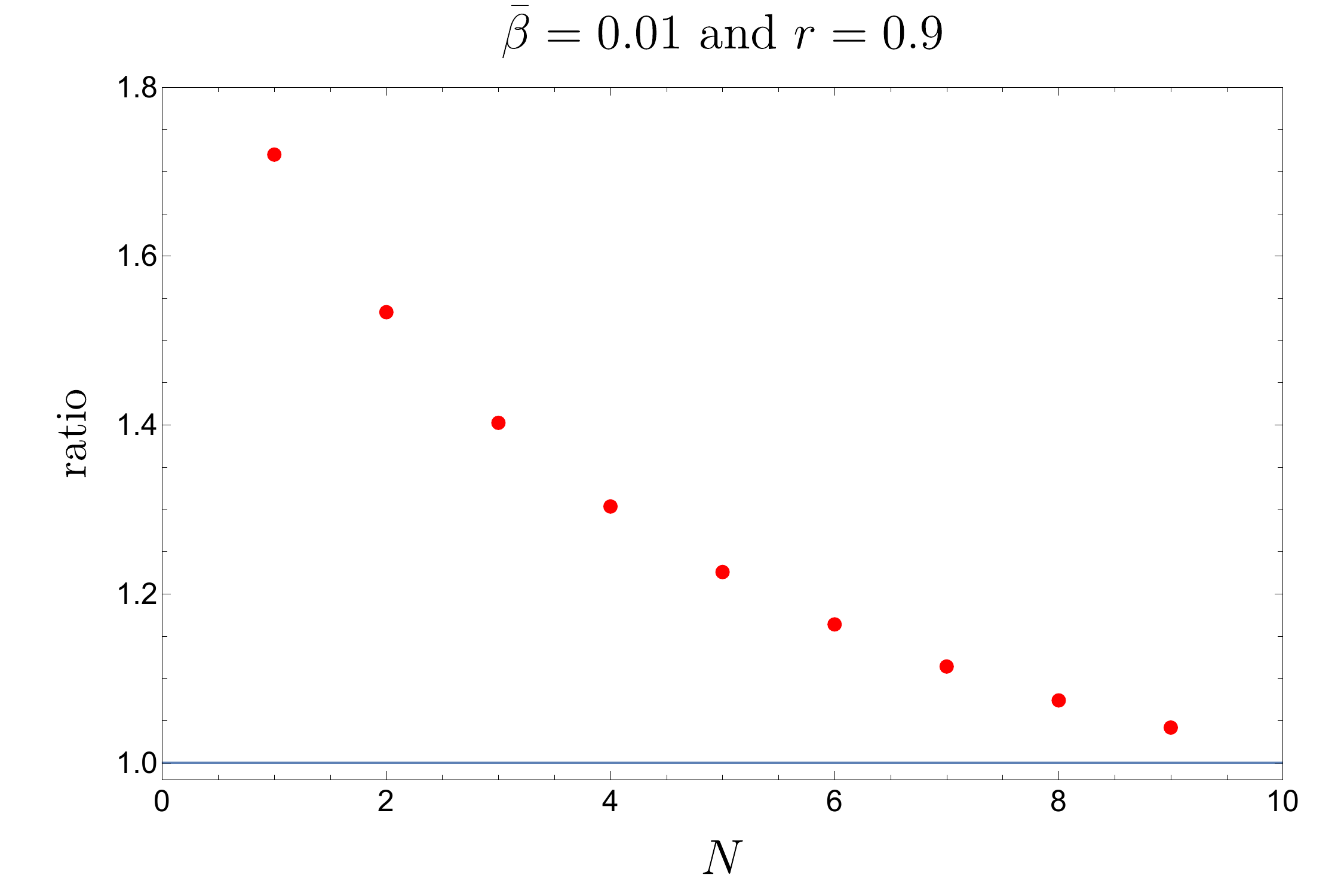}
\end{minipage}
\\
\multicolumn{2}{c}{
\centering
\begin{minipage}[b]{1\hsize}
\includegraphics[width=8.4cm]{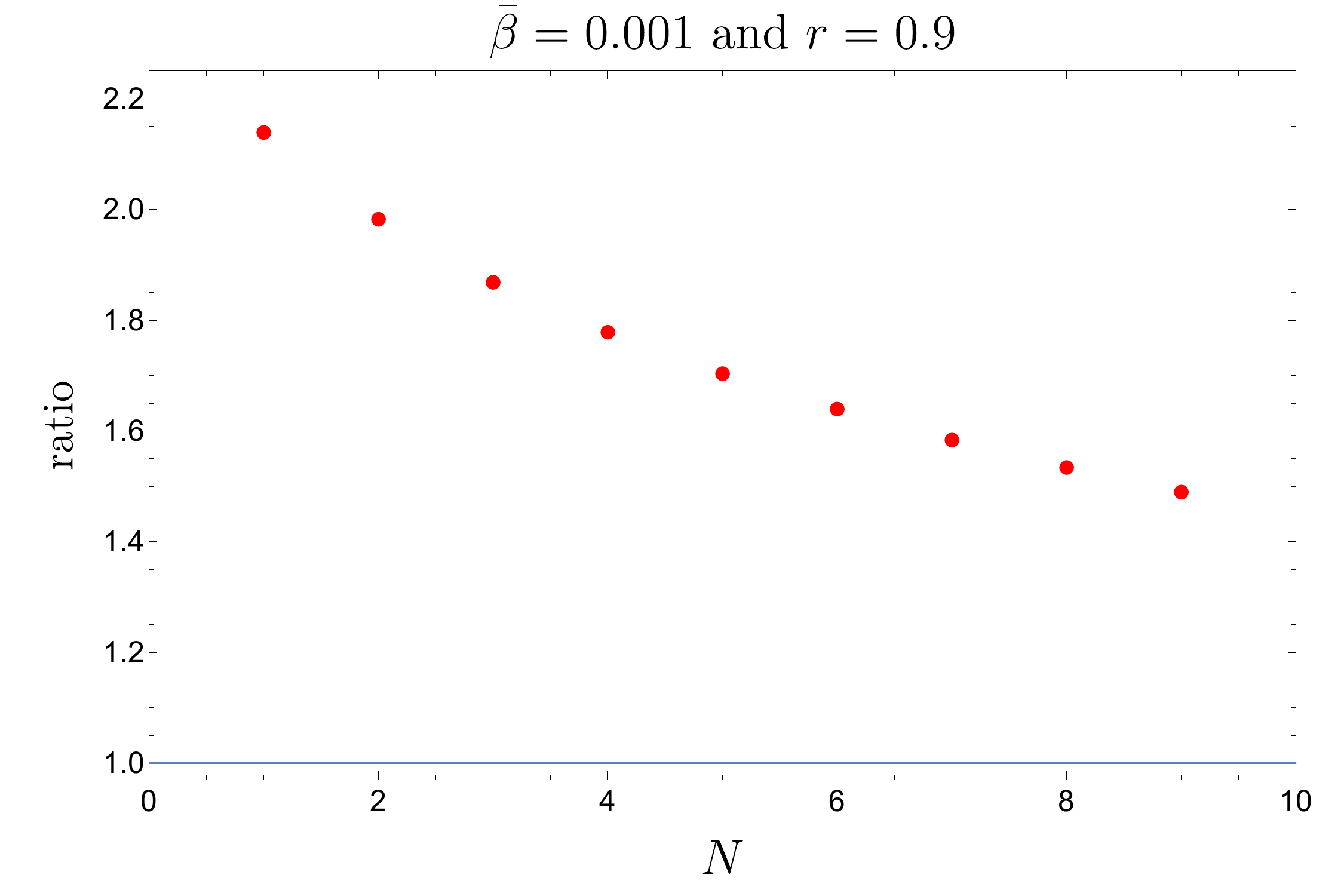}
\end{minipage}
}
\end{tabular}
\caption{The ratio $S^{(2)}\text{ (fixed $N$)}$ over $S^{(2)}\text{ (grand canonical)}$. These figures indicate that the ratio approaches 1 as $N$ increases similar to Fig.~\ref{Ratio_GC_and_exact_beta=0.01_r=0.5} with $\bar\beta=0.01$ and $r=0.5$.}
\label{Ratio_GC_and_exact_beta_other_plot}
\end{figure}
\begin{figure}[H]
\begin{tabular}{cc}
\begin{minipage}[t]{0.5\hsize}
\includegraphics[width=9cm]{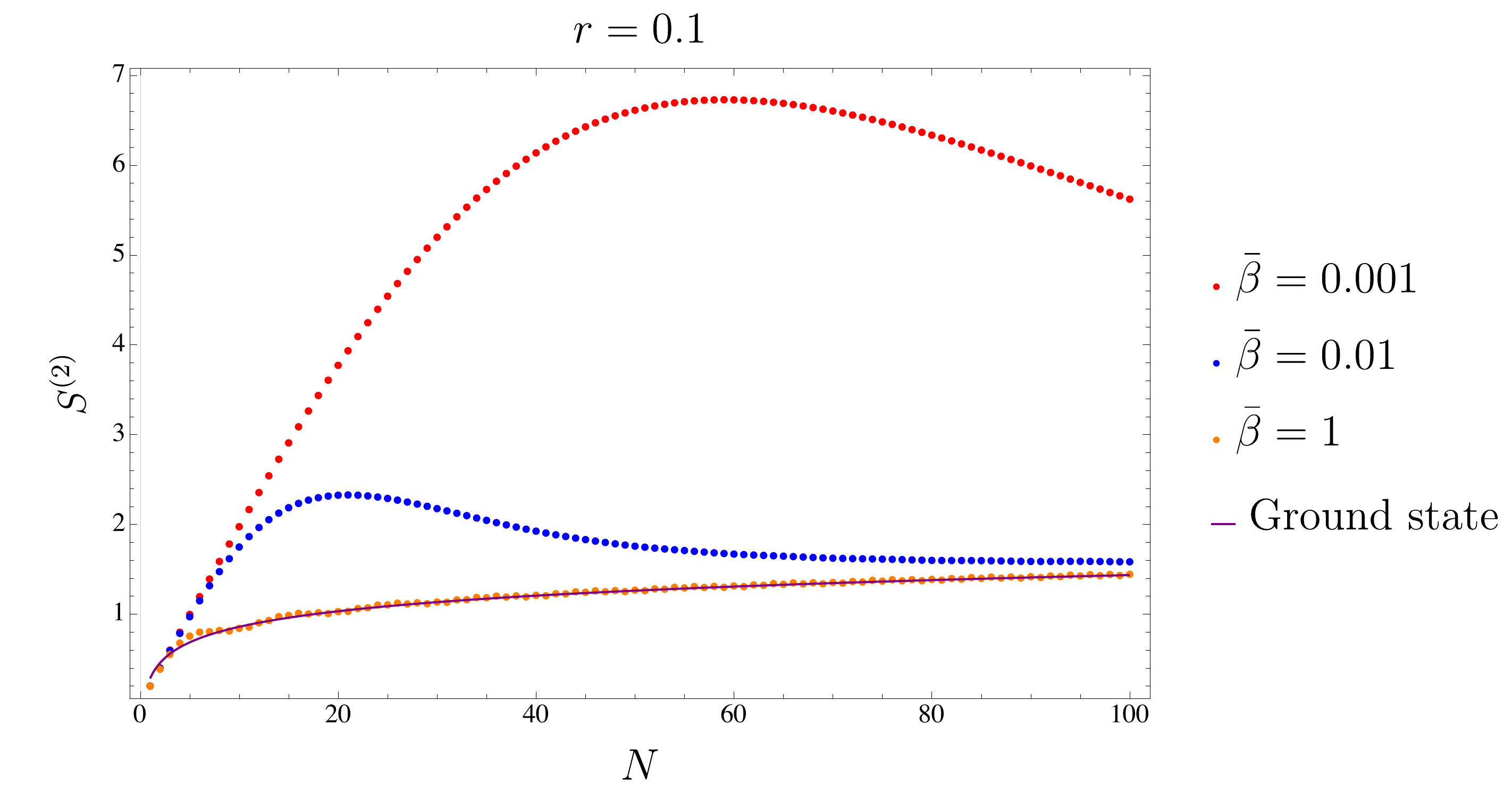}
\end{minipage}
\begin{minipage}[t]{0.5\hsize}
\includegraphics[width=9cm]{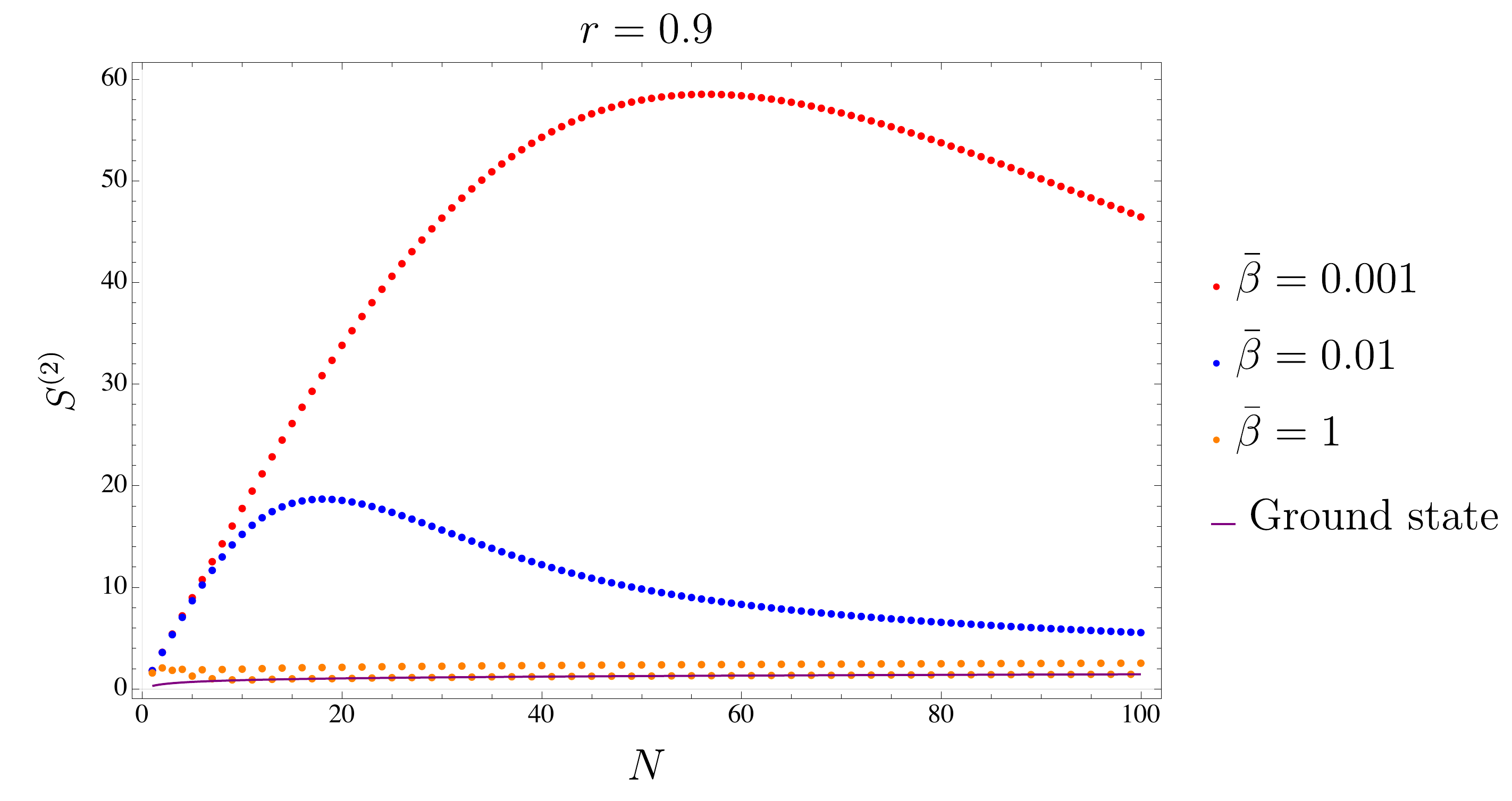}
\end{minipage}
\\
\begin{minipage}[c]{0.5\hsize}
\includegraphics[width=9cm]{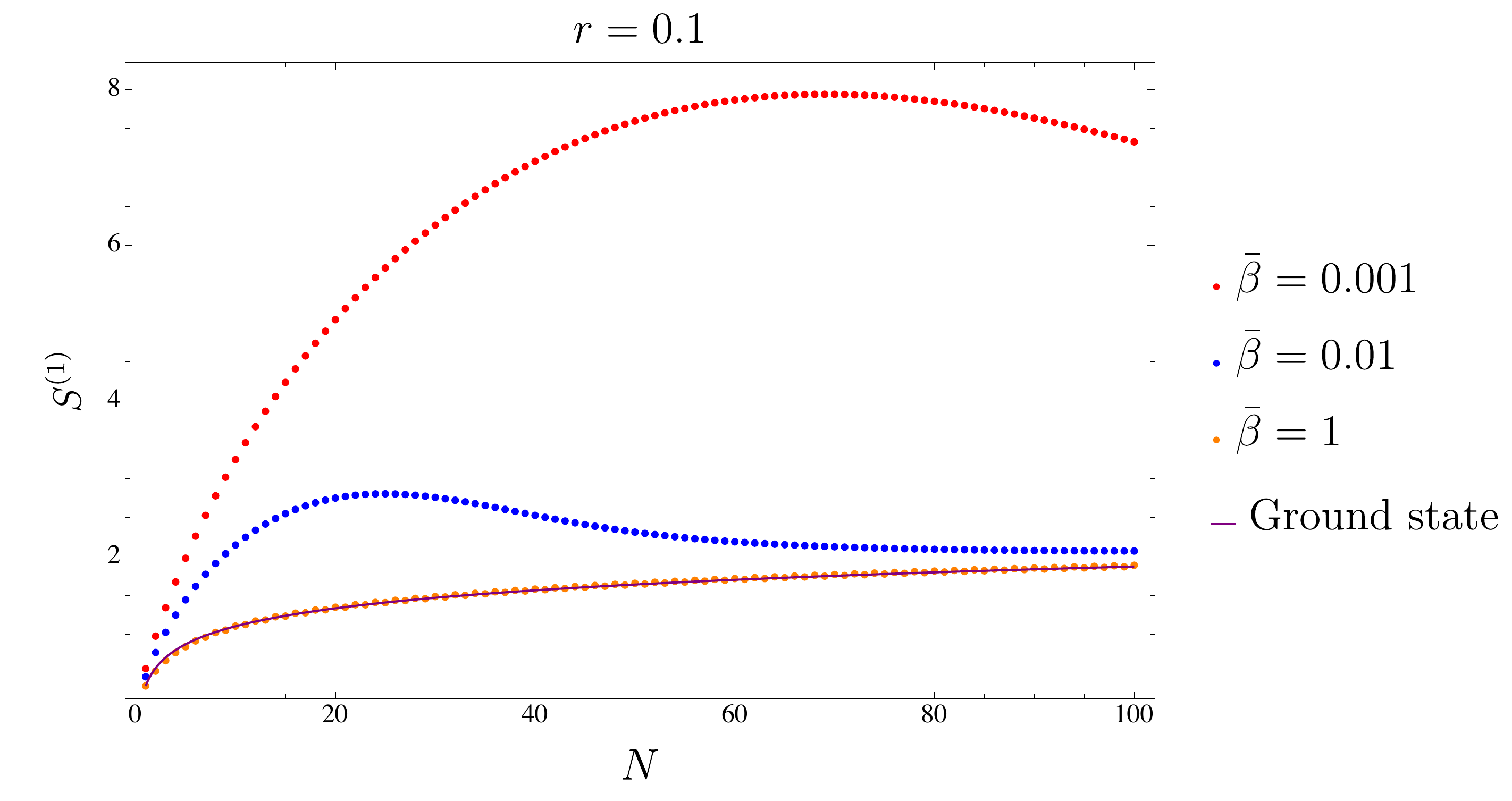}
\end{minipage}
\begin{minipage}[c]{0.5\hsize}
\includegraphics[width=9cm]{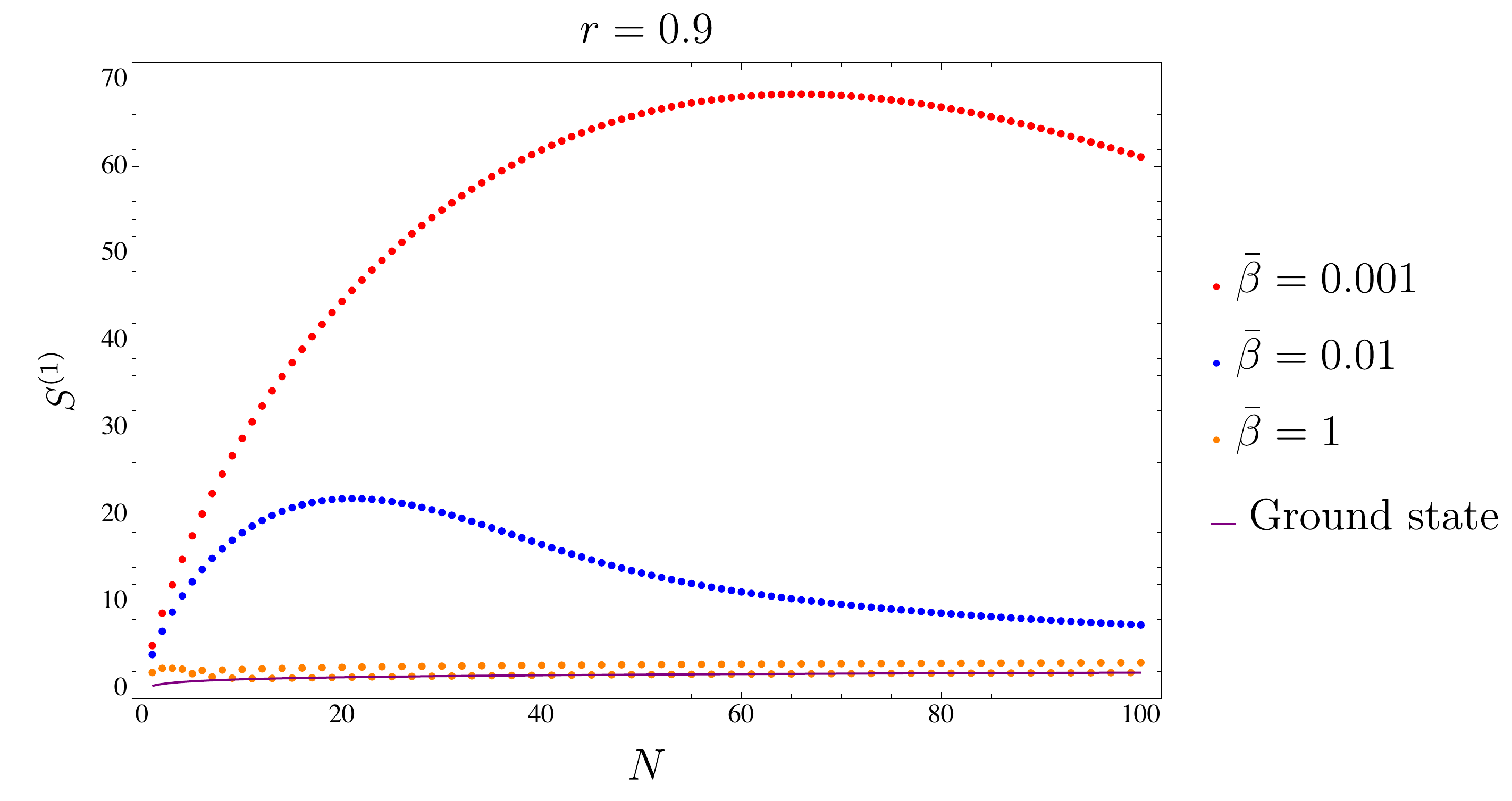}
\end{minipage}
\end{tabular}
\caption{2nd R\'enyi (upper) and  entanglement entropy (lower) for $r=0.1$ (left) and $r=0.9$ (right) for the grand canonical ensemble. They are similar to Figs.~\ref{GC_Renyi_r=0.5}, \ref{GC_Entanglement_r=0.5} with $r=0.5$.}
\label{GC_renyi_entangle_other}
\end{figure}
\begin{figure}[htbp]
\centering
\includegraphics[width=14cm]{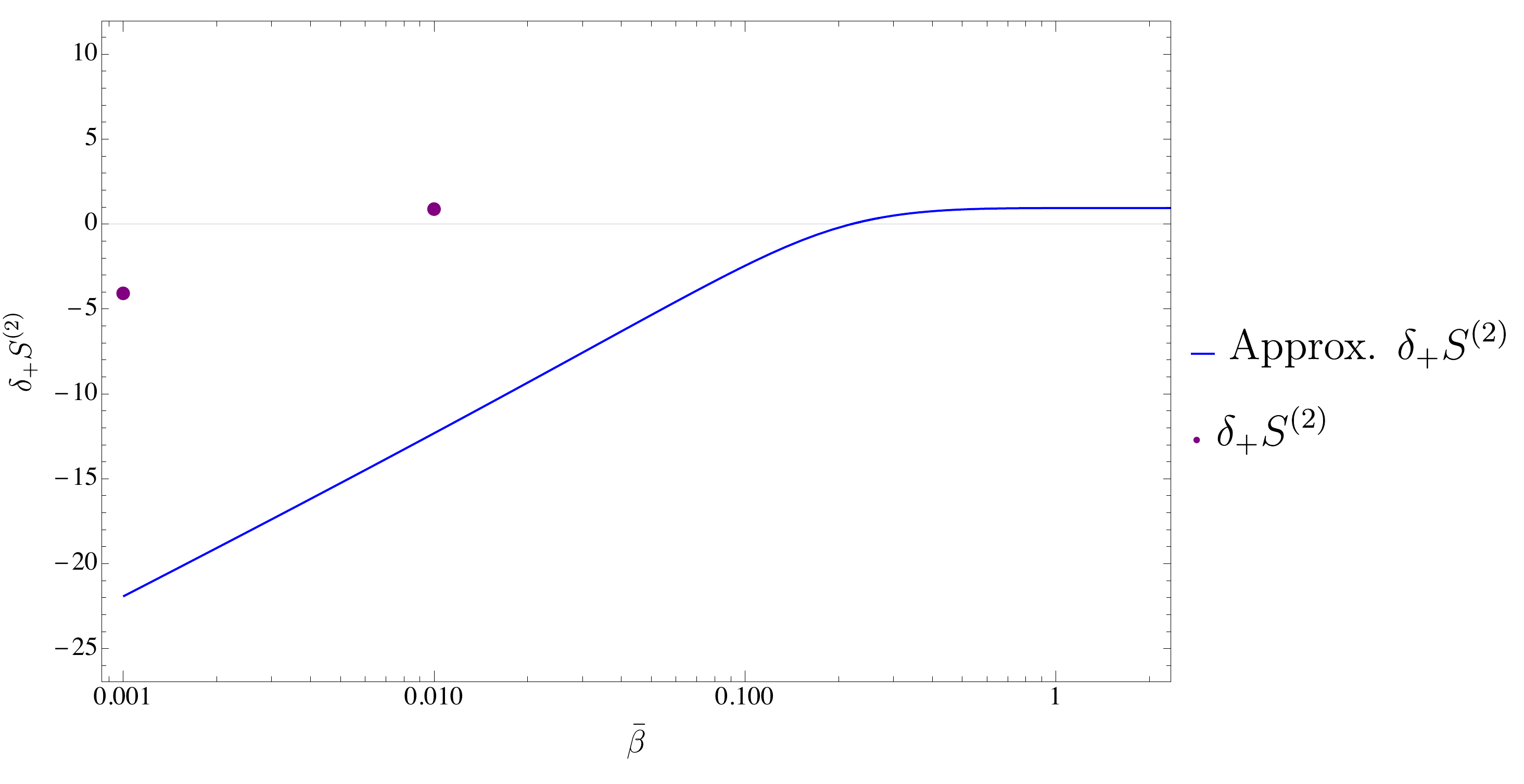}
\caption{$\bar{\beta}$-dependence of $\delta_+ S^{(2)}$ for even $N$. 
It is the plot of $\delta_{+}S^{(2)}=2S_{A(r=0.5)}^{(2)}-S_{th}^{(2)}$ for $N=8$. It is similar to odd $N$ case (Fig.~\ref{Araki-Lieb_N=9}).}
\label{Araki-Lieb_N=8}
\end{figure}
\bibliographystyle{utphys}
\bibliography{ref}

\providecommand{\href}[2]{#2}\begingroup\raggedright\begin{thebibliography}{10}

\bibitem{Sugishita:2021vih}
S.~Sugishita, ``{Target space entanglement in quantum mechanics of fermions and
  matrices},'' \href{http://dx.doi.org/10.1007/JHEP08(2021)046}{{\em JHEP}
  {\bfseries 08} (2021) 046}, \href{http://arxiv.org/abs/2105.13726}{{\ttfamily
  arXiv:2105.13726 [hep-th]}}.

\bibitem{Ryu:2006bv}
S.~Ryu and T.~Takayanagi, ``{Holographic derivation of entanglement entropy
  from AdS/CFT},'' \href{http://dx.doi.org/10.1103/PhysRevLett.96.181602}{{\em
  Phys. Rev. Lett.} {\bfseries 96} (2006) 181602},
  \href{http://arxiv.org/abs/hep-th/0603001}{{\ttfamily arXiv:hep-th/0603001}}.

\bibitem{Ryu:2006ef}
S.~Ryu and T.~Takayanagi, ``{Aspects of Holographic Entanglement Entropy},''
  \href{http://dx.doi.org/10.1088/1126-6708/2006/08/045}{{\em JHEP} {\bfseries
  08} (2006) 045}, \href{http://arxiv.org/abs/hep-th/0605073}{{\ttfamily
  arXiv:hep-th/0605073}}.

\bibitem{Banks:1996vh}
T.~Banks, W.~Fischler, S.~H. Shenker, and L.~Susskind, ``{M theory as a matrix
  model: A Conjecture},''
  \href{http://dx.doi.org/10.1103/PhysRevD.55.5112}{{\em Phys. Rev. D}
  {\bfseries 55} (1997) 5112--5128},
  \href{http://arxiv.org/abs/hep-th/9610043}{{\ttfamily arXiv:hep-th/9610043}}.

\bibitem{Sorkin:2014kta}
R.~D. Sorkin, ``{1983 paper on entanglement entropy: ''On the Entropy of the
  Vacuum outside a Horizon''},'' in {\em {10th International Conference on
  General Relativity and Gravitation}}.
\newblock 1984.
\newblock \href{http://arxiv.org/abs/1402.3589}{{\ttfamily arXiv:1402.3589
  [gr-qc]}}.

\bibitem{Srednicki:1993im}
M.~Srednicki, ``{Entropy and area},''
  \href{http://dx.doi.org/10.1103/PhysRevLett.71.666}{{\em Phys. Rev. Lett.}
  {\bfseries 71} (1993) 666--669},
  \href{http://arxiv.org/abs/hep-th/9303048}{{\ttfamily arXiv:hep-th/9303048}}.

\bibitem{Jacobson:2012yt}
T.~Jacobson, ``{Gravitation and vacuum entanglement entropy},''
  \href{http://dx.doi.org/10.1142/S0218271812420060}{{\em Int. J. Mod. Phys. D}
  {\bfseries 21} (2012) 1242006},
  \href{http://arxiv.org/abs/1204.6349}{{\ttfamily arXiv:1204.6349 [gr-qc]}}.

\bibitem{Bianchi:2012ev}
E.~Bianchi and R.~C. Myers, ``{On the Architecture of Spacetime Geometry},''
  \href{http://dx.doi.org/10.1088/0264-9381/31/21/214002}{{\em Class. Quant.
  Grav.} {\bfseries 31} (2014) 214002},
  \href{http://arxiv.org/abs/1212.5183}{{\ttfamily arXiv:1212.5183 [hep-th]}}.

\bibitem{Myers:2013lva}
R.~C. Myers, R.~Pourhasan, and M.~Smolkin, ``{On Spacetime Entanglement},''
  \href{http://dx.doi.org/10.1007/JHEP06(2013)013}{{\em JHEP} {\bfseries 06}
  (2013) 013}, \href{http://arxiv.org/abs/1304.2030}{{\ttfamily arXiv:1304.2030
  [hep-th]}}.

\bibitem{Balasubramanian:2013lsa}
V.~Balasubramanian, B.~D. Chowdhury, B.~Czech, J.~de~Boer, and M.~P. Heller,
  ``{Bulk curves from boundary data in holography},''
  \href{http://dx.doi.org/10.1103/PhysRevD.89.086004}{{\em Phys. Rev. D}
  {\bfseries 89} no.~8, (2014) 086004},
  \href{http://arxiv.org/abs/1310.4204}{{\ttfamily arXiv:1310.4204 [hep-th]}}.

\bibitem{Myers:2014jia}
R.~C. Myers, J.~Rao, and S.~Sugishita, ``{Holographic Holes in Higher
  Dimensions},'' \href{http://dx.doi.org/10.1007/JHEP06(2014)044}{{\em JHEP}
  {\bfseries 06} (2014) 044}, \href{http://arxiv.org/abs/1403.3416}{{\ttfamily
  arXiv:1403.3416 [hep-th]}}.

\bibitem{Balasubramanian:2014sra}
V.~Balasubramanian, B.~D. Chowdhury, B.~Czech, and J.~de~Boer, ``{Entwinement
  and the emergence of spacetime},''
  \href{http://dx.doi.org/10.1007/JHEP01(2015)048}{{\em JHEP} {\bfseries 01}
  (2015) 048}, \href{http://arxiv.org/abs/1406.5859}{{\ttfamily arXiv:1406.5859
  [hep-th]}}.

\bibitem{Jacobson:2015hqa}
T.~Jacobson, ``{Entanglement Equilibrium and the Einstein Equation},''
  \href{http://dx.doi.org/10.1103/PhysRevLett.116.201101}{{\em Phys. Rev.
  Lett.} {\bfseries 116} no.~20, (2016) 201101},
  \href{http://arxiv.org/abs/1505.04753}{{\ttfamily arXiv:1505.04753 [gr-qc]}}.

\bibitem{Anous:2019rqb}
T.~Anous, J.~L. Karczmarek, E.~Mintun, M.~Van~Raamsdonk, and B.~Way, ``{Areas
  and entropies in BFSS/gravity duality},''
  \href{http://dx.doi.org/10.21468/SciPostPhys.8.4.057}{{\em SciPost Phys.}
  {\bfseries 8} no.~4, (2020) 057},
  \href{http://arxiv.org/abs/1911.11145}{{\ttfamily arXiv:1911.11145
  [hep-th]}}.

\bibitem{Mazenc:2019ety}
E.~A. Mazenc and D.~Ranard, ``{Target Space Entanglement Entropy},''
  \href{http://arxiv.org/abs/1910.07449}{{\ttfamily arXiv:1910.07449
  [hep-th]}}.

\bibitem{Das:2020jhy}
S.~R. Das, A.~Kaushal, G.~Mandal, and S.~P. Trivedi, ``{Bulk Entanglement
  Entropy and Matrices},''
  \href{http://dx.doi.org/10.1088/1751-8121/abafe4}{{\em J. Phys. A} {\bfseries
  53} no.~44, (2020) 444002}, \href{http://arxiv.org/abs/2004.00613}{{\ttfamily
  arXiv:2004.00613 [hep-th]}}.

\bibitem{Das:2020xoa}
S.~R. Das, A.~Kaushal, S.~Liu, G.~Mandal, and S.~P. Trivedi, ``{Gauge invariant
  target space entanglement in D-brane holography},''
  \href{http://dx.doi.org/10.1007/JHEP04(2021)225}{{\em JHEP} {\bfseries 04}
  (2021) 225}, \href{http://arxiv.org/abs/2011.13857}{{\ttfamily
  arXiv:2011.13857 [hep-th]}}.

\bibitem{Hampapura:2020hfg}
H.~R. Hampapura, J.~Harper, and A.~Lawrence, ``{Target space entanglement in
  Matrix Models},'' \href{http://dx.doi.org/10.1007/JHEP10(2021)231}{{\em JHEP}
  {\bfseries 10} (2021) 231}, \href{http://arxiv.org/abs/2012.15683}{{\ttfamily
  arXiv:2012.15683 [hep-th]}}.

\bibitem{Frenkel:2021yql}
A.~Frenkel and S.~A. Hartnoll, ``{Entanglement in the Quantum Hall Matrix
  Model},'' \href{http://dx.doi.org/10.1007/JHEP05(2022)130}{{\em JHEP}
  {\bfseries 05} (2022) 130}, \href{http://arxiv.org/abs/2111.05967}{{\ttfamily
  arXiv:2111.05967 [hep-th]}}.

\bibitem{Tsuchiya:2022ffu}
A.~Tsuchiya and K.~Yamashiro, ``{Target space entanglement in a matrix model
  for the bubbling geometry},''
  \href{http://dx.doi.org/10.1007/JHEP04(2022)086}{{\em JHEP} {\bfseries 04}
  (2022) 086}, \href{http://arxiv.org/abs/2201.06871}{{\ttfamily
  arXiv:2201.06871 [hep-th]}}.

\bibitem{Das:2022mtb}
S.~R. Das, S.~Hampton, and S.~Liu, ``{Entanglement entropy and phase space
  density: lowest Landau levels and 1/2 BPS states},''
  \href{http://dx.doi.org/10.1007/JHEP06(2022)046}{{\em JHEP} {\bfseries 06}
  (2022) 046}, \href{http://arxiv.org/abs/2201.08330}{{\ttfamily
  arXiv:2201.08330 [hep-th]}}.

\bibitem{Gautam:2022akq}
V.~Gautam, M.~Hanada, A.~Jevicki, and C.~Peng, ``{Matrix Entanglement},''
  \href{http://arxiv.org/abs/2204.06472}{{\ttfamily arXiv:2204.06472
  [hep-th]}}.

\bibitem{Klebanov:1991qa}
I.~R. Klebanov, ``{String theory in two-dimensions},'' in {\em {Spring School
  on String Theory and Quantum Gravity (to be followed by Workshop)}},
  pp.~30--101.
\newblock 7, 1991.
\newblock \href{http://arxiv.org/abs/hep-th/9108019}{{\ttfamily
  arXiv:hep-th/9108019}}.

\bibitem{Polchinski:1994mb}
J.~Polchinski, ``{What is string theory?},'' in {\em {NATO Advanced Study
  Institute: Les Houches Summer School, Session 62: Fluctuating Geometries in
  Statistical Mechanics and Field Theory}}.
\newblock 11, 1994.
\newblock \href{http://arxiv.org/abs/hep-th/9411028}{{\ttfamily
  arXiv:hep-th/9411028}}.

\bibitem{Das:1995vj}
S.~R. Das, ``{Geometric entropy of nonrelativistic fermions and two-dimensional
  strings},'' \href{http://dx.doi.org/10.1103/PhysRevD.51.6901}{{\em Phys. Rev.
  D} {\bfseries 51} (1995) 6901--6908},
  \href{http://arxiv.org/abs/hep-th/9501090}{{\ttfamily arXiv:hep-th/9501090}}.

\bibitem{Hartnoll:2015fca}
S.~A. Hartnoll and E.~Mazenc, ``{Entanglement entropy in two dimensional string
  theory},'' \href{http://dx.doi.org/10.1103/PhysRevLett.115.121602}{{\em Phys.
  Rev. Lett.} {\bfseries 115} no.~12, (2015) 121602},
  \href{http://arxiv.org/abs/1504.07985}{{\ttfamily arXiv:1504.07985
  [hep-th]}}.

\bibitem{Dean:2016sug}
D.~S. Dean, P.~Le~Doussal, S.~N. Majumdar, and G.~Schehr, ``{Noninteracting
  fermions at finite temperature in a $d$-dimensional trap: Universal
  correlations},'' \href{http://dx.doi.org/10.1103/PhysRevA.94.063622}{{\em
  Phys. Rev. A} {\bfseries 94} no.~6, (2016) 063622},
  \href{http://arxiv.org/abs/1609.04366}{{\ttfamily arXiv:1609.04366
  [cond-mat.stat-mech]}}.

\bibitem{Klich:2004pb}
I.~Klich, ``{Lower entropy bounds and particle number fluctuations in a Fermi
  sea},'' \href{http://dx.doi.org/10.1088/0305-4470/39/4/L02}{{\em J. Phys. A}
  {\bfseries 39} (2006) L85--L92},
  \href{http://arxiv.org/abs/quant-ph/0406068}{{\ttfamily
  arXiv:quant-ph/0406068}}.

\bibitem{klich2008scaling}
I.~Klich and L.~Levitov, ``Scaling of entanglement entropy and superselection
  rules,'' \href{http://arxiv.org/abs/0812.0006}{{\ttfamily arXiv:0812.0006
  [quant-ph]}}.

\bibitem{Calabrese:2011zzb}
P.~Calabrese, M.~Mintchev, and E.~Vicari, ``{The entanglement entropy of
  one-dimensional gases},''
  \href{http://dx.doi.org/10.1103/PhysRevLett.107.020601}{{\em Phys. Rev.
  Lett.} {\bfseries 107} (2011) 020601},
  \href{http://arxiv.org/abs/1105.4756}{{\ttfamily arXiv:1105.4756
  [cond-mat.stat-mech]}}.

\bibitem{Calabrese:2011vh}
P.~Calabrese, M.~Mintchev, and E.~Vicari, ``{The Entanglement entropy of 1D
  systems in continuous and homogenous space},''
  \href{http://dx.doi.org/10.1088/1742-5468/2011/09/P09028}{{\em J. Stat.
  Mech.} {\bfseries 1109} (2011) P09028},
  \href{http://arxiv.org/abs/1107.3985}{{\ttfamily arXiv:1107.3985
  [cond-mat.stat-mech]}}.

\bibitem{Song:2011gv}
H.~F. Song, S.~Rachel, C.~Flindt, I.~Klich, N.~Laflorencie, and K.~Le~Hur,
  ``{Bipartite Fluctuations as a Probe of Many-Body Entanglement},''
  \href{http://dx.doi.org/10.1103/PhysRevB.85.035409}{{\em Phys. Rev. B}
  {\bfseries 85} (2012) 035409},
  \href{http://arxiv.org/abs/1109.1001}{{\ttfamily arXiv:1109.1001
  [cond-mat.mes-hall]}}.

\bibitem{Mintchev:2022xqh}
M.~Mintchev, D.~Pontello, A.~Sartori, and E.~Tonni, ``{Entanglement entropies
  of an interval in the free Schr\"odinger field theory at finite density},''
  \href{http://arxiv.org/abs/2201.04522}{{\ttfamily arXiv:2201.04522
  [hep-th]}}.

\bibitem{Mintchev:2022yuo}
M.~Mintchev, D.~Pontello, and E.~Tonni, ``{Entanglement entropies of an
  interval in the free Schr\"odinger field theory on the half line},''
  \href{http://arxiv.org/abs/2206.06187}{{\ttfamily arXiv:2206.06187
  [hep-th]}}.

\bibitem{peschel2003calculation}
I.~Peschel, ``Calculation of reduced density matrices from correlation
  functions,'' {\em Journal of Physics A: Mathematical and General} {\bfseries
  36} no.~14, (2003) L205,
  \href{http://arxiv.org/abs/cond-mat/0212631}{{\ttfamily
  arXiv:cond-mat/0212631}}.

\bibitem{Casini:2009sr}
H.~Casini and M.~Huerta, ``{Entanglement entropy in free quantum field
  theory},'' \href{http://dx.doi.org/10.1088/1751-8113/42/50/504007}{{\em J.
  Phys. A} {\bfseries 42} (2009) 504007},
  \href{http://arxiv.org/abs/0905.2562}{{\ttfamily arXiv:0905.2562 [hep-th]}}.

\bibitem{Araki:1970ba}
H.~Araki and E.~H. Lieb, ``{Entropy inequalities},''
  \href{http://dx.doi.org/10.1007/BF01646092}{{\em Commun. Math. Phys.}
  {\bfseries 18} (1970) 160--170}.

\bibitem{Adesso:2012ni}
G.~Adesso, D.~Girolami, and A.~Serafini, ``{Measuring Gaussian quantum
  information and correlations using the R\'enyi entropy of order 2},''
  \href{http://dx.doi.org/10.1103/PhysRevLett.109.190502}{{\em Phys. Rev.
  Lett.} {\bfseries 109} no.~19, (2012) 190502},
  \href{http://arxiv.org/abs/1203.5116}{{\ttfamily arXiv:1203.5116
  [quant-ph]}}.

\bibitem{Das:2019qaj}
S.~R. Das, S.~Hampton, and S.~Liu, ``{Quantum Quench in Non-relativistic
  Fermionic Field Theory: Harmonic traps and 2d String Theory},''
  \href{http://dx.doi.org/10.1007/JHEP08(2019)176}{{\em JHEP} {\bfseries 08}
  (2019) 176}, \href{http://arxiv.org/abs/1903.07682}{{\ttfamily
  arXiv:1903.07682 [hep-th]}}.

\end{thebibliography}\endgroup
\end{document}